\documentclass[12pt,preprint]{aastex} 
 
\usepackage{lscape} 
\usepackage{epstopdf} 
 
\shorttitle{The V471 Tauri System} 
 
\shortauthors{T.R. Vaccaro, R. E. Wilson, W. Van Hamme, \& D. Terrell} 
 
\begin{document} 
 
\title{The V471 Tauri System: A Multi-datatype Probe} 
 
\author{T. R. Vaccaro}  
\affil{Physics Department, St. Cloud State University, St. Cloud, MN 56301} \email{trvaccaro@gmail.com} 
 
\author{R. E. Wilson}
\affil{Astronomy Department, University of Florida, Gainesville, FL 32611 and Astronomy Department, Indiana University, Swain Hall West,
Bloomington, IN 47405} \email{rewilson@ufl.edu}

\author{W. Van Hamme} 
\affil{Department of Physics, Florida 
International University, Miami, FL 33199} \email{vanhamme@fiu.edu} 
 
\and 
 
\author{D. Terrell} 
\affil{Department of Space Studies, Southwest Research Institute, 1050 Walnut Street, Suite 300, Boulder, CO 80302, USA} \email{terrell@boulder.swri.edu} 
 
\begin{abstract} 

V471 Tauri, a white dwarf -- red dwarf eclipsing binary in the Hyades,
is well known for stimulating development of common envelope theory, whereby
novae and other cataclysmic variables form from much wider binaries by catastrophic
orbit shrinkage. Our evaluation of a recent imaging search that
reported negative results for a much postulated third body
shows that the object could
have escaped detection or may have actually been seen.
The balance of evidence continues to favor a brown dwarf companion
about 12 AU from the eclipsing binary.
A recently developed algorithm finds unified solutions from three datatypes.
New radial velocities (RVs) of the red dwarf and $BVR_CI_C$ light curves are
solved simultaneously along with white dwarf and red dwarf RVs from the literature, $uvby$ data,
the MOST mission light curve, and 40 years of eclipse timings. Precision-based weighting is the key
to proper information balance among the various datasets. Timewise variation of modeled starspots
allows unified solution of multiple data eras.
Light curve amplitudes strongly suggest decreasing spottedness
from 1976 to about 1980, followed by approximately constant spot coverage from 1981 to 2005.
An explanation is proposed for lack of noticeable variation in 1981 light curves, in terms of competition
between spot and tidal variations.
Photometric-spectroscopic distance is estimated.
The red dwarf mass comes out larger than normal for a K2\,V star, and even larger than
adopted in several structure and evolution papers. An identified cause for this
result is that much improved red dwarf RVs curves now exist. 

\end{abstract} 

 
\section{Introduction} 
 
V471~Tau, a white dwarf-red dwarf eclipsing binary (EB) in the Hyades with orbit period $0^d.52118$, 
is primarily known for its unique historical role as a stimulus to common 
envelope evolution theory \citep{chau,refsdal,sparks,ostriker,paczynski,alexander,taam}. 
Other properties include a likely brown dwarf companion to the EB, measured white dwarf 
spin, mass loss and exchange in a detached binary,  
differential rotation measured via magnetic spots, spot distributions, accurate white dwarf parameters, 
and photometric-spectroscopic distance measures that help to pin down the binary's 
location within the Hyades. Publications on its properties and 
evolution include the discovery paper and follow-up \citep{nelson70,nelson76}, 
along with many that form background for this paper.  
\citet[][hereafter OBBS]{obrien} 
have interesting discussions of evolutionary 
possibilities, including ideas that would involve a third star. Reasonable 
space limitations restrict our discussions to those most directly 
related to the putative third object's orbital properties, 
along with relative and absolute properties of the EB. 
Issues beyond the scope of this paper include radio eclipses \citep{lim, nicholls}, 
red dwarf atmospheric structure \citep{kim,walter}, 
chemical abundances \citep{martin,still,garcia,shimansky}, a circumbinary shell
\citep{sion89}, red dwarf wind \citep{mullan}, and flares \citep{young83,tunca93}. 

Physical/geometric quantities are derived 
here by simultaneous analysis of radial velocity (RV) curves, differential and absolute light 
curves,\footnote{Here `absolute' refers to observed and computed 
fluxes being in standard physical units. The definition extends to light 
curves that can reliably be converted to absolute flux, such as those in 
accurately calibrated standard magnitudes.} 
and eclipse timings by a unified multi-datatype algorithm \citep{wilson14}
that allows all of the main sources of orbital, stellar, and timing information to be analyzed 
coherently and is built around a general binary star model.
 
V471~Tau has been widely believed to show a 
subtle light-time variation due to the EB's motion about the barycenter of a postulated 
triple system (hereafter 3b system). The semi-amplitude is about 2.3 minutes over an 
orbital period of 30+ years. \citet{guinan} analyzed the variation from 163 eclipse 
timings, finding that the third object's mass corresponds to a brown dwarf 
for inclinations ($i_{3b}$) above about $35\degr$. Here 
the presumed 3b companion's properties are estimated, based on three types of data solved 
simultaneously in several combinations and also separately -- six RV curves, nine light curves, and 224 
eclipse timings. Results of our analysis together with a history of interpretations of timing residual excursions 
are in \S\ref{3body}.

Hierarchical multiple systems are fairly common despite the difficulties of discovery due to long periods,
small light-time amplitudes, and faintness of companions \citep{pribulla,rucin07}.
\citet{guinan} proposed (their Section~5) to observe the EB's astrometric
orbit due to its reflex motion about a triple system barycenter (thus measuring the
outer orbit's inclination and the 3b system's dynamical mass), and also proposed coronographic imaging to
detect the 3b. They included evolution-based estimates of the suspected brown dwarf's magnitude
and orbit-based estimates of its maximum angular distance from the EB in their text, and their Table 2 and Figure 2.
Although the astrometry has not yet begun, an adaptive optics (AO) image from December 11, 2014 by \citet{hardy}
has led to a claim that the 3b's existence has been disproved. 
Our analyses indicate that the 3b has several ways to have avoided detection. 
We estimate significantly fainter minimum brightness
for the 3b than did \citeauthor{hardy}, with our result based on more extensive data than
previously accessed in any V471 Tau analyses. Specifics are discussed in \S\ref{reality}.
We also show that its expected location is closer to the AO image center (i.e. closer to the EB) than
estimated by \citet{hardy}. At least two features at the revised angular distance, in their image of the field, could be 
candidates for the long sought object. 

\section{Observational Input} 
\label{sec-lcs} 
 
RVs of the red dwarf from our 1998 Kitt Peak 
National Observatory (KPNO) observations made with the 0.9-m Coud\'{e} Feed of the 2.1-m telescope
are in Table~\ref{tbl-v471rv}. Their characteristics 
(instrument, integration times, etc.) are in \citet{vaccaro}. 
Other red star RVs by \citet{young76}, \citet{bois88}, \citet{hussain},
and \citet{kaminski}\footnote{
The HJED times in Table~1 of \citet{kaminski} have mid-exposure corrections that are full 
rather than the appropriate half integration times, according to K. Kaminski, who kindly provided the properly corrected times used in this paper.}
also were utilized, along with 
white dwarf RVs by OBBS. The five sets of red star RVs are in approximate mutual agreement 
to visual inspection within their scatter bands. Four of the twelve white dwarf RVs from the 
Hubble Space Telescope tabulated by OBBS 
preceded NASA's Corrective Optics Space Telescope Axial Replacement (COSTAR) 
repair mission. Here the eight OBBS post-COSTAR RVs were used, but not the four pre-COSTAR RVs (see the OBBS paper
for comments on the relative merits of their pre and post COSTAR data). 

The white dwarf Si IV RVs illustrated in Figure 6 of \citet{sion} 
were not fully analyzed here, partly because the physical situation in
which the lines are formed is excessively complicated, and partly because
the RV's lead to an implausible red star mass. The physical intricacy
is made clear by \citeauthor{sion}'s Section 4 that discusses irregular
surface distributions combined with line strength modulations
on the 555 s. white dwarf rotation period, as well as blending of absorption
and emission line components and blending of unresolved Zeeman components.
Funneling of accretion flows into the magnetic poles also can be expected.
Accordingly the measured RVs are not likely to be representative of the
white dwarf surface as a whole, although they may be formed on or near the surface.
Measurements on \citeauthor{sion}'s Figure 6 show that the white dwarf RV 
amplitude is 28 percent larger than in OBBS, necessarily leading to a greater
binary system mass according to Kepler's Third Law, and to a greater
mass ratio ($M_2/M_1$) so that the red star's share of the total mass
is increased over that in OBBS. A Least Squares solution, based on the figure's
dots and all red star RVs, quantified this expectation by finding a red star mass of
$1.68 M_\sun$, which we do not consider realistic for a K2 V star.
Although the idea of
using these lines was resourceful, the implausible RV amplitude probably
indicates that the local physics is too uncertain for extraction of 
reliable masses.
 
Our 1998 $BVR_CI_C$ differential light curves with respect to comparison star BD\,+16 515, observed at KPNO
with the 0.9-m Southeastern Association for Research in Astronomy (SARA) telescope and described in \citet{vaccaro}, are in 
Table~\ref{tbl-BVRI}. Means and standard deviations of 31 newly observed $BVR_CI_C$ magnitudes of BD\,+16 515
are in Table \ref{tbl-comparison}. 
Band to band changes of 
eclipse depth are dramatic, as seen in Figures~\ref{v471_u} to \ref{v471_MOST}. Also entered were 
$uvby$ light curves by \citet{rucinski81} and an unusually extensive one from the MOST 
mission \citep{kaminski}. 
In order to keep (iterated) solution run times within reasonable limits,
normal points of the 57,000+ MOST light curve points were made by
averaging times and magnitudes in groups of 10 points, except for those in the steep 
descending and ascending parts of eclipses where no averaging was done. 
This procedure resulted in the 6001 normal points used here and shown in Figure~\ref{v471_MOST}.
Standard errors of averages were calculated from standard errors of individual observations. 
Normal point weights (inversely proportional to normal point standard errors squared) were 
used as individual weights in all our solutions
that included the MOST light curve.
The MOST differential magnitudes had 9.5 mag added (by us) to simulate actual 
magnitudes, which improved initial solution 
convergence. The additive constant does not affect results since solutions are given with
luminosity ratios, $L_1/(L_1+L_2)$, rather than $L_1$ and $L_2$. 
 
Eclipse minima (total 224) from many observers were entered into our unified 
light-RV-timing solutions, most having been collected and tabulated by 
\citet{ibanoglu05}. A few that are not in the \citeauthor{ibanoglu05} 
online table are in Table \ref{tbl-times}, including four recent ones by \citet{hardy}. 
 
To derive ephemerides on a steady time system, and following a suggestion by 
\citet{bastien}, \citet{guinan} consistently adopted HJED, an astronomical heliocentric 
version of Universal Atomic Time, \footnote{See Sections B6 and B7 of the Astronomical 
Almanac for the Year 2013 \citep{gallaudet}.} and later authors
of V471~Tau timing analyses have also done so. Differences between HJED and 
Coordinated Universal Time (UTC), which essentially tracks Earth's 
decelerating rotation, have accumulated to somewhat over a minute during the 
last four decades and obviously need to be considered in investigations of a 
2.3 minute effect. All times for observations utilized in this paper have 
been converted to HJED or were originally reported in HJED.

\section{Procedural Overview} 
 
Unified binary system solutions, as the term is used here, simultaneously 
process RVs, multiband light curves, and eclipse timings so as to treat stellar, 
orbital, and ephemeris parameters coherently in one conceptual  
step.\footnote{If necessary, the full parameter set can be broken into 
subsets to handle discontinuous period changes, although that 
was not done here.} The solutions are by the well known 
method of differential corrections (DC), as implemented in \citet{wilson71}  
and revised several times, most recently as described in \citet{wilson14}.  
Such solutions allow all considered 
astrophysical effects to be handled without personal intervention, except 
for choice of starting estimates. 
Although the development in \citet{wilson14} 
emphasizes the orbit period ($P$), a reference time, and other timing-related 
quantities 
($dP/dt$, $d\omega/dt$,\footnote{Parameter $d\omega/dt$ is the rotation rate of an 
elliptical orbit within its own plane.} and light-time effect parameters), any chosen model 
parameters can be processed as a coherent set 
with optional automated data weighting. Table~\ref{tbl-sets} summarizes the datasets
used in this paper and includes the time window for each.
 
\subsection{Data Weighting Basics} \label{datawt}
 
The essentials of light curve weighting are in \citet{wilson79}, including 
curve-dependent, level-dependent, and individual data point weights. 
Weighting is important in simultaneous solutions of multiple datasets, 
especially when several data types are included, but can be 
somewhat tedious if done by personal intervention, so it varies among 
publications and often is not mentioned. Without proper weighting the 
weakest observations are likely to have the most influence. RV 
weighting and eclipse timing weighting are not level-dependent but otherwise 
are done in the same way as light curve weighting.
Only weight ratios matter among the points of a given data subset 
such as a light or RV curve, since curve-dependent weights are computed 
subsequently by DC, taking account of individual weights. Accordingly the scaling factor for 
individual weights is arbitrary and was set to unity. 
Level-dependent weights are
generated within DC from applicable statistics that were assumed to be photon counting 
statistics for the light curves. 

\subsection{Curve-dependent Weights for V471~Tau}

Iterative automation of curve-dependent weights is briefly outlined 
and applied to several EBs in \citet{wilson14}, and full explanation is in the
documentation monograph that accompanies the DC program's most recent version.\footnote{
Available at \url{ftp://ftp.astro.ufl.edu/pub/wilson/lcdc2013/}} 
Automated weighting saves an operational step for each input curve and also eliminates
a class of possible mistakes, such as entry of a $\sigma$ from individual data
points when averages of points were the actual input.
Our intention was to apply automated curve-dependent weights in the V471~Tau solutions.
However it became clear that stretches of the
light curves show systematic deviations from modeled curves
due to unknown, possibly transient effects. These anomalous intervals lead to
oversize $\sigma$'s that can significantly misrepresent a curve's precision. 
Accordingly weights were not based on automated curve-dependent $\sigma$'s,
but on fixed $\sigma$'s (Table~\ref{tbl-sigmas}) found separately for each light curve by
Fourier series fits (up to $7\theta$), excluding points near eclipse of the white dwarf.  
These standard deviations served as the (fixed) curve-dependent light curve $\sigma$'s for all solutions.
For the RVs and timings, the $\sigma$'s are from RV-only and timing-only DC solutions, respectively.

\subsection{Treatment of Magnetic Spot Growth and Decay} 
 \label{spots}

At most epochs, spottedness is the largest cause of V471~Tau's 
brightness variation for wavelengths longer than about 0.4 microns and  
cannot be ignored, although the red dwarf's tidal variation is significant, with reflection being a much 
smaller effect. Eclipse depths increase strongly toward short wavelengths
and are the largest cause of variation in the ultraviolet, although of course restricted 
to a small phase range, leaving variation due to spots usually being dominant over the rest of the cycle.
Dark magnetic spots are common on stars that have well 
developed outer convection zones, and especially on rapid rotators. Because 
very close binary components typically have been tidally dragged into 
co-rotation with the (fast) orbital motion, spot activity due to magnetic 
fields that are generated by convective motions and advective flows can be far more 
extensive on short period binaries than on the Sun. The spots may densely 
cover large regions, although not being individually much larger than their 
solar counterparts \citep{rucinski79}.  

Because the spots come and go, they can 
be problematic for light curve solutions that cover long stretches of time. 
Onset and disappearance times may need to be considered carefully where datasets 
extend for long times and where they overlap or nearly overlap in time. The 
conceptual spots\footnote{Here the expression 
``conceptual spots'' refers to the option of regarding a model spot as 
representing either an individual spot or a spotted area.} can be varied in area
by setting start and stop times and assigning times of maximum size.  
Even where input data eras are well separated in time, the 
start--stop facility for spots is essential for allowing individual curves 
to be fitted within a common overall dataset (i.e. since real spots come and go, 
model spots must come and go). Considerable experimentation 
may be needed to attain approximate representations of spots or 
spotted regions. If the trial experiments essentially succeed, an impersonal  
algorithm can refine locations and maximum sizes, as well as time 
markers for growth and decay. 
V471 Tau solutions require parameters for many conceptual spots in addition to the other parameters 
(3 spots per epoch $\times$ 3 epochs = 9 adjusted spots), yet
iterations involving parameter subsets ensured good convergence.
A recent scheme for improved accuracy in spot 
representation \citep{wilson12} was applied. 
In the interest of simplicity, the modeled timewise 
profile is, in general, an asymmetric trapezoid, with a symmetric trapezoid 
or triangle as special cases. Results are in Table~\ref{tbl-spots}. 
 
Doppler imaging, also known as eclipse mapping \citep[e.g.][]{vogt,vogt87} 
is based on spectral line profiles and 
gives far more detailed spot information than the light curves 
now at hand, so one cannot expect to learn much about V471~Tau's 
spot behavior beyond that in \citet{ramseyer} and \citet{hussain}. However, 
photometry is more readily done in quantity than is spectroscopy, so its 
usefulness for spots could improve significantly if a tradition developed 
whereby light curves were published rather than only illustrated. 
About 15 other whole light curves (counting multi-band data 
observed together as one curve) have been illustrated in papers and could 
have helped to fill timewise gaps if they had been published.\footnote{See 
comments on the importance of 
publishing observations in useful form by D. Helfand, posted at 
\url{http://aas.org/posts/news/2013/06/presidents-column-making-excellent-journals-even-better}.}
There was  some success in attempts to obtain 
such observations that added one RV curve and one light curve to our 
database (see acknowledgments). 

The parameters for a given spot are: 
\begin{enumerate}
\item spot center `latitude' ($0$ at the +z pole, $\pi$ radians at the $-$z pole);  
\item spot center longitude ($0$ to $2\pi$ radians, starting from the binary system line 
of centers, increasing counter-clockwise as seen from above the +z pole); 
\item spot angular radius at the time of maximum size (in radians), defined at the star center; 
\item ratio of local spot temperature to underlying (i.e. no-spot) local 
temperature; 
\item spot appearance time (and start time of linear growth in area);  
\item ending time of area growth (and start time of constant area); 
\item ending time of constant area (and start time of linear area decay); 
\item spot disappearance time;  
\item ratio of spot angular drift rate to mean orbital angular revolution rate on star 1; and
\item ratio of spot angular drift rate to mean orbital angular revolution rate on star 2.
\end{enumerate}
The last two parameters were set to unity so the spots stay at fixed 
longitudes on the synchronously rotating red dwarf.
The start and ending times for constant spot area were taken to coincide with the 
start and ending times of the corresponding light curves.
 
\subsection{Fractional Spot Coverage over Recent Decades}

The last column of Table~\ref{tbl-spots} is fractional spot area for the entire surface, 
which declined from 13\% for the 1976 \citet{rucinski81}
observations, to 10\% for our 1998 KPNO observations, and to 1.7\% for the 2005 \citet{kaminski} MOST observations.
The red dwarf may have cyclic spot activity similar to the Sun's, although neither
periodicity nor timewise trends can be established from just three epochs. Naturally periodic spot behavior could 
cause cyclic variations of small amplitude in the eclipse timing residuals (see \S\ref{otherper}). 
So is there really a trend in total spot coverage or is too much being read from just three points?
If there is a trend, is its form simple (say linear or bi-linear) or more complicated?
A natural plan is to examine light curve amplitude vs. time, which requires many more epochs
than the three for which digital light curves are available, with amplitude as a rough proxy for total spot coverage. 
All of the light curve papers that lacked 
the actual observations \textit{did} have light curve plots whose amplitudes could be measured graphically.
The references are \citet{cester}, \citet{tunca79}, \citet{skillman}, \citet{tunca93}, \citet{ibanoglu05}, and \citet{miranda}.
There are more light curve epochs than papers because some authors observed at more than one epoch.
Of course amplitudes from papers that did contain or had online digital light 
curves \citep{rucinski81,kaminski}, and from this paper, also were measured. 
Utilization of those light curves involved computation of
theoretical light curves corresponding to the spot parameters in Table~\ref{tbl-spots} and 
reading of their amplitudes, which are plotted along with the graphically determined amplitudes in Figure~\ref{ampl}. 
The left panel has only bands that are close to $V$ in effective wavelength ($V$, $y$, and MOST) -- those most abundant in the literature -- while the right panel has  
all bands. For the left panel there should be little concern with possible band dependence of amplitude since
the MOST effective wavelength is near 5250\,\AA, so not very far from $V$ and $y$.
A related issue concerns how strongly amplitude varies from band to band. The theoretical light curves
mentioned above show that the band dependence is modest compared to observational scatter in the
observed curves. The right panel, with amplitudes for all reasonably standard bands,
helps by having 31 observed and 9 analytic points compared to the 17 observed and
3 analytic points of the left panel. The overall inference reinforces the 
trend seen in spot coverage, which is represented by only 3 points,
and shows that it is not a linear trend by filling timewise gaps. Instead an interval of fast 
decline is followed by one of nearly constant amplitude. One must keep in mind that the variation is not
entirely due to spots but has a significant tidal component. Note that the computational points at a given 
epoch fall rather close to corresponding graphical points, giving confidence in consistency of the overall process.

\subsection{Curious October 1981 Light Curves: a Proposed Explanation} \label{curious}

Figures~\ref{rucin_b_y} and \ref{rucin_u_v} illustrate October 1981 $uvby$ light curves of V471~Tau \citep{rucinski83} 
that look basically level and flat. Although only 11 points are in each band, the 
points are about as well distributed in phase as can be expected and
show little or no variation except for one point in each band that lies within the eclipse of the white dwarf.
Rucinski's Figure 1 shows, in addition to $y$ band points, an idealized rendition of a $y$ light
curve from 5 years earlier that has an amplitude of about 0.20 mag. The substantial variation seen in the 1976
observations obviously disappeared in 1981. 
Can the absence of variation in 1981 be due to absence of spots? One must consider that tidal (i.e. ellipsoidal) variation 
of the red dwarf is surely present since, with the system parameters (star masses, sizes, etc.) being well
known, the tidal variation can be computed reliably from well established equipotential theory.  
Computations show the tidal amplitude to be of order 0.1 mag, although band-dependent, which should readily be seen in 
the light curves, so complete absence of spots should leave a pure tidal variation, not an essentially
flat curve. Furthermore, given that spots had been the dominant cause of brightness variation, from system discovery until about 1980, 
their complete vanishing would be a major surprise. No suggestion of a reason for the 1981 curves'
lack of variation seems to have been published, but there is a simple possibility -- that the surface spot distribution was just that
needed to cancel the ellipsoidal effect. A few numerical trials, followed by a simultaneous $uvby$ solution of 
the 44 \citet{rucinski83} points quickly found a spot configuration that approximately satisfies our conjecture.
However, the actual $uvby$ fits showed a somewhat complicated result whereby the impersonal Least Squares
solution exploited phase gaps (again -- only 11 points per band) to weave among the dots with appreciable
non-zero variation. Accordingly the apparent lack of variation in 1981 seems mainly due to a temporary spot distribution
that nearly cancels ellipsoidal variation, combined with overall low spottedness. A contributing factor is the small
number of data points, with inevitable phase gaps that make eye-assessments difficult. 
The analytic $y$ band amplitude is $\approx 0.04$\,mag and a corresponding point
is in both panels of Figure~\ref{ampl} at JD 2444892.
Figures~\ref{rucin_b_y} and \ref{rucin_u_v} also contain the fitted curves and theoretical pure tidal curves (i.e. without spots).

\section{Solutions: Character and Outcome} 

Only about $10^{-4}$ of the red dwarf's face can be eclipsed by the 
white dwarf so only one eclipse per orbit cycle is realistically observable 
-- a nearly rectangular notch as the very hot white dwarf is occulted by the 
red star. The notch is increasingly prominent toward short wavelengths and 
nearly disappears in the red and infra-red bands. The partial phases last only  
$\approx 1\%$ of an eclipse so, even at the best time resolution to date, 
they measure little more than the length of the projected chord traversed by 
the partially occulted white dwarf. A consequence is near-degeneracy between 
the dimensionless white dwarf radius, $R_1/a$, and EB orbit inclination, $i$. Although the 
solutions directly or indirectly accessed $\approx60,000$ light curve 
points, the radii remain a problem due to the nearly rectangular 
eclipse form that compromises the inclination's accuracy, and thereby that 
of $R_1/a$ and $R_2/a$. 
For given $i$, the partial phase duration essentially measures  
$(R_2-R_1)/a$ while the full
eclipse width, again for given $i$, 
basically measures the sum of the relative radii, $(R_1+R_2)/a$. 
The sum and difference together determine $R_1/a$ and $R_2/a$. Because the ratio 
$R_1/R_2$ is very small (of order 0.01), the eclipse duration is 
affected only slightly by $R_1/a$ so the white dwarf radius can be nearly 
lost in the noise. 
Actual radii $R_1$ and $R_2$ are further affected by the $a$ vs. $i$ 
correlation. The red star's slight tidal and rotational distortions are 
properly computed and should not cause significant uncertainty. 
In the 3b orbit, only the product $a_{3b}\sin i_{3b}$ is measurable, not $a_{3b}$ and $i_{3b}$ separately, since none of the 
input data can distinguish outer orbit inclination effects from those of 
outer orbit size.  
 
\subsection{Solution Results} \label{numcon}

Both absolute \citep{wilson08} and traditional solutions were carried out. Briefly the distinction
is that absolute solutions have their observed and computed fluxes in standard physical 
units. Naturally only absolute solutions can
produce distance as direct solution output, with a standard error.
Rucinski's $uvby$ light curves can be put on an absolute flux basis reliably, since he
provided all the necessary information. 
They were processed in two ways -- as part of the multi-datatype non-absolute solutions
and also in separate absolute solutions.

Table~\ref{tbl-sol} has  results from five input datatype combinations:
\begin{enumerate}
\item all data except the MOST light curve; 
\item timings only;
\item the MOST light curve only; 
\item all data except MOST but with the stellar radii fixed at the MOST-only solution values, 
thereby taking advantage of MOST's high temporal density to resolve the fast partial eclipse jumps; 
\item all data, including MOST.
\end{enumerate}
Auxiliary parameters and absolute dimensions are in Table~{\ref{tbl-aux}. Although our derived masses are
larger than those in OBBS (see \S\ref{redmass}), the red dwarf and white dwarf radii are virtually 
the same as theirs.
Based on the 3b mass function in Table~\ref{tbl-sol}
(essentially independent of solution type), 3b masses for three orbit inclinations ($i_{3b}$) are
in Table~\ref{tbl-3bmass}. The 3b mass, $M_3$, exceeds the core hydrogen-burning threshold of $0.07\,M_\sun$ 
only for $i_{3b}$ below $30\degr$, which makes the third body a likely brown dwarf candidate, in agreement with the conclusions of   
\citet{guinan} and \citet{ibanoglu05}.
Selected observed and computed RV curves are compared in Figures~\ref{rv1} to \ref{rv3}.
Figures~\ref{v471_u} to \ref{v471_MOST} include computed light curves based on the Table~\ref{tbl-sol} results.
 
\section{Timing Residual Excursions Historically Interpreted as a Light-Time Effect} 
\label{3body} 

\citet{herczeg} tentatively considered a 3b light-time interpretation of the 
timing residuals with a period around 5 years -- much shorter than 
recent estimates of 30+ years. Existing data in 1975 covered only about five 
years around the first observed maximum in the timing residual diagram, from about JD 2,440,500 to 2,442,300.
Rather than turn upward, as expected for a 5 year periodicity, 
the residual curve continued downward. 
Other early interpretations in terms of true period changes \citep{young75,tunca79} 
mentioned implausibly large $dP/dt$'s, perhaps due to mis-stated units.
The idea of successive changes in 
the EB period, separated by intervals of constancy \citep{oliver}, was then about as  
well regarded as the light-time hypothesis. 
 
A decade later, \citet{beavers} 
re-examined the 3b hypothesis with timings that extended nearly to JD 
2,446,100 and found a period of 24.6 years, although the descending 
residuals had not yet turned upward. 
\citet{skillman} argued against a 3b light-time 
interpretation on grounds that it does not explain all wiggles in the 
residual diagram and that residual variations similar to V471~Tau's appear 
in binaries that are thought not to have third star companions. 
\citet{bois91} repeated the \citeauthor{beavers} 
analysis, with some procedural changes and the same data, finding again 
a $P_{3b}$ of 24.6 years and a slightly different light-time amplitude. 
With nearly another decade of timings in hand -- seemingly most 
of a cycle -- \citet{ibanoglu94} came up with a $P_{3b}$ of -- again -- 24.6 years. 
Then from essentially a full cycle, \citet{guinan} found a period of 30.5 years -- 
considerably longer than any before. The increase is likely due to near
completion of the diagram's rising branch, which is less steep than had been 
estimated from substantially less than a cycle, thereby taking longer than 
expected to reach the peak. Next, \citet{ibanoglu05} published similar 
results from eclipse times that continued the trend to longer period, with  
$P_{3b}= 32.4$ years. \citet{kaminski} and \citet{hric} then found 3b periods of 33.7 and 33.2 years, respectively, 
with other parameters roughly similar to 
those of \citet{guinan} and \citet{ibanoglu05}. Our all-data $P_{3b}$ of $30.11\pm0.16$ years (Table~\ref{tbl-sol}, column 6)
is the same as the $30.5\pm1.6$ years by \citeauthor{guinan} within the uncertainty, although $e_{3b}\approx 0.39$ 
and $\omega_{3b}\approx 1.37$ radians, both parameters being somewhat larger than found by \citeauthor{guinan} 
(respectively 0.31 and 1.09 radians) from timings only and for a shorter timewise baseline.

Although \citet{kaminski}
computed 3b parameters, they considered two explanations for the 
timing diagram that did not involve a third star -- apsidal motion in a slightly eccentric 
orbit and actual orbital period changes for the EB. 
In any case, the overall picture has been 
one of increasing period estimates for the light-time effect, although the situation may now have 
stabilized, as the long-awaited second historical downturn in the timing 
residuals has finally occurred, as shown by Figure~\ref{eclresno3b}.  
Another 12 years of timings have arrived since the \citeauthor{guinan} paper and unification solutions now can include whole 
light curves and RVs, with proper weighting.
Our various solutions that properly separate true period 
change from light-time effect find quite small $dP/dt$'s 
of order $+3\times 10^{-11}$, typically differing from zero by about $20\sigma$. 
Timing residual diagrams for the  
all-data solutions of Table~\ref{tbl-sol} look essentially the same as Figure~\ref{eclresno3b} 
so they are not shown.

More can now be said about the reality of the 3b component with the help of another decade and a half of timing observations. 
Residuals from a linear ephemeris have the form expected for a 3b light-time effect, 
as shown by Figure~\ref{eclresno3b}. Actually the form was already right in 
Figure 2 of \citeauthor{guinan}, but the additional timings make the agreement with expectation significantly clearer.
More than one cycle is now in hand and the data points are repeating!
There are two disturbances, but that is a familiar experience with EB timing diagrams, which only rarely 
repeat with all `desired' accuracy. Mainly the observed and computed timing residual curves agree. 
However some EB timing diagrams have repeated for part of a second cycle, but not after that,
so a stronger timing-based decision on V471~Tau's light-time effect may have to await several more decades.

The outer orbit's orientation within its own plane, specified by argument of periastron $\omega_{3b}$, 
affects eclipse times via the light-time effect and 
affects results from the other two data types to some extent. The outer orbit is too large to show 
significant rotation over the 40+ years of observation, so one cannot expect 
to measure $d\omega _{3b}/dt$. 

\section{The Red Dwarf's Mass} \label{redmass}
 
The all-data $M_2$ of Table \ref{tbl-aux} that taps into virtually all 
existing and relevant V471~Tau mass information has $M_2=0.9971\pm0.0012\,M_\sun$.
How can a K2\,V star be as massive as the Sun? Although the envelope would likely 
be chemically contaminated by passage through a giant star envelope, a normal view of common 
envelope evolution would see the red dwarf as sufficiently close to chemical uniformity to be a
well adjusted main sequence star. Given the K2 spectral type and radius within main 
sequence limits, the red dwarf should have the mass of a K2 main sequence star, around 0.70 to 0.80\,$M_\sun$.
Section 5 of OBBS thoroughly examined this K2\,V mass issue in the context of
their $M_2=0.93\pm0.07\,M_\sun$ result, finding the star over-massive for the main sequence,
and now our solutions produce the same outcome with a smaller formal uncertainty. 
The apparent discrepancy between observational and expected $M_2$ requires serious investigation, as it bears upon all V471~Tau evolutionary
contributions and on the understanding of common envelope evolution.
The $M_2$ determined in OBBS has been adopted for 
various applications \citep[e.g.][]{guinan,chatz,parsons,zoro,hardy}, although one should remember
that it is based on the only V471~Tau white dwarf RV curve ever published (also true of our $M_2$ result)
and on one of only two red dwarf RV curves of the system in print at the time). Note also
that the OBBS $M_2$'s $1\sigma$ uncertainty of $0.07\,M_\sun$ allows a rather wide range
of actual values, yet is sometimes adopted without mention of that fact.

\subsection{Investigation of the Apparent Red Dwarf Mass Anomaly}

OBBS devoted two pages to discussion of measurement difficulties of their white dwarf RVs. The only useful white dwarf spectral
line in their Hubble Space Telescope spectra was Ly$_\alpha$ which is \emph{very} wide (about 40\,\AA), thus undermining its usefulness as a sharp
radial velocity marker. OBBS mention that the pre- and post-COSTAR 
systemic radial velocities ($V_\gamma$), determined independently of the red star RVs, differ by 60 km\,s$^{-1}$. 
They also write that ``\ldots we feel that the absolute velocity zero point for  Ly$_\alpha$ is 
untrustworthy because of the steep sensitivity function.'' Here the `steep sensitivity function' refers
to the detector response being a steep function of wavelength across the wide Ly$_\alpha$ line, thereby
rendering the recorded line profile asymmetric and introducing a false wavelength (and velocity) shift.
Presumably such a shift would be statistically the same for all of their RV measures. If so
it might not much affect the RV amplitude.
However, line width consequences for RV amplitude are possible, although that issue was not included
in the OBBS discussion.

\subsubsection{Some Possibilities} \label{poss}

Five ideas to account for the discrepancy between the OBBS mass results and ours 
were then checked computationally. All seemed unlikely to account for a significant part of the $M_2$ difference prior
to the actual testing, but \emph{something} must cause the anomaly, so the tests were made.
Parameters that are essentially photometric were fixed at the all-data values of Table \ref{tbl-sol}.
The tests were:
\begin{enumerate}
\item allowing for a slight orbital eccentricity by solving with parameters $e$ and $\omega$ adjusted;
\item searching for a local minimum in the parameter space of our solutions that might be deeper than that initially found;
\item checking on the effect of correlations between the ``untrustworthy'' $V_\gamma$ of the white dwarf RVs and parameters $a$ and $M_2/M_1$; 
\item comparing RV-only solutions done with and without a 3b light-time effect (OBBS did not include a light-time effect);
\item examining results of our five RV-only solutions to see if the red dwarf mass based on the \citet{bois88} data (the only red dwarf RVs used by OBBS) is typical of those from the other four sets.
\end{enumerate}
Items 1 to 4 indeed turned out to be of little or no importance with 
regard to derived masses. Specifics are omitted in the interest of brevity.
Item 5 led to some understanding, as the \citet{bois88} observations gave the lowest 
masses of the five red dwarf RV curves, as seen in Table~\ref{tbl-rvind}.

Because the computed masses depend on the cube of the RV amplitude via Kepler's Third Law,
small amplitude disagreements can have substantial consequences.
This is clearly a radial velocity problem, as light curves and timings of a well-detached EB carry almost no
mass information. Therefore separate RV-only solutions were carried out for the
five red dwarf RV datasets, each solved together with the one white dwarf dataset. Results are in Table \ref{tbl-rvind}. 
The purpose was to see if the widely adopted $M_2$ result of OBBS is specifically characteristic
of the (only) red dwarf RVs \citep{bois88} entered into the OBBS analysis.
Rather than apply a binary star model, OBBS simply fitted a sine curve to 
their white dwarf velocities to obtain two parameters, an amplitude and a 
(subsequently discarded) $V_\gamma$. Because our binary model analysis 
requires observations with a meaningful zero point, while OBBS reject any 
such zero point in their RV1s as ``untrustworthy'' (and did not mention a resulting $V_\gamma$), 
a preliminary exercise here was to fit the white dwarf RVs in the same way as stated by OBBS.
Our thus fitted sine curve led to an $M_2$ value close to that by OBBS when matched with
an amplitude from the \citet{bois88} red star RVs,\footnote{A precise match cannot be expected since the OBBS
data weighting may have differed from ours.} also found from a fitted sine wave. $V_\gamma$ for the white dwarf
differed by 10 km\,s$^{-1}$ from that of the all-data solution in Table \ref{tbl-sol}. Accordingly,
10 km\,s$^{-1}$ was subtracted from each white dwarf velocity entered into our five RV-only
solutions (one for each published red dwarf velocity curve) so as to allow essential
consistency between star 1 and star 2 RVs 
(our binary star analyses find one $V_\gamma$ that serves for stars 1 and 2).
Resulting mass differences among RV datasets in the test solutions are very much
larger than the corresponding $\sigma$'s of the masses in Table~\ref{tbl-aux}, which are \emph{internal} uncertainties that
do not account for systematic dataset errors, most notably in RV amplitudes. Apparently such systematic errors, although 
not striking, are significant in most or perhaps all of the RV curves (cf. Figure \ref{rv2res-all}).
The $M_2$'s in Table~\ref{tbl-rvind} range from 0.90\,$M_\sun$ (Bois et al. RVs)
to 1.01\,$M_\sun$ (Hussain et al. RVs). 
With two of the more precise red dwarf RVs, $M_2$ is 0.04\,$M_\sun$ and 0.11\,$M_\sun$ higher than with the \citet{bois88} RVs, 
thus accounting respectively for about half (\citeauthor{kaminski}) or slightly more than all (\citeauthor{hussain}) of the discrepancy between our $M_2$ and that of OBBS.
Since the OBBS $M_2$ utilized only the 
\citeauthor{bois88} RVs for the red dwarf (those giving the lowest $M_2$ of the five data sources),
while the largest $M_2$ is from the recent highly precise and accordingly highly weighted \citeauthor{hussain} dataset,
our high $M_2$ from the all-data solution is probably explained. 

\subsubsection{A Summary of the Investigation on Mass Results}

A brief summary of mass findings may be useful. 
First, a preliminary exercise showed that the OBBS masses are essentially recovered 
when their procedure is followed and the \citeauthor{bois88} RVs are adopted.
Then solution experiments dismissed items 1 to 4 (\S \ref{poss}) as unimportant.
Next the dataset dependence was isolated via RV-only solutions for the five red dwarf
(plus one white dwarf) RV datasets. 
One can examine the mass entries of Table \ref{tbl-rvind} to see how the derived red dwarf and white dwarf masses
change from dataset to dataset. The RVs from \citet{bois88} give the lowest masses.
We find that the red dwarf mass of $0.93\,M_{\sun}$ derived by OBBS and properly called high for a K2\,V star,
actually comes out somewhat higher (now $M_2\approx 1.00\,M_{\sun}$) if all of the now existing RV data are 
utilized, properly weighted. A revised evolutionary explanation seems needed unless
new RVs of the white dwarf can give an observational way out. Such an explanation cannot be developed now
since, even if the K2 star did spiral through a giant star envelope, the
specifics of the encounter (giant star's chemical profile, etc.) are uncertain, as is subsequent
radial redistribution by convection and advection within the resulting (now K2) star.  
A very simple assumption, that the K2 star became uniformly enriched in metals, goes in the
right sense to account for the mass anomaly, as the red star would then lie to the right of the
solar-composition main sequence, explaining its late spectral type in terms of
composition rather than mass. However, uniform enrichment seems unlikely.
In the alternative (observational) fix, the much cited mass
results may need revision based on new white dwarf RVs,
although that will be difficult due to the white dwarf's faintness.

\section{Photometric-Spectroscopic Distance} 

EB light curves allow measurement of relative radii, $R_{1,2}/a$, while 
RVs set the geometric scale via the orbital semi-major axis parameter, $a$. 
These steps are combined in simultaneous light/RV solutions.
With a full physical model that specifies local radiative behavior at all
points on both stars, observable bandpass flux at given aspect and distance
can be computed in standard physical units and compared with observed fluxes
in those units. Numerical inversion of this \emph{parameters to flux} problem
can yield distance in parsecs if distance is one of the parameters.
Specifics and examples of the process are in \citet{wilson08}, with further examples
in \citet{wilson09}, \citet{vaccaro10}, and \citet{wilson11}. A separate distance estimation step with
spherical star assumptions and other simplifications is no longer needed.

The best current V471~Tau distance estimate is likely that by \citet{debruijne} 
from secular parallax ($48.64\pm0.78$ pc), while EB
photometric-spectroscopic distance as well as trigonometric parallax 
distance have larger uncertainties for V471~Tau, but give welcome checks. 
Interstellar extinction can be a major cause of photometric-spectroscopic distance inaccuracy but 
probably is negligible for Hyades objects in optical and infrared bands. Only one 
spectroscopic temperature is needed for favorable EBs where eclipse 
depths for both stars establish a relation between the two temperatures. 
However, eclipses of V471~Tau's red star by its white dwarf companion are not realistically 
observable at present, so the distance cannot be estimated in the usual way from light and RV curves 
without prior temperature knowledge for both stars. Good spectroscopic 
temperatures are known for both V471~Tau components so a photometric-spectroscopic 
distance can be computed in an absolute solution, as only one distance will be compatible
with the observed absolute fluxes and other system parameters. Those other parameters 
are already known from the non-absolute solutions of Table \ref{tbl-sol}
and can be taken from the table and applied to the (absolute) distance solution.
Alternatively (not done in this paper), all adjustable parameters 
can be determined in an (absolute) distance solution. 
 
The MOST data were not in the distance solutions because their absolute photometric 
calibration is not well known. V471~Tau's distance was found from the Rucinski light curves,
as they are on the standard $uvby$ system and the comparison star magnitudes are known. 
Its distance was also found from the KPNO photometry of Table~\ref{tbl-BVRI} after conversion 
of the differential instrumental magnitudes, $\Delta m_B$ and $\Delta m_V$, to standard $B$ and $V$ magnitudes. 
The transformations are based on our 1998 KPNO instrumental magnitudes of BD\,+16 515 and BD\,$-3$ 5358, 
on BD\,+16 515's standard $B$ and $V$ means of Table~\ref{tbl-comparison}, and on standard $B$ and $V$ measures 
of BD\,$-3$ 5358 made from September 8, 2011 to November 11, 2012, and provided in Table \ref{bd5358}.
BD\,+16 515 and BD\,$-3$ 5358 have an acceptable color difference and thereby give satisfactory 
transformations as BD\,+16 515 is redder than BD\,$-3$ 5358 by about 0.53 mag in $B-V$.
The tabulated $\Delta m_B$'s and $\Delta m_V$'s were converted to $B$ and $V$, respectively, via
\begin{equation}
B=B_{\rm comp}+\Delta m_B+0.2695 \left(\Delta m_B - \Delta m_V\right) \nonumber
\end{equation}
and
\begin{equation}
V=V_{\rm comp}+\Delta m_V-0.1090 \left(\Delta m_B - \Delta m_V\right), \nonumber
\end{equation}
where $B_{\rm comp}$ and $V_{\rm comp}$ are from the means in Table~\ref{tbl-comparison}.
Required flux calibrations for $u$, $v$, $b$, $y$, $B$, and $V$ are from Table 1 of \citet{wilson10}.
Individual single band curves were solved for distance with stepped input of surface temperature
for both the white dwarf and red dwarf, with results 
in Table~\ref{tbl-abssol}. The reasons for stepping the temperatures are (a) to give a realistic impression
of how strongly distance results depend on temperature input, and (b) to allow interpolation if
temperature estimates improve (although they already seem rather consistent among authors).
Distances for the six $T_1$, $T_2$ combinations and six photometric bands sprinkle from 
about 41.4 pc to 54.6 pc, which is a larger range than for normal EBs
that have two observable eclipses per cycle. We are reminded that V471~Tau is a difficult object.
 
Distance accuracy for V471~Tau is affected to some extent by uncertainty in 
spottedness (spot latitudes, longitudes, sizes, temperatures, growth and decay, 
and drift motions), but bolometric luminosity (and, to first order, bandpass 
luminosity) should not be changed very much since convective
energy flow that is blocked by the magnetic fields of spots must come out elsewhere. 
The distances of Table~\ref{tbl-abssol} and others in the literature
are remarkably close in aggregate to Hyades distance estimates, considering that V471~Tau does not project upon
the cluster core, being approximately $10\degr$ ($\approx 8$ pc) from the center, which is well outside
the tidal radius ($\approx 3$ pc) and core radius. At that projected separation, it would not have 
been surprising if V471~Tau were a similar linear distance nearer or farther than the Hyades center.
However the V471~Tau distances, taken as a whole, seem reasonably well determined so the close match may just be a 
coincidence. Anyway the system's location within the Hyades seems well established. 
 
\section{Other Periodic Light and Velocity Changes} \label{otherper}

\citet{ibanoglu94,ibanoglu05} mention a periodic fluctuation in                                  
the timing residuals of about 5 or 5.5 yr, possibly modulated by a concurrent periodic change in average brightness.                                  
\citet{kaminski} briefly discuss a 10 yr cycle and suggest that it may be due to an activity cycle on the red dwarf surface.
A Lomb-Scargle \citep{lomb,scargle} period search in
our timing-only residuals found three (all of low power) peaks near 13, 9 and 5 years. Sinusoids
of the form
\begin{equation}
 \mathcal{A}+\mathcal{B}\sin\left[ (2\pi/P)(t-t_0)\right]  \label{eqnsin}
\end{equation}
were fitted to the residuals from the timing solution of column 3 in Table~\ref{tbl-sol}. 
Derived parameters for Equation~\ref{eqnsin} are in Table~\ref{tbl-ressinfit}
and corresponding curves are in Figures~\ref{eclresp13} to \ref{eclresp5}.
Possible mechanisms could include the magnetic phenomenon proposed by \citet{applegate}.

A close look at the residuals in many of the photometric bands and the red dwarf velocity curves reveals a much shorter
periodicity, $\approx 0.26$ d, a value nearly indistinguishable from half the orbital period. This
variation is clearly visible in the residual velocity graph in Figure~1 of \citet{hussain}, who attribute it
to perturbations by spots. Many of the light curves show the phenomenon, although our
modeled spots should have removed spot effects from the residuals, at least in first approximation.
Assuming synchronization with the orbit period, the fundamental of a variation 
that arises from a variable tide in a slightly eccentric orbit would be very difficult to
detect in competition with ordinary proximity effects (reflection and ellipsoidal variation), but its harmonics may be easier to find. 
Accordingly searches were made via the Lomb-Scargle and CLEAN \citep{roberts} algorithms for signatures of harmonics near 2, 3, and 4 times the orbital frequency.
Because of its high density of points and absence of major gaps,
the MOST residual curve is the one best suited to a search for such frequencies. The Lomb-Scargle and CLEAN periodograms 
of the MOST residuals are
in Figure~\ref{LS-CLEAN}. Their more prominent peaks
coincide, although not with the same order of relative height. Results of
Least-squares fits of Equation~\ref{eqnsin} to the MOST-only residuals
(column 4 of Table~\ref{tbl-sol}), starting with the periods of the six
highest periodogram peaks, are in 
Table~\ref{tbl-lcrvressinfit} and
Figures~\ref{mostres1} to \ref{mostres4}. Similar fits for periods close to 0.26 d were made for the 
KPNO residual light curves and
the red dwarf radial velocity residuals from the all-data solution in the last column of Table~\ref{tbl-sol}, 
with parameters in Table~\ref{tbl-lcrvressinfit} and waveforms in 
Figures~\ref{Bres} to \ref{hussainres}. The overall outcome is that there \textit{are} close matches between  
harmonics of the orbit period and the identified periodicities, but the latters' very small $\sigma$'s indicate
that the matches are not formally valid. 
Continued light curve monitoring could further quantify these correspondences. 

As a causal candidate for the 0.26-d periodicity, one might consider the first harmonic of a tidal oscillation, whose fundamental 
frequency would be $1/P_{\rm orb}$ s$^{-1}$. The slight orbital eccentricity needed to drive the tide may have
already been found by \citet{kaminski}, whose Table 4 gives $e=0.0121\pm0.0006$ from eclipse timings, and whose Table 6 
gives $0.012\pm0.003$ from the red dwarf RVs. While caution is the watchword
for acceptance of such small eccentricities from RVs \citep{lucy71,lucy73}, these measures from two unrelated data types are formally $20\sigma$ and $4\sigma$ results!
To explore the possibility of a non-zero eccentricity, both the \citeauthor{kaminski} and \citeauthor{hussain} red dwarf radial velocities were solved, allowing
eccentricity, $e$, and argument of periastron, $\omega$, to adjust, together with parameters $a$, $V_\gamma$ 
and ephemeris zero-epoch, $T_0$. The results (Table~\ref{tbl-rv2sol}) do not agree with those 
in \citet{hussain} or \citet{kaminski}. Our RV-only solutions find zero eccentricity within its uncertainty from the \citeauthor{kaminski} 
data. The \citeauthor{hussain}
RVs give a small non-zero $e$, although one should note that the solution may be affected by the strong $P_{\rm orb}/2$ residual signal. 

\section{On the Reality of the Third Star} \label{reality}

\subsection{Recent Adaptive Optics Results in Perspective}

Longstanding and widespread acceptance of a 3b light-time effect as the cause of the apparent period variation
has recently been challenged \citep{hardy} by means of AO observations in the photometric $H$ band made on Decenber 11, 2014 that hopefully mark
the beginning of imaging of the system's near field. Figure 3 of \citeauthor{hardy} contains a composite rendering 
of the field and a plot of estimated contrast vs. angular separation, where contrast means magnitude difference 
between the 3b and EB. The EB lies in the center of a circular ring 
of radius 260 mas, the \citeauthor{hardy} estimated separation for the date of the AO observation. The search area is a ring 
because the 3b's position angle cannot be determined from light-time variations. 
The \citeauthor{hardy} separation estimate came from an analysis of eclipse timings -- essentially the same timings analyzed here
as one type of input to our unified light-RV-timing solutions.
No obvious features are seen in their search ring, although there are a few barely visible wisps.
\citeauthor{hardy} conclude that existence of the light-time 3b is disproved by their AO image,
as they predict the 3b to be at least 3 mag above their computed marginal detection level.
As we are now armed with the most comprehensive and thoroughly checked set of analytic results ever
assembled on V471 Tau, based not only on eclipse timings but on those timings plus nine light curves and six RV curves, 
the predicted EB to 3b separation and $H$ band magnitude difference can be newly computed.
The key issues will now be examined.

\subsection{The 3b Minimum Mass and Maximum Age} 

Categories for the prospective 3b except brown dwarf clearly seem ruled out,
as lower mass categories are eliminated by the minimum mass estimates in several papers, while
ordinary stars of even the lowest masses would have been discovered easily. A compact
exotic object in a very low inclination orbit may not be excluded with complete 
certainty, but can be excluded realistically. At issue with respect to detectability
is the estimated faint limit for an observationally determined minimum 3b mass.

Brown dwarf bandpass luminosities increase with mass and decrease with age, with some dependence
on metallicity and surface gravity, so predicted minimum luminosity mainly follows from estimates
of minimum mass and maximum age. Our minimum mass of 
$0.0350\pm0.0005\,M_{\sun}$ in Table \ref{tbl-3bmass} that corresponds to $i_{3b}=90\degr$
is 20\% lower than the \citeauthor{hardy} value of $0.044\pm0.001\,M_{\sun}$. $H$-band
luminosity depends steeply on mass, so the difference in minimum mass is significant.
Maximum plausible age for the 3b is that of the Hyades cluster, which has typically
been taken to be about 625 Myr, although recent preprints \citep{brandta,brandtb}
have estimated about 800 Myr after including rotation in evolutionary models.
Accordingly Table \ref{tbl-brown} gives 3b minus EB magnitude differences ($\delta H$ indicates contrast) for both ages,
based on tables of brown dwarf models at the website of F. Allard.\footnote{http://perso.ens-lyon.fr/france.allard/}
A descriptive review paper is \citet{allard}. The results for ages
625 and 800 Myr are $\delta H=9.65$ mag and $9.96$ mag, respectively, making the brown dwarf 
about $0.5$ mag and $0.8$ mag fainter than estimated by \citet{hardy}, and thus
closer to the marginal detection level by those amounts, although still 
respectively about $2.4$ mag and $2.1$ mag above 
the \citeauthor{hardy} detection limit for the 260 mas search ring.
A possible reason for the difference in minimum mass (\citeauthor{hardy} versus here)
is that \citeauthor{hardy} adopted the OBBS masses that are based on the \citet{bois88} RVs that give outlier masses among results from the five red dwarf RV curves.
All red star RV curves except the one used by OBBS, including the much more precise RVs from \citet{kaminski} and from \citet{hussain}, give higher masses for both EB components.

\subsection{A Binary Brown Dwarf?}

A plausible circumstance not mentioned by \citeauthor{hardy} is that the 3b may be a \emph{pair} of
brown dwarfs that could be much fainter than one brown dwarf with the total mass of the little binary. 
Such a binary is not hard to believe, given the well known theoretical difficulties of forming
single stars and the existence of hierarchical multiple star systems.
Table \ref{tbl-brown} therefore also gives $\delta H$ for examples of such a pair (two
$0.0175\,M_\sun$ brown dwarfs) extracted from the Allard tables. The $\delta H$ then becomes $11.21$ mag
at 625 Myr and $11.71$ mag at 800 Myr, which places them about $0.9$ mag and $0.4$ mag
above marginal detection in \citeauthor{hardy}'s Figure~3. So even in this case the object could perhaps be
seen, assuming the \citeauthor{hardy} marginal AO detection curve is accurate, although with a margin as little as 0.4 mag 
rather than the 3.0 mag by \citeauthor{hardy} or our 2.1 mag margin estimate for the case of one brown dwarf. However, the solution
outcome with most potential consequence comes next.

\subsection{The Angular Separation on December 11, 2014}

   The account up to this point is of two sets of brightness estimates (\citeauthor{hardy} and here) that roughly
agree (within $0.8$ mag) if the 3b is one brown dwarf but not if it is a binary of two equal brown dwarfs.
\citeauthor{hardy} predict sure detection if the 3b exists while this paper allows 
for less confident detection or perhaps non-detection
if the 3b is really a binary of two brown dwarfs. The story now becomes
more interesting -- the object may have been seen. 

Starting from the hypothesis that a 3b is responsible for the $\approx 30$-year periodicity via a light-time
effect, as did \citeauthor{hardy}, we computed the 3b-EB sky separation vs. time, based on three of the solutions
in Table \ref{tbl-sol} (all data, all data except MOST light curve, and timing-only solutions).
The observational input had three kinds of precision-based
weights according to precepts in \S\ref{datawt}. 
Representative sky path and angular separation (vs. time) curves
are shown in Figure~\ref{sep_vs_time}, while separation on the date of the \citeauthor{hardy} AO observing is shown vs.
the (unknown) outer orbit inclination in Figure~\ref{sep_vs_i3b}. Since none of the three datatypes convey
information on the 3b orbit's rotational state about the line of sight, the position angle
of the ascending node was set arbitrarily to $\pi/2$ radians, clockwise from north. 
In plane of the sky rectangular coordinates, angular EB-3b displacements\footnote{The displacements are referenced 
to the orbit's focus, as in usual practice, rather than to its center as in
the often adopted work by \citet{irwin52,irwin59} on the light-time effect. 
That is why an unnecessary term in $e\cos\omega_{3b}$ is absent from our displacement equations and our light-time equation 
lacks the term in $e\sin\omega_{3b}$ seen in Irwin's papers and in papers that cite his work. This `center vs. focus'
distinction also explains why V471 Tau light-time plots in \citet{ibanoglu94}, \citeauthor{guinan}, and \citeauthor{hardy} extend equally above and below
zero for an eccentric orbit whose major axis is not in the plane of the sky. Maximum excursions above and below zero light-time,
with respect to the system barycenter, are necessarily unequal for such an orbit.} in arcseconds are
\begin{equation}
\delta_y=\frac{a_{3b} \cos(\upsilon_{3b} + \omega_{3b}) (1 - e_{3b}^2)}{(1+e_{3b} \cos \upsilon_{3b}) d_{pc}} \label{delta_y},
\end{equation}
and
\begin{equation}
\delta_z=\frac{a_{3b} \cos i_{3b} \sin(\upsilon_{3b} + \omega_{3b}) (1 - e_{3b}^2)}{(1+e_{3b} \cos \upsilon_{3b}) d_{pc}}. \label{delta_z}
\end{equation}
Quantities $\upsilon_{3b}$ and $\omega_{3b}$ are respectively the true
anomaly and argument of periastron in the outer (3b) orbit.
The semi-major axis of the outer relative orbit, $a_{3b}$, is in AU
and $d_{pc}$ is the system's distance in parsecs, for which 49 pc was adopted, as that is the mean of
distances from our several solutions. It is nearly the same distance adopted by \citeauthor{hardy} (50 pc). 

Our estimates of the angular separation $\delta_\rho=(\delta_y^2+\delta_z^2)^{1/2}$ at the date of the \citeauthor{hardy} observation, 
obtained from Equations~\ref{delta_y} and \ref{delta_z}, 
differ from those of \citeauthor{hardy}, likely due to differing orbit parameters. 
Our values for $e_{3b}$, $\omega_{3b}$, $a_{3b}\sin i_{3b}$, $P_{3b}$, $dP/dt$, and ${T_0}_{3b}$ (time of superior conjunction of the 3b) are
in Table 8. Those by \citeauthor{hardy} are not yet in print to our knowledge, so a discussion of reasons for
differences between this paper's predicted sky separation and that by \citeauthor{hardy} cannot be in terms
of parameter results but only in terms of the quantity, precision, and 
variety of the basic input data, in concert with applied analysis strategies.

\subsection{Why a Different Estimate of Angular Separation Now?}

Without knowing the input parameter values for the \citeauthor{hardy} separation calculation, one cannot be sure
of the reason or reasons for our different result. 
The difference may arise due to the overall comprehensiveness of our solutions, whose input
has information from
all previous eclipse timings, including the four recent \citeauthor{hardy} points, and 
also from all published light curves and RV curves. It has been done
in various data combinations in addition to the all-data solution (see Table \ref{tbl-sol}) so as to explore possible systematic
variations among datasets. 
We have not just relied on the enormously weighty MOST data (57000+ points) 
being fully satisfactory, but have experimented with various combinations, some of which exclude MOST.
All such solutions except all-data give a ring radius about 80 percent of that of \citeauthor{hardy}, who
analyzed the same eclipse timings, although not any light curves or RV curves. The all-data ring is even smaller. 

A specific reason for the difference in predicted separation could be our
inclusion of $dP/dt$ as a solution parameter, while \citeauthor{hardy} assumed that
$dP/dt=0$. Table \ref{tbl-sol} shows that the dimensionless $dP/dt$ is non-zero and a 25$\sigma$
result, consistently over Table \ref{tbl-sol}'s five solutions, with a value of $+0.295\pm0.012 \times 10^{-10}$
for the all-data solution. Indeed we reproduce the \citeauthor{hardy} prediction of
about 260 mas for December 11, 2014 if we do a timing-only solution with $dP/dt=0$ and apply the
resulting parameters to a calculation of projected sky motion, which suggests that
most of the difference in separation may be due
to our allowance for period change, which has not been done in previous V471~Tau papers.
This outcome is surprising, as the  small $dP/dt$ integrates to phase differences of only about 0.00343 cycles,
or 2.6 minutes of time, over the 40 years of timings, so the match with the \citeauthor{hardy}
computed separation in the no-$dP/dt$ case may be a coincidence, although it is suggestive.
A likely cause of the slow period increase is transfer of angular
momentum from white dwarf spin to the orbit via magnetic coupling. 
The white dwarf is in fast rotation
with a period of 9.25 minutes \citep{jensen,barstow92,barstow97,robinson,sion92,stanghellini,clemens,wheatley}. 

\subsection{Could the 3b Have Already Been Seen?} \label{seen}

The proposed 3b would have passed the point of maximum projected separation and made a good start
toward minimum separation when the AO
observations were made. From our timing-only solution (for most direct comparison with \citeauthor{hardy}) we find $\approx 207$ mas separation for that date, with small variations 
according to which of our solutions is applied, but essentially predicting a radius for the search ring about 80 percent of that by \citet{hardy}. 
This reduction in 
ring radius is a game changer for two crucial reasons. One is that 
the AO detection limit is a magnitude brighter for the smaller ring (see Figure 3 of \citeauthor{hardy}), so our former 2.4 mag
margin for detection of a $0.0350\,M_\sun$ brown dwarf at 625 Myr age becomes 1.4 mag, and the 2.1 mag detection 
margin at 800 Myr becomes 1.1 mag. For the binary brown dwarf at 625 and 800 Myr
the respective margins drop from 0.9 mag and 0.4 mag to negative (i.e. undetectable) values of $-0.1$ mag
and $-0.6$ mag. The second crucial reason is that features \emph{are} seen within the smaller ring,
for example strong ones at about 4 and 10 o'clock. So \citeauthor{hardy} may actually have 
discovered the 3b. Of course continued imaging observations that look for Keplerian motion, 
so as to distinguish real star-like objects from artifacts of the contrast enhancement process,
will be needed to check on this possibility. Success would tell the position angle of the ascending node
and have a definite measurement of the 3b orbit inclination, thus also the 3b mass rather than only mass
as a function of inclination.

\subsection{The Thirty-Year Waveform}

The timing diagram (Figure~\ref{eclresno3b}, left panel)
has translational symmetry -- a shift of about 30 years brings the two maxima essentially into register.
The symmetry \emph{could} result from three discontinuous period changes of about the same
magnitude, alternate sign, and equal spacings -- but
would require matched behavior at three epochs, which seems unlikely.
A 3b light-time effect represents the eclipse timings well, 
apart from two relatively small dips, although full
consensus on the existence of the third star from timing data may require further timings over
several decades or more, as periodicity needs to be established for a slowly varying phenomenon. 

The light-time waveform agrees with Keplerian motion over 1.3 cycles, for the most part. We do 
not see the two brief dips as a serious concern, since few timing diagrams lack 
some irregularity, as can be seen by examining those in \citet{kkn}. 
\citet{hardy} suggest a magnetic phenomenon \citep{applegate}
that has no waveform prediction except that the form can change from cycle to cycle (i.e. that there 
is no definite waveform) and therefore the phenomenon cannot be tested from timing variations.

Our bottom-line estimate of the minimum margin for detection of the light-time 3b is $+0.4$ mag
if it is one brown dwarf and $-0.6$ mag if it is a pair of equal brown dwarfs of the same total mass,
where negative margins correspond to non-detectability. 
Of course these are minimum estimates, showing that the 3b \emph{could} have escaped detection, 
although the 3b could be much brighter and may already have appeared in the AO observation, as
noted in \S\ref{seen}.

Confirmed detection may come from further AO observations that look for a feature in Keplerian motion, 
or from other direct imaging (speckle?, above-atmosphere?) with one or more large telescopes, 
as proposed by \citet{guinan} and now begun by \citet{hardy}. 
With good estimates of the object's minimum brightness and its angular separation from the EB
along with deep images, a positive outcome is likely if the object exists
and would bring closure to a controversy of long standing. 

\acknowledgments 
 
We are pleased to thank K. Kaminski for sending the MOST mission light 
curves, as well as answering numerous questions about the data and resolving the issue
related to mid-exposure time corrections. We thank G. 
Hussain for sending her RVs and answering questions related to the times of observation. 
We also thank the several authors 
who published digital V471~Tau light and RV curves, as cited above. We made very extensive use of 
the Simbad and NASA ADS internet sites. Thanks are also due to the referee for a thorough report
and helpful suggestions.

\clearpage 

\begin{deluxetable}{ccrcccrcccrc}
\tabletypesize{\tiny}
\tablewidth{0pt}
\rotate
\tablecaption{Red Dwarf 1998 Radial Velocities \label{tbl-v471rv}}
\tablehead{\colhead{HJD} & \colhead{HJED} & \colhead{RV} & \colhead{Error} 
& \colhead{HJD} & \colhead{HJED} & \colhead{RV} & \colhead{Error} 
& \colhead{HJD} & \colhead{HJED} & \colhead{RV} & \colhead{Error} \\
\colhead{} & \colhead{} & \colhead{(km\,s$^{-1}$)} & \colhead{(km\,s$^{-1}$)} 
& \colhead{} & \colhead{} & \colhead{(km\,s$^{-1}$)} & \colhead{(km\,s$^{-1}$)} 
& \colhead{} & \colhead{} & \colhead{(km\,s$^{-1}$)} & \colhead{(km\,s$^{-1}$)} }
\startdata
2451115.74334 & 2451115.74407 & $ 150.75$ &  2.43 & 2451119.97287 & 2451119.97360 & $  71.80$ &  2.86 & 2451123.94066 & 2451123.94139 & $ 101.86$ &  3.08  \\
2451115.76646 & 2451115.76720 & $ 127.03$ &  3.36 & 2451119.99392 & 2451119.99465 & $  29.40$ &  3.51 & 2451123.96675 & 2451123.96748 & $ 142.83$ &  2.48  \\
2451115.80814 & 2451115.80887 & $  67.28$ &  3.76 & 2451120.69401 & 2451120.69474 & $ -86.62$ &  3.68 & 2451123.98779 & 2451123.98852 & $ 166.47$ &  3.48  \\
2451115.82916 & 2451115.82989 & $  17.70$ &  4.71 & 2451120.71505 & 2451120.71578 & $ -63.05$ &  3.37 & 2451124.00883 & 2451124.00956 & $ 179.75$ &  3.01  \\
2451115.85965 & 2451115.86038 & $ -34.01$ &  2.68 & 2451120.73609 & 2451120.73682 & $ -18.14$ &  3.31 & 2451124.67755 & 2451124.67828 & $  41.20$ &  2.82  \\
2451115.88069 & 2451115.88142 & $ -62.08$ &  3.57 & 2451120.76867 & 2451120.76940 & $  36.50$ &  2.33 & 2451124.69859 & 2451124.69933 & $   8.17$ &  2.72  \\
2451115.90463 & 2451115.90536 & $ -91.47$ &  3.47 & 2451120.78972 & 2451120.79045 & $  72.97$ &  3.26 & 2451124.71965 & 2451124.72038 & $ -32.14$ &  3.13  \\
2451115.92567 & 2451115.92640 & $-104.07$ &  2.54 & 2451120.81076 & 2451120.81149 & $ 105.76$ &  2.91 & 2451124.74071 & 2451124.74144 & $ -67.55$ &  4.62  \\
2451115.94671 & 2451115.94745 & $-121.51$ &  4.56 & 2451120.83180 & 2451120.83253 & $ 129.20$ &  2.90 & 2451124.76902 & 2451124.76976 & $-101.68$ &  3.23  \\
2451115.97263 & 2451115.97336 & $-115.26$ &  4.21 & 2451120.85942 & 2451120.86015 & $ 160.30$ &  3.67 & 2451124.79041 & 2451124.79114 & $-116.14$ &  3.74  \\
2451115.99366 & 2451115.99439 & $ -90.57$ &  2.21 & 2451120.88047 & 2451120.88120 & $ 180.57$ &  2.25 & 2451124.81163 & 2451124.81236 & $-117.86$ &  3.38  \\
2451117.72782 & 2451117.72855 & $ 162.68$ &  4.13 & 2451120.90150 & 2451120.90223 & $ 183.59$ &  2.43 & 2451124.83267 & 2451124.83340 & $-112.96$ &  2.84  \\
2451117.75008 & 2451117.75081 & $ 177.66$ &  3.20 & 2451120.93026 & 2451120.93099 & $ 177.45$ &  3.82 & 2451124.85370 & 2451124.85443 & $ -95.83$ &  2.97  \\
2451117.77166 & 2451117.77240 & $ 177.51$ &  4.10 & 2451120.95131 & 2451120.95204 & $ 157.39$ &  2.73 & 2451124.87474 & 2451124.87548 & $ -72.88$ &  2.69  \\
2451117.79334 & 2451117.79408 & $ 174.68$ &  2.96 & 2451120.97235 & 2451120.97308 & $ 133.49$ &  2.33 & 2451124.89750 & 2451124.89823 & $ -36.92$ &  2.81  \\
2451117.82558 & 2451117.82632 & $ 153.90$ &  2.90 & 2451120.99339 & 2451120.99412 & $ 106.38$ &  2.57 & 2451124.91853 & 2451124.91926 & $   1.59$ &  2.48  \\
2451117.84533 & 2451117.84606 & $ 134.35$ &  2.33 & 2451121.01774 & 2451121.01847 & $  70.74$ &  3.95 & 2451124.93956 & 2451124.94029 & $  36.96$ &  2.56  \\
2451118.70485 & 2451118.70558 & $  73.23$ &  2.26 & 2451121.69333 & 2451121.69406 & $-116.54$ &  3.27 & 2451124.96061 & 2451124.96135 & $  72.47$ &  3.76  \\
2451118.73872 & 2451118.73945 & $ 119.48$ &  1.86 & 2451121.72619 & 2451121.72692 & $ -99.96$ &  2.44 & 2451124.98164 & 2451124.98238 & $ 102.90$ &  2.03  \\
2451118.75977 & 2451118.76050 & $ 146.35$ &  3.02 & 2451121.75320 & 2451121.75394 & $ -65.87$ &  2.86 & 2451125.00268 & 2451125.00341 & $ 136.19$ &  2.65  \\
2451118.80495 & 2451118.80568 & $ 186.83$ &  1.82 & 2451121.77425 & 2451121.77498 & $ -25.82$ &  3.08 & 2451125.02912 & 2451125.02985 & $ 166.97$ &  3.26  \\
2451118.83388 & 2451118.83461 & $ 179.38$ &  3.00 & 2451121.79528 & 2451121.79601 & $   7.29$ &  1.77 & 2451125.67631 & 2451125.67704 & $ 120.26$ &  2.60  \\
2451118.85516 & 2451118.85589 & $ 168.58$ &  2.16 & 2451121.81632 & 2451121.81705 & $  42.97$ &  3.03 & 2451125.69736 & 2451125.69809 & $  89.11$ &  4.06  \\
2451118.87621 & 2451118.87694 & $ 150.92$ &  3.76 & 2451121.85086 & 2451121.85159 & $ 101.58$ &  2.76 & 2451125.71840 & 2451125.71913 & $  52.17$ &  3.87  \\
2451118.89896 & 2451118.89969 & $ 124.91$ &  2.72 & 2451121.87269 & 2451121.87342 & $ 129.19$ &  1.92 & 2451125.73943 & 2451125.74016 & $   8.11$ &  2.36  \\
2451118.92000 & 2451118.92073 & $  90.37$ &  2.09 & 2451121.89372 & 2451121.89445 & $ 156.94$ &  2.74 & 2451125.77829 & 2451125.77903 & $ -61.57$ &  4.18  \\
2451118.94103 & 2451118.94176 & $  53.01$ &  2.94 & 2451121.91477 & 2451121.91550 & $ 178.24$ &  2.27 & 2451125.79933 & 2451125.80006 & $ -91.21$ &  3.46  \\
2451118.96209 & 2451118.96282 & $  11.96$ &  3.53 & 2451121.93584 & 2451121.93657 & $ 180.96$ &  2.82 & 2451125.82038 & 2451125.82111 & $-107.14$ &  2.62  \\
2451118.98479 & 2451118.98552 & $ -32.13$ &  2.29 & 2451121.96619 & 2451121.96692 & $ 176.30$ &  3.64 & 2451125.84144 & 2451125.84218 & $-118.25$ &  3.47  \\
2451119.00760 & 2451119.00833 & $ -69.95$ &  4.08 & 2451121.99232 & 2451121.99305 & $ 158.11$ &  2.73 & 2451125.86539 & 2451125.86612 & $-115.76$ &  2.93  \\
2451119.70071 & 2451119.70144 & $  -7.77$ &  3.07 & 2451122.01363 & 2451122.01436 & $ 140.11$ &  2.49 & 2451125.88834 & 2451125.88907 & $-104.00$ &  3.16  \\
2451119.72174 & 2451119.72247 & $  27.63$ &  4.15 & 2451123.68673 & 2451123.68746 & $ -52.03$ &  3.88 & 2451125.91125 & 2451125.91198 & $ -78.14$ &  2.51  \\
2451119.74277 & 2451119.74350 & $  61.31$ &  2.51 & 2451123.70781 & 2451123.70854 & $ -82.69$ &  3.43 & 2451125.93230 & 2451125.93303 & $ -48.05$ &  3.15  \\
2451119.76382 & 2451119.76455 & $  95.76$ &  3.67 & 2451123.72884 & 2451123.72957 & $-102.70$ &  2.48 & 2451125.95336 & 2451125.95409 & $ -14.92$ &  2.43  \\
2451119.79115 & 2451119.79189 & $ 135.49$ &  3.52 & 2451123.74988 & 2451123.75061 & $-110.56$ &  3.13 & 2451125.98089 & 2451125.98162 & $  36.66$ &  2.24  \\
2451119.81218 & 2451119.81292 & $ 156.45$ &  3.11 & 2451123.77711 & 2451123.77784 & $-116.17$ &  2.46 & 2451126.00221 & 2451126.00294 & $  67.12$ &  3.65  \\
2451119.83323 & 2451119.83396 & $ 178.16$ &  2.17 & 2451123.79814 & 2451123.79888 & $-106.79$ &  2.66 & 2451126.02324 & 2451126.02397 & $ 106.17$ &  2.77  \\
2451119.85427 & 2451119.85500 & $ 179.75$ &  3.63 & 2451123.83267 & 2451123.83340 & $ -68.75$ &  3.69 & 2451126.68505 & 2451126.68578 & $ 158.59$ &  3.37  \\
2451119.88811 & 2451119.88884 & $ 170.94$ &  2.99 & 2451123.85431 & 2451123.85505 & $ -37.82$ &  1.91 & 2451126.70609 & 2451126.70682 & $ 134.94$ &  1.78  \\
2451119.90916 & 2451119.90989 & $ 158.35$ &  3.00 & 2451123.87544 & 2451123.87617 & $  -0.19$ &  2.61 & 2451126.73635 & 2451126.73708 & $  90.36$ &  2.26  \\
2451119.93020 & 2451119.93093 & $ 133.01$ &  2.83 & 2451123.89651 & 2451123.89725 & $  39.15$ &  3.68 & 2451126.78502 & 2451126.78575 & $  -4.84$ &  2.77  \\
2451119.95124 & 2451119.95197 & $ 101.90$ &  2.86 & 2451123.91765 & 2451123.91838 & $  74.56$ &  3.50 & 2451126.80638 & 2451126.80712 & $ -39.00$ &  2.73  \\
\enddata
\tablecomments{Errors are individual RV standard errors.}
\end{deluxetable}

\clearpage

\begin{deluxetable}{cccccccccccc}
\tabletypesize{\tiny}
\tablewidth{0pt}
\rotate
\tablecaption{$BVR_C I_C$ 1998 Light Curves \label{tbl-BVRI}}
\tablehead{\colhead{HJD} & \colhead{HJED} & \colhead{$\Delta m_B$} &  \colhead{HJD} & \colhead{HJED} & \colhead{$\Delta m_V$}  
& \colhead{HJD} & \colhead{HJED} & \colhead{$\Delta m_{R_C}$} & \colhead{HJD} & \colhead{HJED} & \colhead{$\Delta m_{I_C}$} }
\startdata
2451115.82767 & 2451115.82840 & $-0.347$ & 2451115.82530 & 2451115.82603 & $-0.027$ & 2451115.81963 & 2451115.82036 & 0.159 & 2451115.82234 & 2451115.82307 & 0.425 \\
2451115.82989 & 2451115.83062 & $-0.345$ & 2451115.82643 & 2451115.82716 & $-0.026$ & 2451115.82155 & 2451115.82228 & 0.163 & 2451115.82317 & 2451115.82390 & 0.427 \\
2451115.87319 & 2451115.87392 & $-0.339$ & 2451115.87142 & 2451115.87215 & $-0.017$ & 2451115.86929 & 2451115.87002 & 0.174 & 2451115.82400 & 2451115.82473 & 0.427 \\
2451115.93012 & 2451115.93085 & $-0.375$ & 2451115.92841 & 2451115.92914 & $-0.053$ & 2451115.87038 & 2451115.87111 & 0.175 & 2451115.85564 & 2451115.85637 & 0.433 \\
2451115.93806 & 2451115.93879 & $-0.379$ & 2451115.93350 & 2451115.93423 & $-0.057$ & 2451115.92742 & 2451115.92815 & 0.139 & 2451115.85623 & 2451115.85696 & 0.435 \\
\enddata
\tablecomments{Differential magnitudes are in the sense $Variable - Comparison$, with BD\,+16 515 as the comparison star.
Table~\ref{tbl-BVRI} is published in its entirety in the electronic edition of the journal. A portion is shown here
for guidance regarding its form and content.}
\end{deluxetable}

\clearpage

\begin{deluxetable}{cc}
\tablewidth{400pt}
\tablecaption{Comparison Star BD\,+16 515 Standard Magnitudes \label{tbl-comparison}}
\tablehead{\colhead{Band} & \colhead{Magnitude} }
\startdata
$B$ & $10.718\pm0.011$ \\
$V$ & $9.433\pm0.011$  \\
$R_C$ & $8.803\pm0.026$ \\
$I_C$ & $8.246\pm0.019$ \\
\enddata
\tablecomments{The mean HJED of 31 observations in each band is 2456337.2525, 
ranging from 2456328.6090 to 2456350.6354.}
\end{deluxetable}

\clearpage

\begin{deluxetable}{lclclc} 
\tabletypesize{\scriptsize} 
\tablewidth{0.0pt} 
\tablecaption{Eclipse Timings Not in Table 1 of \citet{ibanoglu05} \label{tbl-times}} 
\tablehead{\colhead{Timing (HJED)} & \colhead{Ref.} & \colhead{Timing (HJED)} & \colhead{Ref.} & \colhead{Timing (HJED)} & \colhead{Ref.}} 
\startdata 
2443053.89343 & 1 & 2446006.91705 & 2 & 2454810.22895 & 4 \\ 
2443113.82994 & 1 & 2446025.67955 & 2 & 2454884.23699 & 4 \\ 
2443485.95520 & 1 & 2446798.59438 & 3 & 2455064.56636 & 4 \\ 
2444195.80596 & 1 & 2446823.61126 & 3 & 2455075.51140 & 4 \\ 
2444226.55528 & 1 & 2454028.45413 & 4 & 2455076.55364 & 4 \\ 
2445671.79622 & 2 & 2454055.55555 & 4 & 2455512.78406 & 5 \\
2455532.58899 & 5 & 2455545.61859 & 5 & 2455547.70336 & 5 \\
\enddata 
\tablerefs{(1) \citet{miranda}; (2) \citet{beach}; (3) \citet{eitter}; (4) \citet{hric}; \citet{hardy} } 
\end{deluxetable} 

\clearpage 

\begin{deluxetable}{lcccc}
\tabletypesize{\scriptsize}
\tablewidth{0pt}
\tablecaption{V471~Tau Data Sets \label{tbl-sets}}
\tablehead{\colhead{Reference} & \colhead{Data Type} & \colhead{Time Range (JD)} & \colhead{Number of Points} & \colhead{Calendar Range} }
\startdata
\citet{rucinski81} &  $u$, $v$, $b$, $y$ lc's &  2443076-2443085 & 232, 237, 237, 237 & Oct. - Nov. 1976 \\
This paper & $B$, $V$, $R_{C}$, $I_{C}$ lc's & 2451115-2451127 & 876, 850, 798, 643 & Oct. - Nov. 1998\\
\citet{kaminski} & MOST lc & 2453708-2453718 & 6001\tablenotemark{a} & Dec. 2005 \\
\citet{obrien} & RV1 & 2449643-2449651 & 7 & Oct. 1994 \\
\citet{obrien} & RV1 & 2449792.9 & 1 & March 1995 \\
\citet{young76} &  RV2 & 2440517-2441283 & 37 & Oct. 1969 - Nov. 1971 \\
\citet{bois88} &  RV2 & 2442797-2445648 & 202 & Jan. 1976 - Nov. 1983 \\
This paper &  RV2 & 2451115-2451127 & 126 & Oct. - Nov. 1998 \\
\citet{hussain} &  RV2 & 2452601-2452605 & 93 & Nov. 2002 \\
\citet{kaminski} &  RV2 & 2453717-2453724 & 37 & Dec. 2005 \\
\citet{ibanoglu05}; Table~\ref{tbl-times} & Eclipse Timings & 2440612-2455547 & 224 & Jan. 1970 - Dec. 2010 \\
\enddata
\tablenotetext{a}{Normal points; see comment in \S\ref{sec-lcs}}
\end{deluxetable}

\clearpage 
 
\begin{deluxetable}{cc}
\tablewidth{0pt}
\tablecaption{Curve-dependent $\sigma$'s \label{tbl-sigmas}}
\tablehead{\colhead{Curve} & \colhead{$\sigma$} }
\startdata
RV1 & 1.70 km/s \\
RV2 & 2.00 km/s \\
$u$ & $0.205 \times 10^{-6}$ \\
$v$ & $0.349 \times 10^{-6}$ \\
$b$ & $0.517 \times 10^{-6}$ \\
$y$ & $0.794 \times 10^{-6}$ \\
$B$ & $0.419 \times 10^{-6}$ \\
$V$ & $1.04 \times 10^{-6}$ \\
$R_{C}$ & $1.67 \times 10^{-6}$ \\
$I_{C}$ & $1.95 \times 10^{-6}$ \\
MOST normals & $0.484 \times 10^{-6}$ \\
Timings & 0.000175 d \\
\enddata
\end{deluxetable}

\clearpage

\begin{deluxetable}{ccccc}
\tabletypesize{\scriptsize}
\tablewidth{0pt}
\tablecaption{Spot Parameters \label{tbl-spots}}
\tablehead{\colhead{Co-latitude (rad)} & \colhead{Longitude (rad)} & \colhead{Radius (rad)} & \colhead{$T_{\rm spot}/T2$} & \colhead{Fractional Spot Area} }
\startdata
  0.632 & 0.021 & 0.514 & 0.779 & 0.065 \\ 
  2.352 & 5.128 & 0.222 & 0.903 & 0.012 \\ 
  0.582 & 1.888 & 0.448 & 0.853 & 0.049 \\ 
 & & & \\
  0.471 & 4.101 & 0.355 & 0.798 & 0.031 \\ 
  2.342 & 2.405 & 0.373 & 0.930 & 0.035 \\ 
  2.132 & 5.636 & 0.391 & 0.879 & 0.038 \\ 
 & & & \\
  0.329 & 1.705 & 0.223 & 0.599 & 0.012 \\ 
  0.575 & 3.549 & 0.117 & 0.727 & 0.003 \\ 
  2.040 & 0.165 & 0.053 & 0.988 & 0.001 \\ 
\enddata
\tablecomments{The first three spots are for epochs HJED 2442076 to 2444086, with maximum spot radii between
HJED 2443076 and HJED 2443086, the observational window for the \citet{rucinski81} light curves. 
The second triad of spots is for epochs HJED 2450115-2452127, with maximum spot radiii between HJED 2451115 and 2451127
(the KPNO observational window). The last triad is for epochs HJED 2452713-2454724, with the spots 
reaching maximum radii between HJED 2453713 and 2453724 (the MOST observational window). Spot parameters are
defined in \S\ref{spots}}
\end{deluxetable}

\clearpage

\begin{deluxetable}{lccccc}
\rotate
\tabletypesize{\scriptsize}
\tablewidth{0.0pt}
\tablecaption{V471~Tau Multi-data-type Solutions \label{tbl-sol}}
\tablehead{\colhead{Parameter} &  \colhead{All Data Excluding MOST} & \colhead{Times Only} & \colhead{MOST Light Curve Only}
 & \colhead{All Data Excluding MOST} & \colhead{All Data} \\
 &  &  &  & \colhead{with Fixed Radii from MOST} &}
\startdata
$a$ ($R_\sun$) & $3.3827\pm0.0024$ & \nodata & 3.3827 & $3.3628\pm0.0024$ & $3.3608\pm0.0012$ \\
$V_\gamma$ (km\,s$^{-1}$) & $36.91\pm0.13$ & \nodata & \nodata & $36.91\pm0.13$ & $36.88\pm0.11$  \\
$i$ (deg) & $78.796\pm0.059$ & \nodata & 78.796 & $78.809\pm0.055$ & $78.755\pm0.030$  \\
$T_{1}$ (K) & 34500 & \nodata & 34500 & 34500 & 34500 \\
$T_{2}$ (K) & $5019\pm8$ & \nodata & 5019 & $5084\pm8$ & $5066\pm4$ \\
$\Omega_1$ & $338.1\pm3.1$ & \nodata & $315.50\pm0.86$ & $315.50$ & $319.1\pm1.4$  \\
$\Omega_2$ & $5.0931\pm0.0068$ & \nodata & $5.0691\pm0.0018$ & $5.0691$ & $5.0646\pm0.0033$  \\
$M_2/M_1$ & $1.1386\pm0.0025$ & \nodata & 1.1386 & $1.1332\pm0.0026$ & $1.1360\pm0.0016$ \\
$T_{0}$ (${\rm HJED}-2445821.0$) & $0.898269\pm0.000036$ & $0.898293\pm0.000043$ & $0.898963\pm0.000038$ & $0.898270\pm0.000036$ & $0.898291\pm0.000030$ \\
$P_{0}$\tablenotemark{a} (d) & $0.5211833840(32)$ & $0.5211833822(39)$ & $0.5211833840$ & $0.5211833844(32)$ & $0.5211833875(27)$ \\
$dP/dt$ & $(+0.295\pm0.012) \times 10^{-10}$ &  $(+0.297\pm0.015) \times 10^{-10}$ &  $+0.295 \times 10^{-10}$ &  $(+0.293\pm0.013) \times 10^{-10}$ & $+0.286\pm0.011 \times 10^{-10}$ \\
$L_{1}/\left( L_1+L_2\right) _{u}$ & $0.2533\pm0.0014$ & \nodata & \nodata & $0.2511\pm0.0014$ & $0.2531\pm0.0011$ \\
$L_{1}/\left( L_1+L_2\right) _{v}$ & $0.07699\pm0.00046$ & \nodata & \nodata & $0.07868\pm0.00046$ & $0.07876\pm0.00032$ \\
$L_{1}/\left( L_1+L_2\right) _{b}$ & $0.03159\pm0.00025$ & \nodata & \nodata & $0.03316\pm0.00026$ & $0.03296\pm0.00014$ \\
$L_{1}/\left( L_1+L_2\right) _{y}$ & $0.01775\pm0.00016$ & \nodata & \nodata & $0.01883\pm0.00016$ & $0.018660\pm0.000084$   \\
$L_{1}/\left( L_1+L_2\right) _{B}$ & $0.05239\pm0.00036$  & \nodata & \nodata & $0.05435\pm0.00036$ & $0.05419\pm0.00022$  \\
$L_{1}/\left( L_1+L_2\right) _{V}$ & $0.01829\pm0.00016$ & \nodata & \nodata & $0.01935\pm0.00017$ & $0.019193\pm0.000085$  \\
$L_{1}/\left( L_1+L_2\right) _{R_{C}}$ & $0.01041\pm0.00011$ & \nodata & \nodata & $0.01117\pm0.00011$ & $0.011035\pm0.000054$ \\
$L_{1}/\left( L_1+L_2\right) _{I_{C}}$ & $0.006190\pm0.000070$ & \nodata & \nodata & $0.006700\pm0.000073$ &$0.006604\pm0.000035$ \\
$L_{1}/\left( L_1+L_2\right) _{\rm MOST}$ & \nodata  & \nodata & $0.02133\pm0.00011$ & \nodata & $0.019859\pm0.000090$ \\
$a_{3b}$ ($R_{\sun}$) & $2588\pm11$  & $2576\pm13$  & 2588 & $2576\pm11$ & $2582\pm10$ \\
$P_{3b}$ (d) & $10932\pm71$ & $10960\pm83$ & $10932$ & $10949\pm72$ & $10996\pm60$ \\
$i_{3b}$ (deg) & 90  & 90 &  90 & 90 & 90 \\
$e_{3b}$  & $0.418\pm0.022$ & $0.435\pm0.026$ & $0.418$ & $0.417\pm0.022$ & $0.392\pm0.018$ \\
$\omega_{3b}$ (radians) & $1.194\pm0.053$  & $1.180\pm0.060$ & $1.194$ & $1.187\pm0.053$ & $1.369\pm0.047$ \\
${T_0}_{3b}$ (HJED) & $2441781\pm53$ & $2441784\pm61$ & $2441781$ & $2441773\pm53$ & $2441910\pm48$ \\
$f(m_3)$ ($M_\sun$) & $1.14 \times 10^{-5}$ & \nodata & $1.14 \times 10^{-5}$ & $1.15 \times 10^{-5}$ & $1.17 \times 10^{-5}$ \\
$a$ vs. $a_{3b}$ correlation & 0.128 & \nodata & \nodata & 0.127 & $-0.040$ \\
\enddata
\tablenotetext{a}{Printed to two significant figures in its standard error, with the number in parentheses the standard error in the last printed digits.}
\end{deluxetable}

\clearpage

\begin{deluxetable}{lcc}
\tabletypesize{\scriptsize}
\tablewidth{0pt}
\tablecaption{V471 Tau Auxiliary Parameters and Absolute Dimensions \label{tbl-aux} }
\tablehead{\colhead{Parameter} & \colhead{Star 1} & \colhead{Star 2} }
\startdata
\cutinhead{All Data Solution Excluding MOST}
$r$(pole) & $0.002978\pm0.000027$ & $0.27195\pm0.00053$ \\
$r$(point) & $0.002978\pm0.000027$ & $0.28857\pm0.00071$ \\
$r$(side) & $0.002978\pm0.000027$ & $0.27731\pm0.00057$ \\
$r$(back) & $0.002978\pm0.000027$ & $0.28449\pm0.00065$ \\
$<r>$\tablenotemark{a} & $0.002978\pm0.000027$ & $0.27738\pm0.00047$ \\
$K$ (km\,s$^{-1}$) & 171.5 & 150.6 \\
$M$ ($M_{\sun}$) & $0.8939\pm0.0018$ & $1.0177\pm0.0021$ \\
$R$ ($R_{\sun}$) & $0.010040\pm0.000092$ & $0.9383\pm0.0019$ \\
$\log g$ (cm\,s$^{-2}$) & $8.3860\pm0.0080$ & $4.5011\pm0.0017$ \\
$\mathrm{M_{bol}}$ & $6.982\pm0.021$ & $5.5007\pm0.0091$ \\
\cutinhead{All Data Solution with Fixed Radii from MOST}
$r$(pole) & $0.003052\pm0.000026$ & $0.27109\pm0.00051$ \\
$r$(point) & $0.003052\pm0.000026$ & $0.28761\pm0.00069$ \\
$r$(side) & $0.003052\pm0.000026$ & $0.27641\pm0.00056$ \\
$r$(back) & $0.003052\pm0.000026$ & $0.28357\pm0.00063$ \\
$<r>$\tablenotemark{a} & $0.003052\pm0.000028$ & $0.27790\pm0.00047$ \\
$K$ (km\,s$^{-1}$) & 170.1 & 150.1 \\
$M$ ($M_{\sun}$) & $0.8804\pm0.0018$ & $0.9977\pm0.0020$ \\
$R$ ($R_{\sun}$) & $0.010697\pm0.000096$ & $0.9345\pm0.0018$ \\
$\log g$ (cm\,s$^{-2}$) & $8.3243\pm0.0078$ & $4.4960\pm0.0016$ \\
$\mathrm{M_{bol}}$ & $6.844\pm0.021$ & $5.4536\pm0.0091$ \\
\cutinhead{All Data Solution}
$r$(pole) & $0.003137\pm0.000014$ & $0.27240\pm0.00028$ \\
$r$(point) & $0.003137\pm0.000014$ & $0.28920\pm0.00038$ \\
$r$(side) & $0.003137\pm0.000014$ & $0.27780\pm0.00031$ \\
$r$(back) & $0.003137\pm0.000014$ & $0.28507\pm0.00035$ \\
$<r>$\tablenotemark{a} & $0.003137\pm0.000014$ & $0.27883\pm0.00025$ \\
$K$ (km\,s$^{-1}$) & 170.2 & 149.8 \\
$M$ ($M_{\sun}$) & $0.8778\pm0.0011$ & $0.9971\pm0.0012$ \\
$R$ ($R_{\sun}$) & $0.010571\pm0.000047$ & $0.93709\pm0.00093$ \\
$\log g$ (cm\,s$^{-2}$) & $8.3333\pm0.0038$ & $4.49331\pm0.00087$ \\
$\mathrm{m_{bol}}$ & $6.870\pm0.010$ & $5.4630\pm0.0051$ \\
\enddata
\tablenotetext{a}{Brackets indicate equal-volume radii. Relative radius, $r$, is $R/a$.}
\end{deluxetable}

\clearpage

\begin{deluxetable}{cc}
\tablewidth{0pt}
\tablecaption{Third Body Mass \label{tbl-3bmass}}
\tablehead{\colhead{$i_{3b}$} & \colhead{$M_3$ ($M_\sun$)}}
\startdata
$90\degr$ & $0.03498\pm0.00046$ \\
$60\degr$ & $0.04047\pm0.00054$ \\
$30\degr$ & $0.07083\pm0.00096$ \\
\enddata
\end{deluxetable}

\clearpage

\begin{deluxetable}{lccccc}
\tabletypesize{\scriptsize}
\rotate
\tablewidth{0pt}
\tablecaption{V471~Tau Individual Radial Velocity Solutions \label{tbl-rvind}}
\tablehead{\colhead{Parameter} & \colhead{\citet{young76}} & \colhead{\citet{bois88}} &
\colhead{Table~\ref{tbl-v471rv}} & \colhead{\citet{hussain}} & \colhead{\citet{kaminski}} }
\startdata
$a$ ($R_{\sun}$) & $3.31\pm0.12$ & $3.263\pm0.048$  & $3.320\pm0.043$ & $3.384\pm0.057$ & $3.317\pm0.045$ \\
$V_{\gamma}$ (km\,s$^{-1}$) & $38.4\pm1.4$ & $37.03\pm0.36$  & $35.54\pm0.36$ & $37.58\pm0.20$ & $35.94\pm0.25$  \\
$M_2/M_1$ & $1.150\pm0.081$ & $1.101\pm0.031$ & $1.109\pm0.028$ & $1.114\pm0.036$ & $1.089\pm0.029$ \\
$K_1$ (km\,s$^{-1}$) & 168.6 & 162.8 & 166.2 & 169.7 & 164.6 \\
$M_1$ ($M_{\sun}$) & $0.83$ & $0.82$ & $0.86$ & $0.91$ & $0.86$ \\
$R_1$ ($R_{\sun}$) & $0.0104$ & $0.0.0103$ & $0.0104$ & $0.0106$ & $0.0104$ \\
$K_2$ (km\,s$^{-1}$) & 146.6 & 147.8 & 149.9 & 152.4 & 151.2 \\
$M_2$ ($M_{\sun}$) & $0.96$ & $0.90$ & $0.95$ & $1.01$ & $0.94$ \\
$R_2$ ($R_{\sun}$) & $0.93$ & $0.88$ & $0.91$ & $0.93$ & $0.89$  \\
\enddata
\tablecomments{Solutions are for the \citet{obrien} white dwarf radial velocities reduced by 10\,km\,s$^{-1}$ (see text for explanation),
combined with each of the red dwarf velocity sets. There is no 3b light-time effect in the tabulated solutions. Separate solutions
with a light-time effect gave very nearly the same results.
Standard errors do not include uncertainties of parameters held fixed.
Radial velocity semi-amplitudes ($K_1$, $K_2$) are provided for readers who prefer to solve radial velocities
in terms of semi-amplitudes instead of direct astrophysical parameters.}
\end{deluxetable}

\clearpage

\begin{deluxetable}{ccc}
\tabletypesize{\scriptsize}
\tablewidth{0pt}
\tablecaption{BD\,$-3$ 5358 Magnitudes \label{bd5358} }
\tablehead{\colhead{HJD} & \colhead{HJED} & \colhead{Standard Magnitude}  \\
\colhead{($- 2450000.0$)} & \colhead{($- 2450000.0$)} & }
\startdata
 5812.6603 & 5812.6611 & $B= 11.036\pm0.002$ \\
 5833.6032 & 5833.6040 & $B= 11.032\pm0.002$ \\
 6199.7300 & 6199.7308 & $B= 11.054\pm0.008$ \\
 6235.6286 & 6235.6294 & $B= 10.992\pm0.006$ \\
 5812.6600 & 5812.6608 & $V= 10.355\pm0.001$ \\
 6199.7301 & 6199.7309 & $V= 10.370\pm0.006$ \\
 6235.6286 & 6235.6293 & $V= 10.396\pm0.005$ \\
 6199.7344 & 6199.7352 & $g'= 10.665\pm0.005$ \\
 6235.6330 & 6235.6338 & $g'= 10.664\pm0.004$ \\
 5833.6041 & 5833.6049 & $r'= 10.322\pm0.001$ \\
 6199.7305 & 6199.7312 & $r'= 10.160\pm0.005$ \\
 6235.6289 & 6235.6297 & $r'= 10.192\pm0.004$ \\
 5812.6627 & 5812.6634 & $i'= 10.008\pm0.002$ \\
 5833.6054 & 5833.6062 & $i'= 10.127\pm0.002$ \\
 6199.7308 & 6199.7316 & $i'= 10.036\pm0.010$ \\
 6235.6293 & 6235.6301 & $i'= 10.025\pm0.008$ \\
\enddata
\tablecomments{Bands $g'$, $r'$, and $i'$ are Sloan Digital Sky Survey systems. Transformations from
Sloan to $R_C$ and $I_C$ have been developed by U. Munari (Henden, private communication). The transformations are 
\[
R_C  =  r'-0.17122 - 0.07747(V-i') - 0.02902 (V-i')^2, 
\]
and
\[
I_C  =  i' - 0.37313 -  0.11431 (V-i') + 0.01066 (V-i')^2. 
\]
}
\end{deluxetable}

\clearpage

\begin{deluxetable}{cccccccc}
\rotate
\tabletypesize{\scriptsize}
\tablewidth{0.0pt}
\tablecaption{Absolute Solutions and Distance \label{tbl-abssol}}
\tablehead{\colhead{$T_1$ (K)} & \colhead{$T_2$ (K)}
& \colhead{$u$}
& \colhead{$v$}
& \colhead{$b$}
& \colhead{$y$}
& \colhead{$B$}
& \colhead{$V$}}
\startdata
34500 & 5000 & $49.73\pm0.45$ & $53.50\pm0.41$ & $54.45\pm0.42$ & $54.59\pm0.66$ & $52.14\pm0.15$ & $52.13\pm0.28$ \\
34500 & 4900 & $45.57\pm0.43$ & $49.27\pm0.37$ & $50.96\pm0.39$ & $51.46\pm0.62$ & $48.43\pm0.13$ & $49.05\pm0.25$ \\
34500 & 4800 & $41.79\pm0.42$ & $45.22\pm0.34$ & $47.53\pm0.36$ & $48.36\pm0.57$ & $44.84\pm0.12$ & $45.99\pm0.22$ \\
33500 & 5000 & $49.42\pm0.45$ & $53.41\pm0.40$ & $54.41\pm0.42$ & $54.57\pm0.66$ & $52.08\pm0.15$ & $52.11\pm0.28$ \\
33500 & 4900 & $45.23\pm0.42$ & $49.17\pm0.37$ & $50.92\pm0.38$ & $51.44\pm0.61$ & $48.37\pm0.13$ & $49.03\pm0.25$ \\
33500 & 4800 & $41.42\pm0.41$ & $45.11\pm0.34$ & $47.48\pm0.35$ & $48.34\pm0.57$ & $44.77\pm0.12$ & $45.97\pm0.22$ \\
\enddata
\tablecomments{Distance in parsecs from single-band Direct Distance
  Estimation (DDE) solutions is given for stepped input of two white
  dwarf and three red dwarf temperatures, so as to allow interpolation
  if knowledge of the star temperatures improves. The standard errors
  are from individual solutions that also adjusted the full set of
  parameters from our all-data solution. Starting values of other parameters
  are from the all-data non-absolute solution of
  Table~\ref{tbl-sol}.}
\end{deluxetable} 
\clearpage

\begin{deluxetable}{ccccc} 
\tabletypesize{\scriptsize} 
\tablewidth{0.0pt} 
\tablecaption{Timing Residuals Fitted Sinusoids \label{tbl-ressinfit}}
\tablehead{\colhead{$\mathcal{A}$ (d)} & \colhead{$\mathcal{B}$ (d)} & \colhead{$t_0$ (HJED)} & \colhead{$P$ (yr)} & \colhead{$\sigma$ (d)} } 
\startdata 
$(0.02\pm0.12) \times 10^{-4}$ & $(0.81\pm0.17) \times 10^{-4}$ & $2443797\pm158$ & $13.01\pm0.43$ & 0.0001669 \\
$(0.02\pm0.11) \times 10^{-4}$ & $(0.70\pm0.16) \times 10^{-4}$ & $2443989\pm77$ & $5.486\pm0.093$ & 0.0001685 \\
$(0.10\pm0.12) \times 10^{-4}$ & $(0.66\pm0.17) \times 10^{-4}$ & $2445573\pm125$ & $9.00\pm0.26$ & 0.0001694 \\
\enddata 
\end{deluxetable} 
 
\clearpage

\begin{deluxetable}{ccccccc}
\rotate
\tabletypesize{\scriptsize}
\tablewidth{0.0pt}
\tablecaption{Magnitude and Velocity Residual Fitted Sinusoids \label{tbl-lcrvressinfit}}
\tablehead{\colhead{Band} & \colhead{$\mathcal{A}$ (mag)} & \colhead{$\mathcal{B}$ (mag)} & \colhead{$t_0$ (HJED)} & \colhead{$P$ (d)} & \colhead{$P_{harmonic}$ (d)} & \colhead{$\sigma$ (mag)} }
\startdata
\cutinhead{Periodicities from Light Curve Residuals}
MOST & $(-0.12\pm0.37) \times 10^{-4}$ & $0.001913\pm0.000052$ & $2453713.4518\pm0.0011$ & $0.256040\pm0.000097$ & $0.26059$ & $0.002849$ \\
MOST & $(-0.17\pm0.37) \times 10^{-4}$ & $0.001878\pm0.000052$ & $2453713.3206\pm0.0012$ & $0.26568\pm0.00011$ & $0.26059$ & $0.002853$ \\
MOST & $(-0.14\pm0.40) \times 10^{-4}$ & $0.000954\pm0.000056$ & $2453713.3726\pm0.0016$ & $0.172546\pm0.000095$ & $0.17373$ & $0.003079$ \\
MOST & $(-0.09\pm0.40) \times 10^{-4}$ & $0.000818\pm0.000058$ & $2453713.4947\pm0.0014$ & $0.130151\pm0.000062$ & $0.13030$ & $0.003100$ \\
$B$ & $(+0.26\pm0.26) \times 10^{-3}$ & $0.00391\pm0.00036$ & $2451122.3797\pm0.0040$ & $0.26294\pm0.00038$ & $0.26059$ & $0.0077$ \\
$V$ & $(-0.019\pm0.25) \times 10^{-3}$ & $0.00176\pm0.00038$ & $2451122.3968\pm0.0083$ & $0.27644\pm0.00081$ & $0.26059$ & $0.0072$ \\
$R_{C}$ & $(-0.080\pm0.25) \times 10^{-3}$ & $0.00144\pm0.00035$ & $2451122.421\pm0.011$ & $0.2734\pm0.0010\phantom{0}$ & $0.26059$ & $0.0071$ \\
$I_{C}$ & $(-0.21\pm0.26) \times 10^{-3}$ & $0.00267\pm0.00038$ & $2451122.6890\pm0.0058$ & $0.27675\pm0.00055$ & $0.26059$ & $0.0065$ \\
\cutinhead{Periodicities from Red Dwarf Radial Velocity Residuals}
\colhead{RV Set} & \colhead{$\mathcal{A}$ (km\,s$^{-1}$)} & \colhead{$\mathcal{B}$ (km\,s$^{-1}$)} & \colhead{$t_0$ (HJED)} & \colhead{$P$ (d)} & \colhead{$P_{harmonic}$} & \colhead{$\sigma$ (km\,s$^{-1}$)}  \\
\tableline
RV2 (Bois) & $-0.08\pm0.49$ & $2.35\pm0.67$ & $2445646.188\pm0.012$ & $0.2546\pm0.0025$ & $0.2606$ & 3.1421 \\
RV2 (KPNO) & $-2.00\pm0.33$ & $2.95\pm0.47$ & $2451121.7252\pm0.0066$ & $0.26029\pm0.00053$ & $0.2606$ & 3.6878 \\
RV2 (Hussain) & $0.42\pm0.16$ & $1.69\pm0.23$ & $2452603.1660\pm0.0057$ & $0.2593\pm0.0012$ & $0.2606$ & 1.5842 \\
\enddata
\end{deluxetable}

\clearpage

\begin{deluxetable}{lcc}
\tabletypesize{\scriptsize}
\tablewidth{0.0pt}
\tablecaption{Eccentric V471~Tau Red Dwarf Radial Velocity Solutions \label{tbl-rv2sol}}
\tablehead{\colhead{Parameter} &  \colhead{RVs from \citet{hussain}} & \colhead{RVs from \citet{kaminski}} }
\startdata
$a$ ($R_{\sun}$) & $3.4692\pm0.0026$  & $3.437\pm0.012$ \\ 
$V_{\gamma}$ (km\,s$^{-1}$) & $37.28\pm0.08$ & $35.79\pm0.34$  \\ 
$e$ & $0.00842\pm0.00074$  & $0.0019\pm0.0031$ \\

$\omega$ (radians) & $5.273\pm0.086$ &  $3.9\pm1.6$ \\
$T_{0}$ (${\rm HJED}-2445821.0$) & $0.90028\pm0.00012$ & $0.89831\pm0.00029$ \\ 
\enddata
\tablecomments{Other parameters were fixed at values for the ``all-data, no-MOST, MOST-radii'' solution (column 5 of Table~\ref{tbl-sol}).}
\end{deluxetable}

\clearpage

\begin{deluxetable}{lcccccc}
\tabletypesize{\scriptsize}
\rotate
\tablewidth{0pt}
\tablecaption{Estimated Brown Dwarf (BD) Magnitude Limits and Related Quantities \label{tbl-brown}}
\tablehead{\colhead{$M/M_\sun$} & \colhead{$T_{eff}$ (K)} & \colhead{$R/R_\sun$} &
\colhead{$\log_{10} g$ (cgs)} & \colhead{Surface $m_H$} & \colhead{$m_H$ (49 pc)} & \colhead{$M_{H}({\rm BD}) - M_{H}({\rm EB})$} }  
\startdata
\cutinhead{Age 625 Myr}
0.0350 & 1325 & 0.0984 & 4.992 & $-34.730$ & 16.99 & 9.65 \\
$2 \times  0.0175$\tablenotemark{a} & 812 & 0.1042 & 4.640 & $-32.296$ & 18.54 & 11.21 \\
\cutinhead{Age 800 Myr}
0.0350 & 1221 & 0.0963 & 5.010 & $-34.461$ & 17.30 & 9.96 \\
$2 \times 0.0175$\tablenotemark{a} & 751 & 0.1027 & 4.655 & $-31.820$ & 19.05 & 11.71 \\
\enddata
\tablenotetext{a}{
The third body is a binary of equal-mass objects with a total mass of $0.0350\,M_\sun$.}
\end{deluxetable}

\clearpage

\begin{figure}
\plottwo{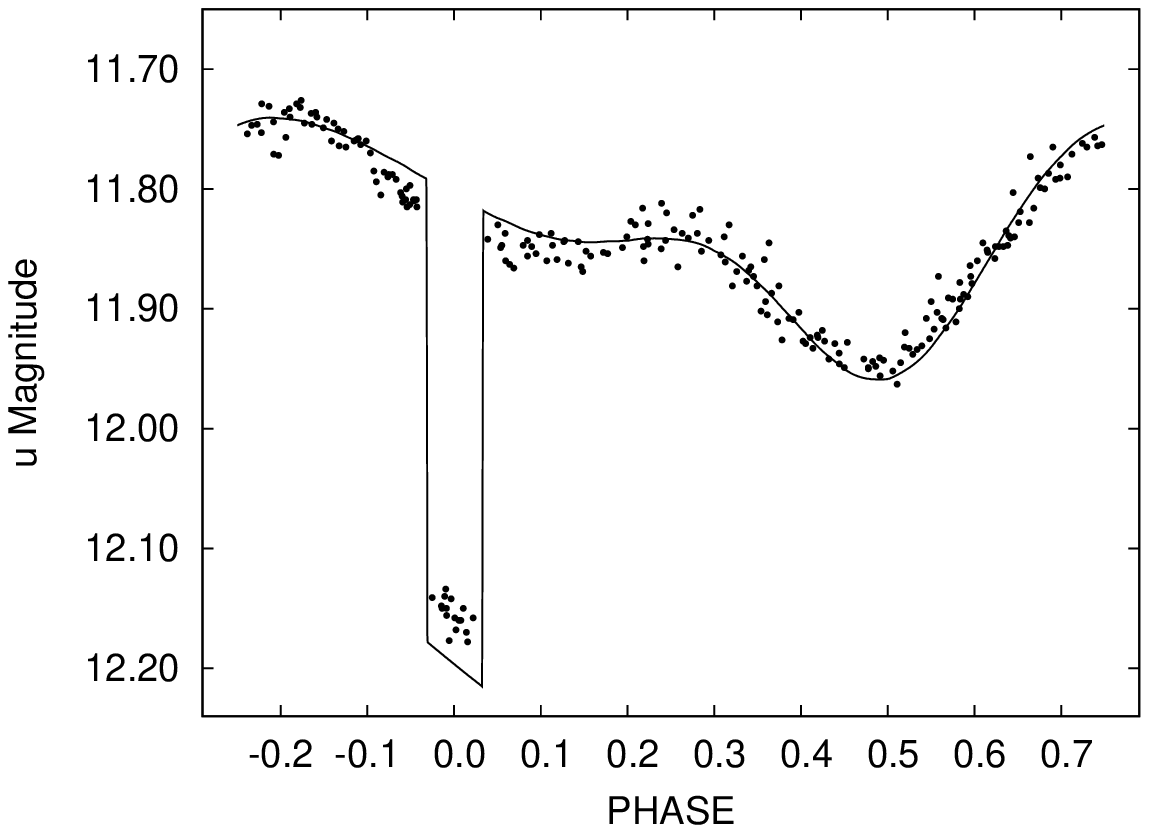}{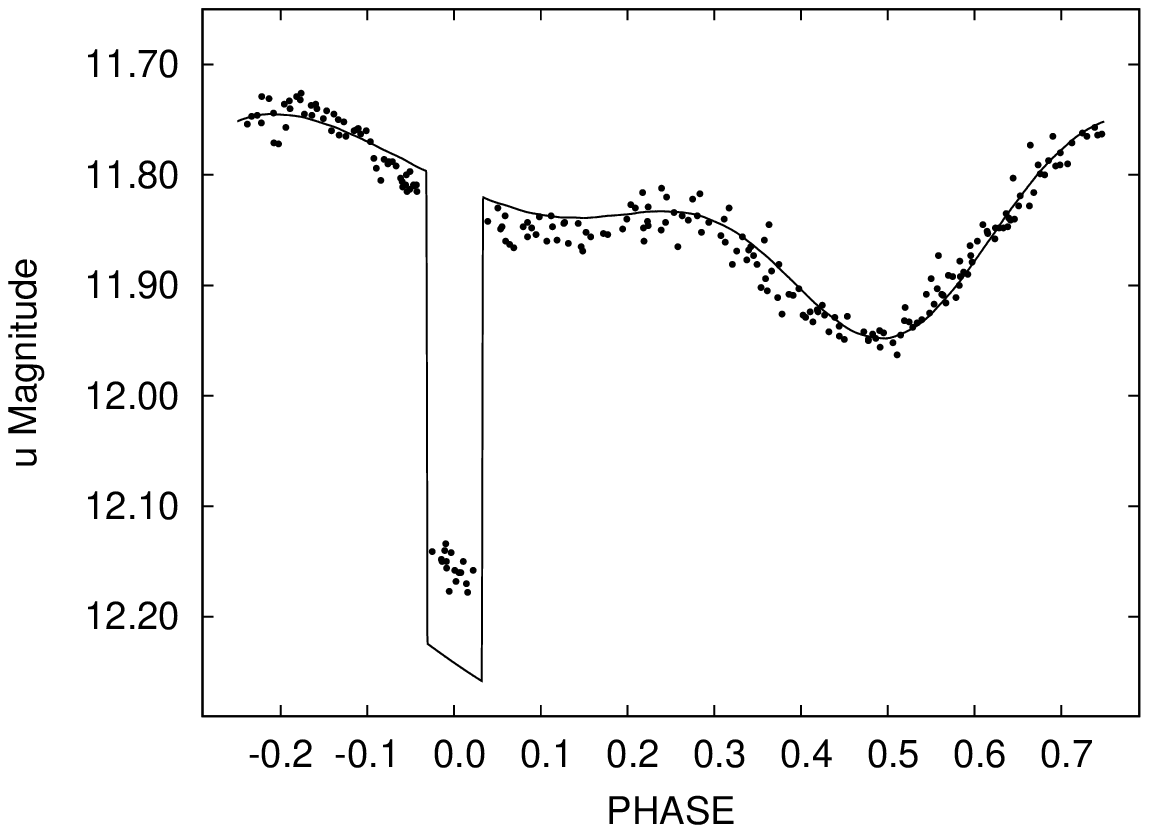}
\caption{V471~Tau phased $u$ band data of Nov. -- Dec., 1976 by \citet{rucinski81} with 
solution curves. The left and right panels are based on the All Data Excluding MOST solution (Table~\ref{tbl-sol} second column)
and the All Data Excluding MOST with MOST-fixed-radii solution (Table~\ref{tbl-sol} fifth 
column) parameters, respectively. Variation due to spots is largest here among this paper's analyzed epochs.} \label{v471_u} 
\end{figure}

\begin{figure}
\plottwo{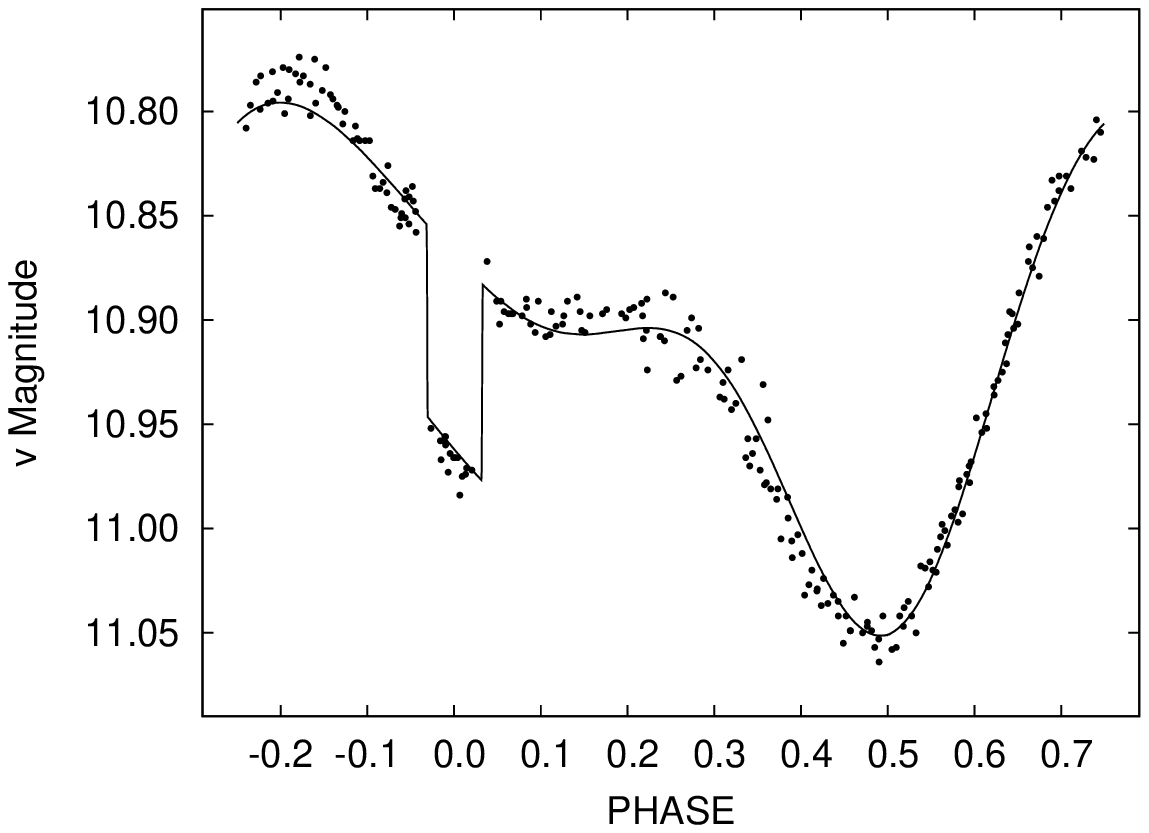}{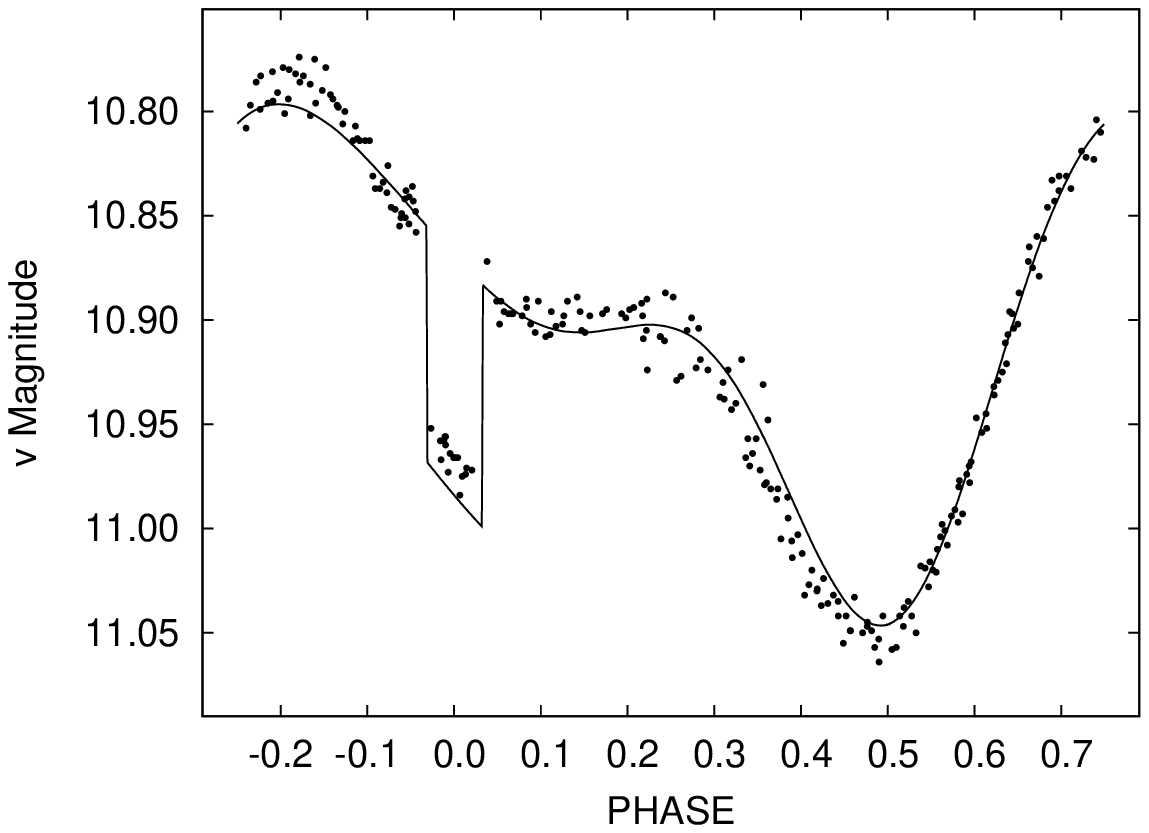}
\caption{Same as Figure~\ref{v471_u} but for the $v$ band.} \label{v471_v}
\end{figure}

\begin{figure}
\plottwo{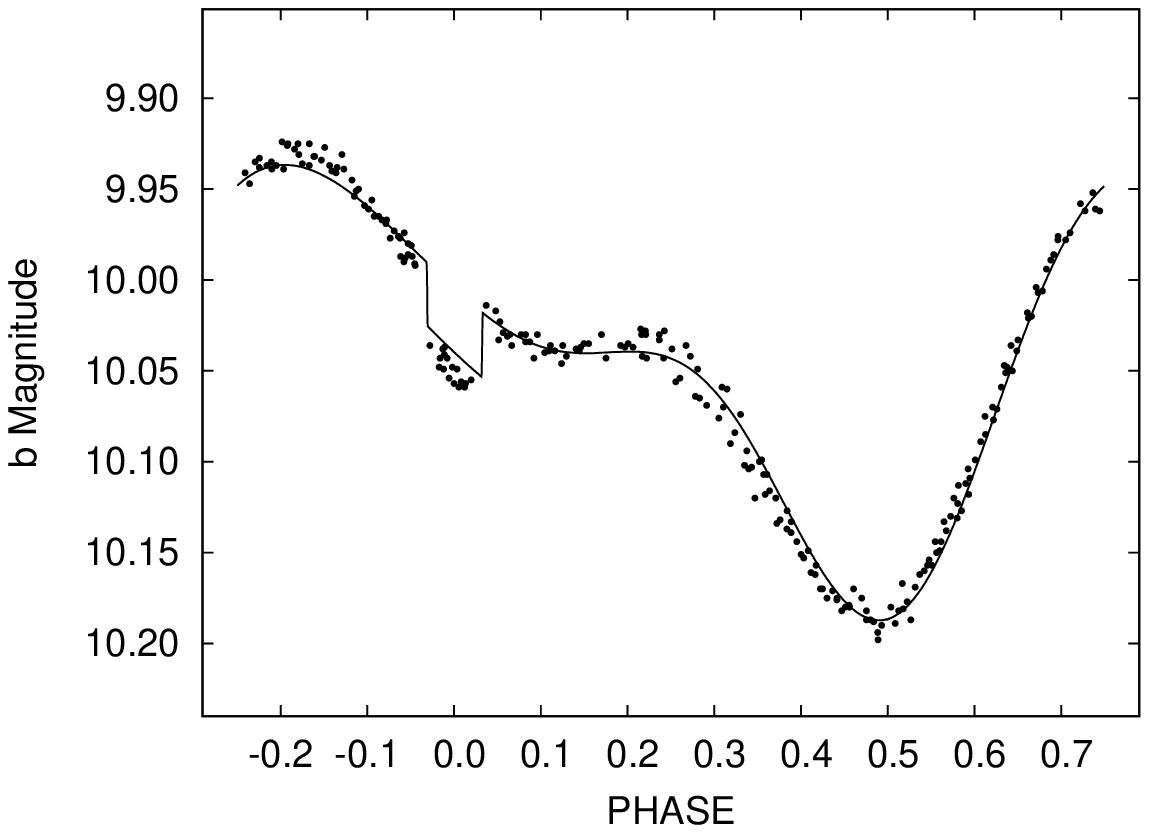}{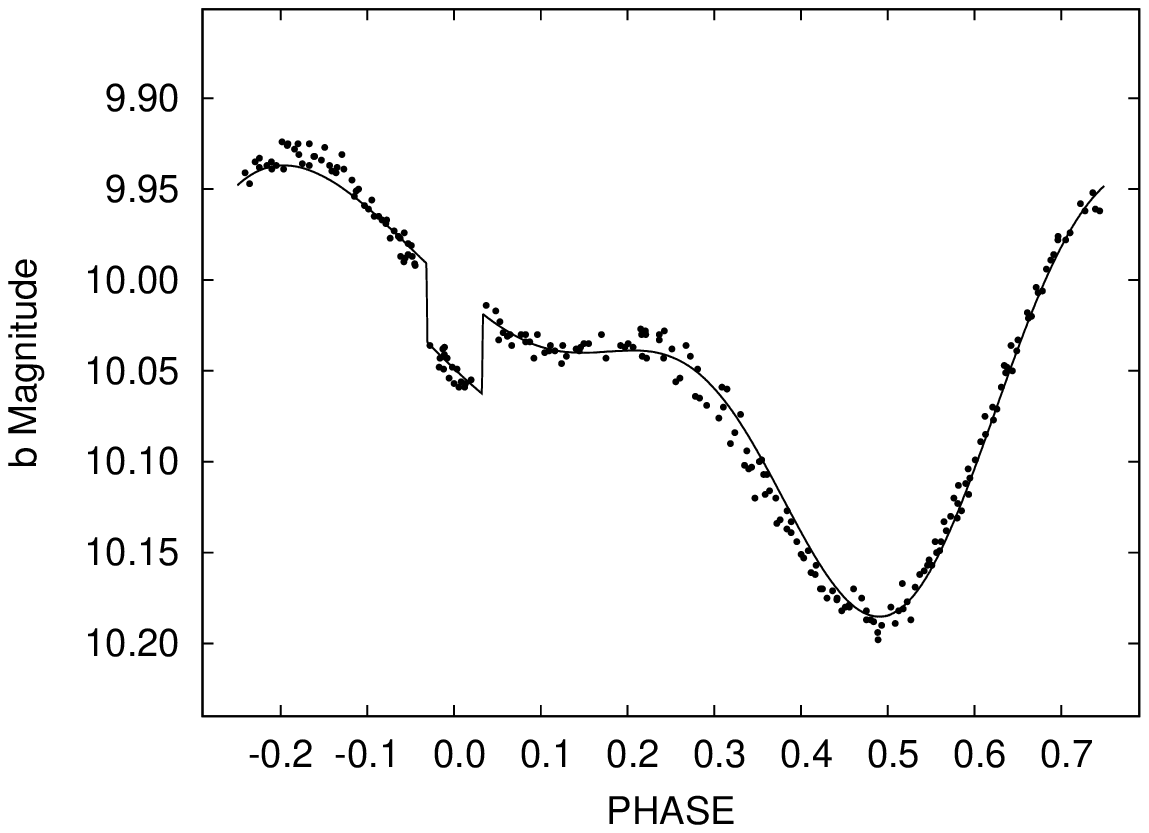}
\caption{Same as Figure~\ref{v471_u} but for the $b$ band.} \label{v471_b}
\end{figure}

\begin{figure}
\plottwo{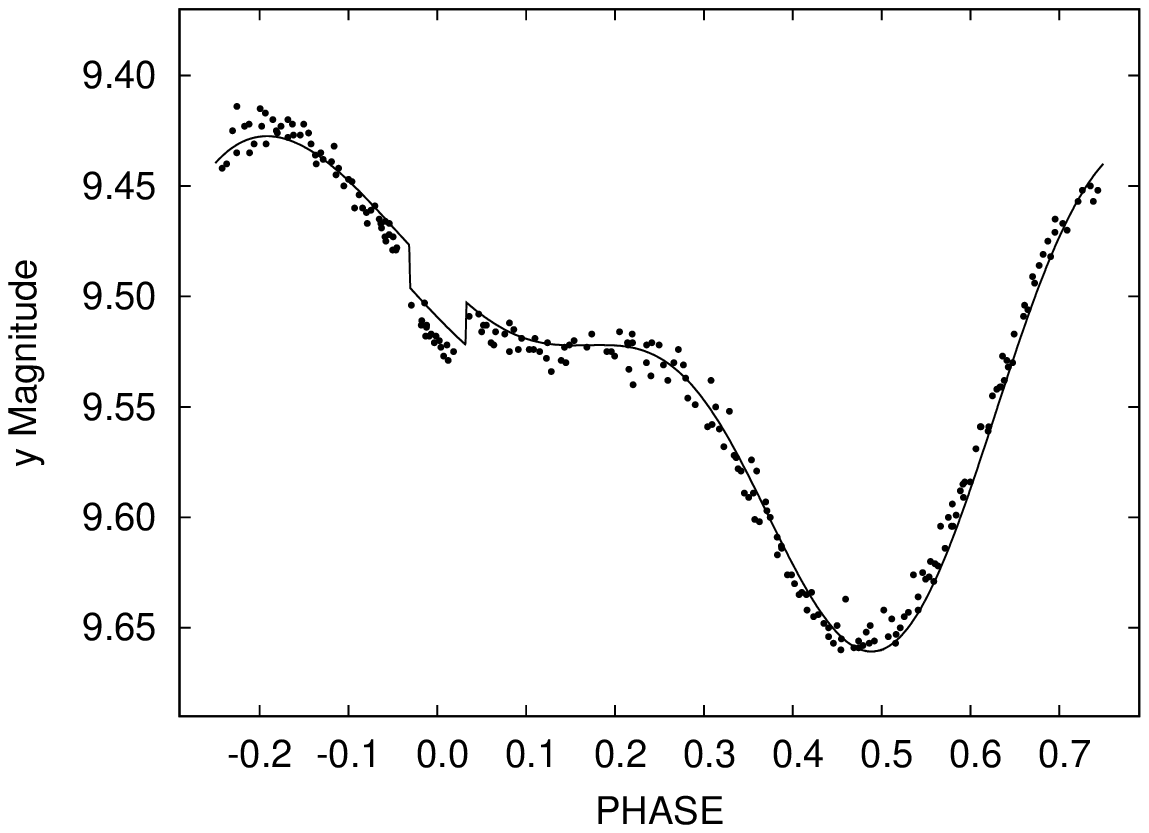}{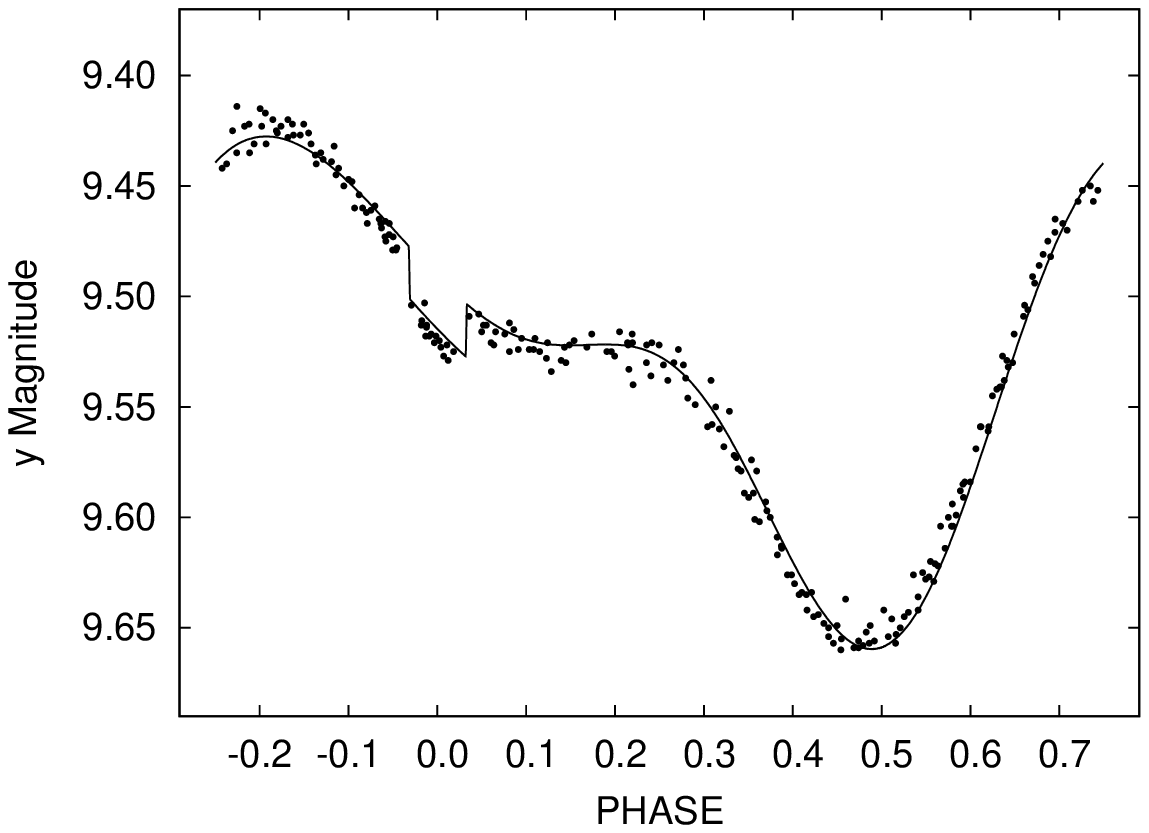}
\caption{Same as Figure~\ref{v471_u} but for the $y$ band.} \label{v471_y}
\end{figure}

\begin{figure}
\plottwo{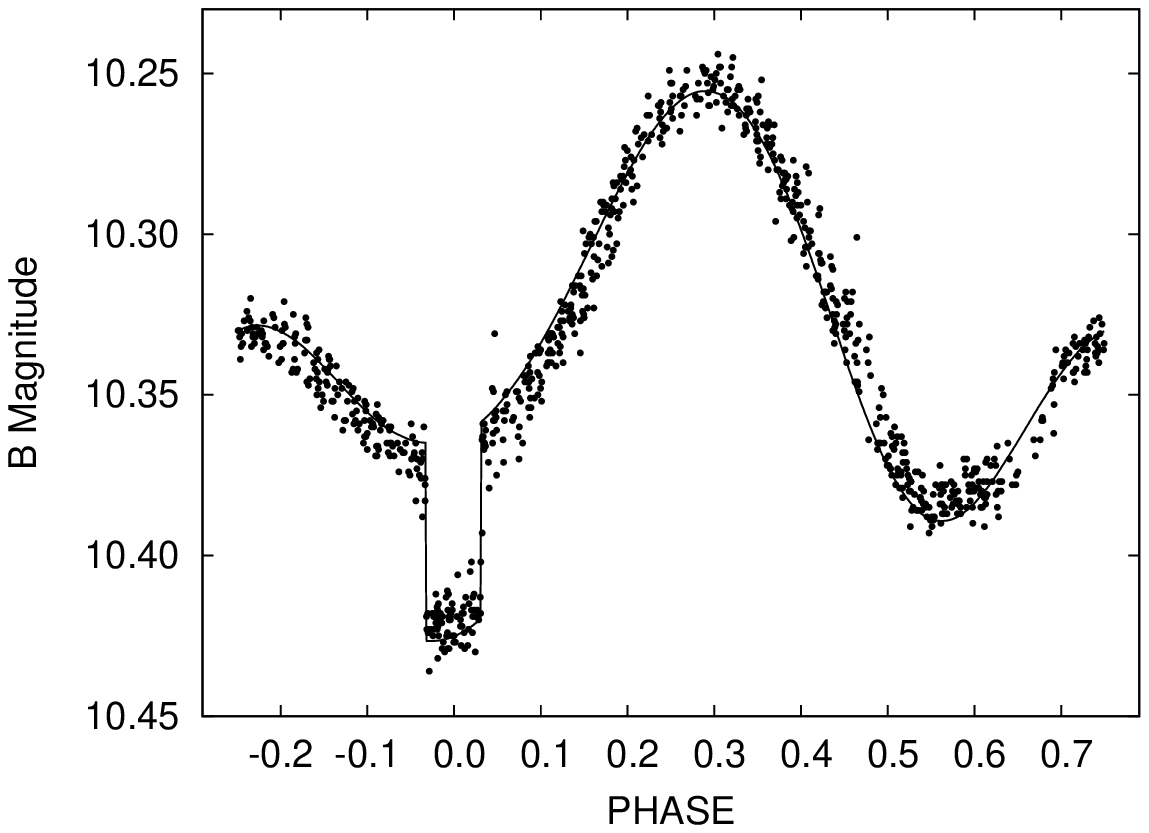}{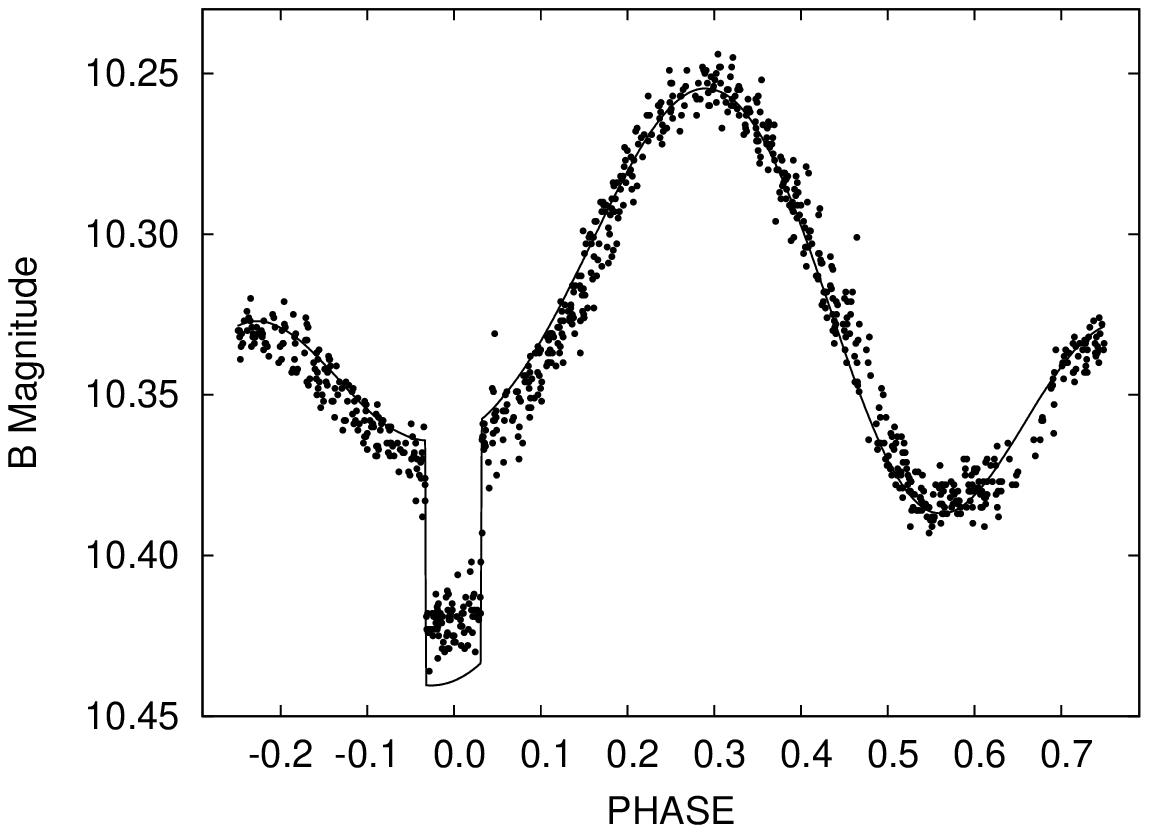}
\caption{Same as Figure~\ref{v471_u} but for the Nov. 1998 $B$ band observations of this paper. 
Note that the amplitude is much smaller than in the Oct.-Nov. 1976 observations of 
Figs. 1 to 4, allowing for the difference in vertical scales.} \label{v471_B}
\end{figure}

\begin{figure}
\plottwo{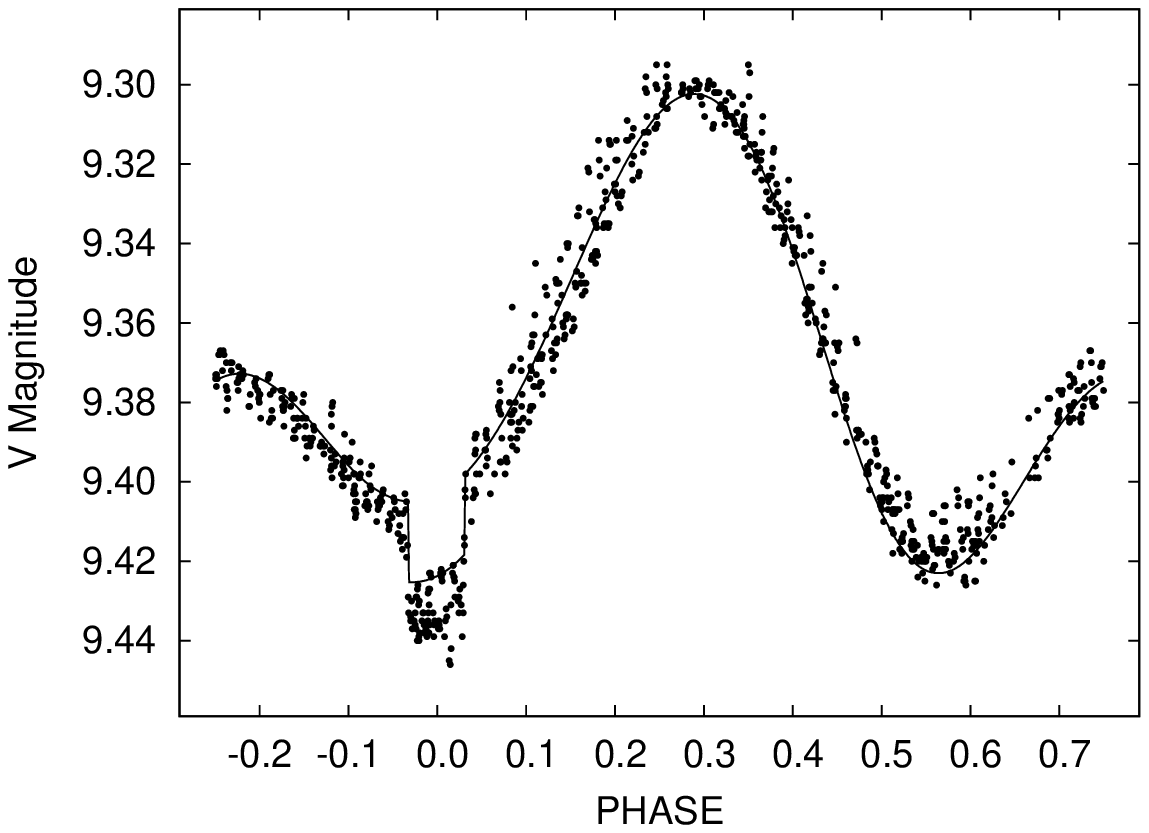}{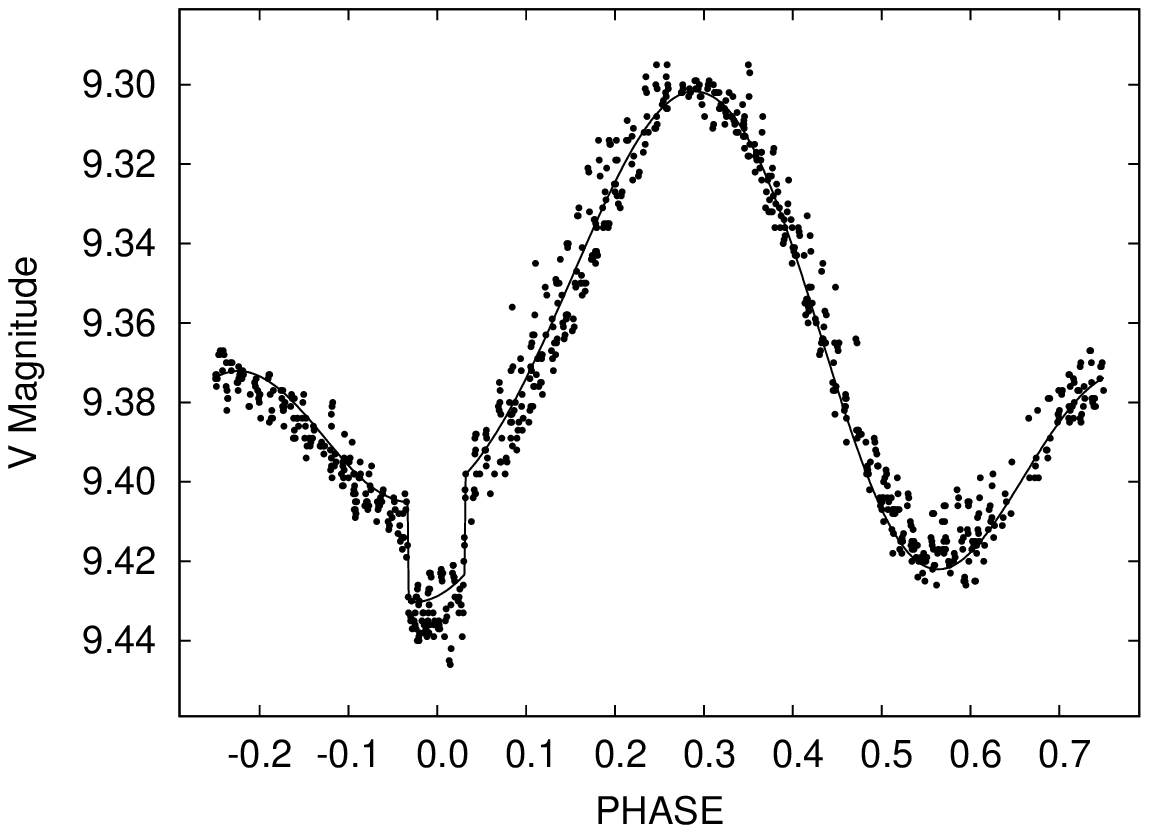}
\caption{Same as Figure~\ref{v471_u} but for the $V$ band observations of this paper.} \label{v471_V}
\end{figure}

\begin{figure}
\plottwo{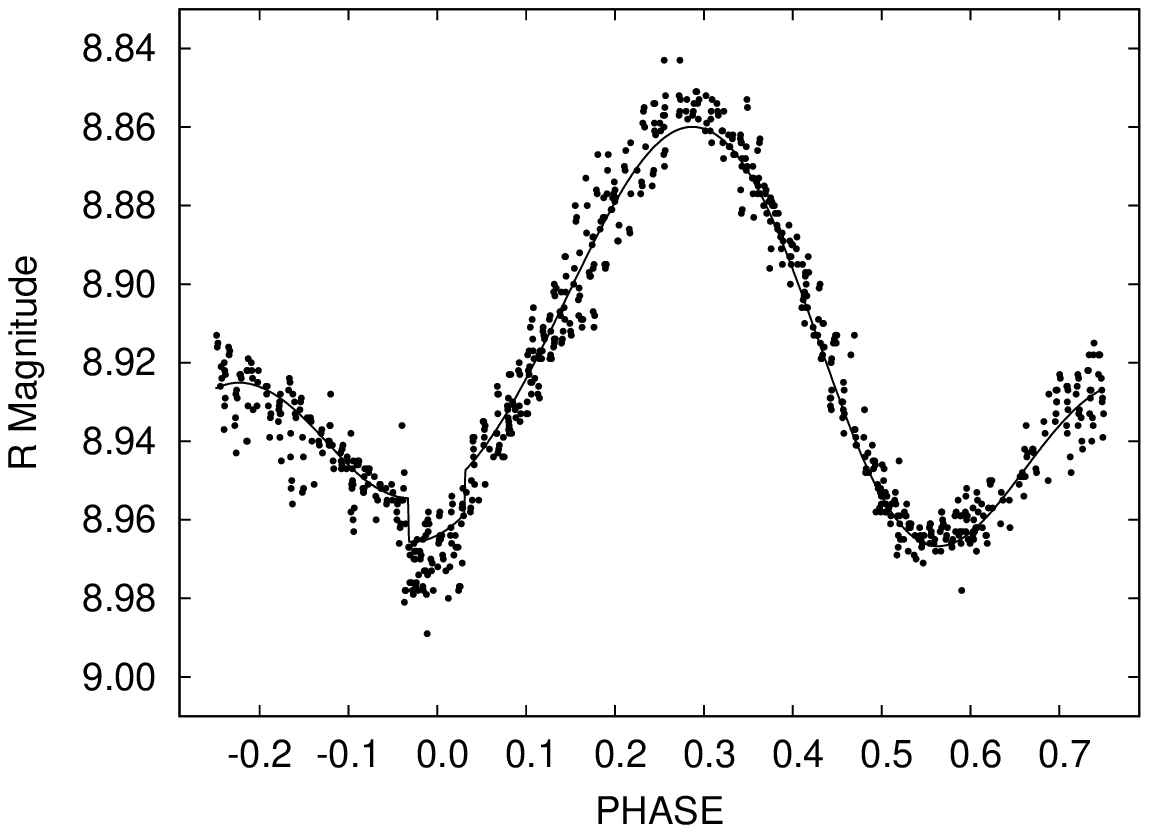}{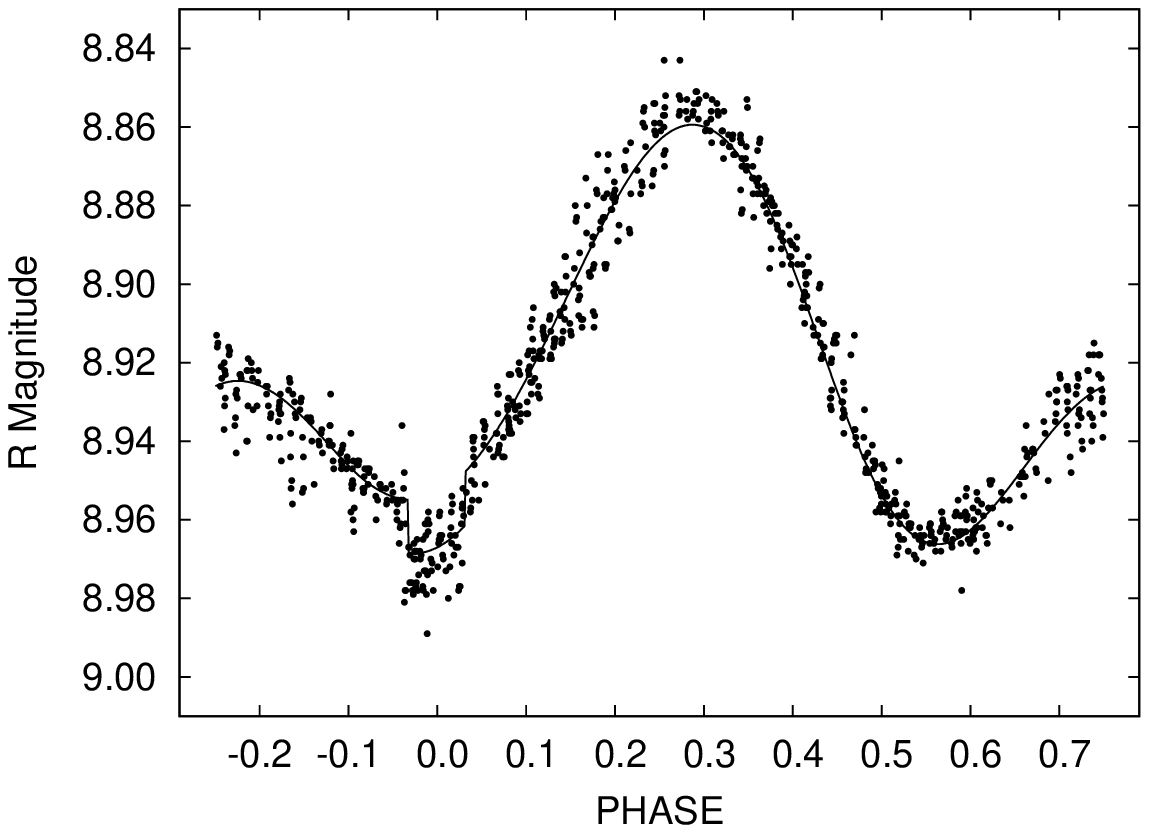}
\caption{Same as Figure~\ref{v471_u} but for the $R_{C}$ band observations of this paper.} \label{v471_R}
\end{figure}

\begin{figure}
\plottwo{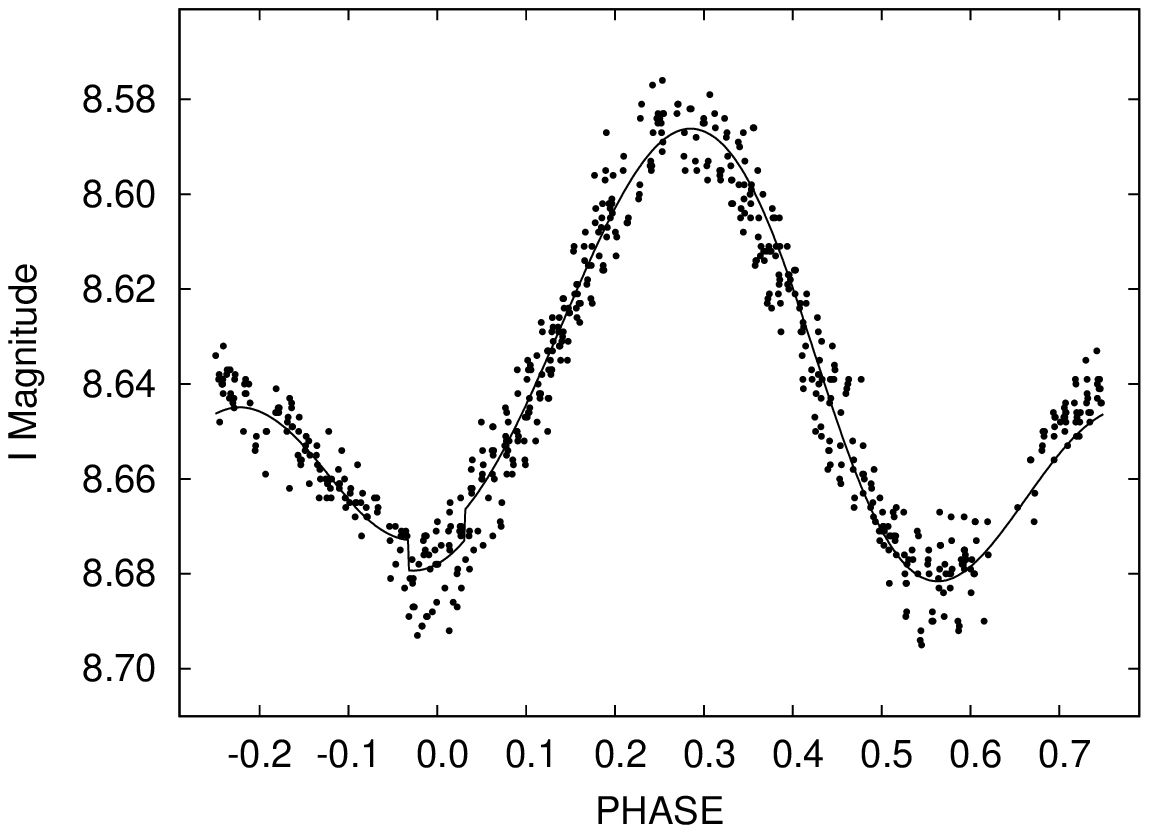}{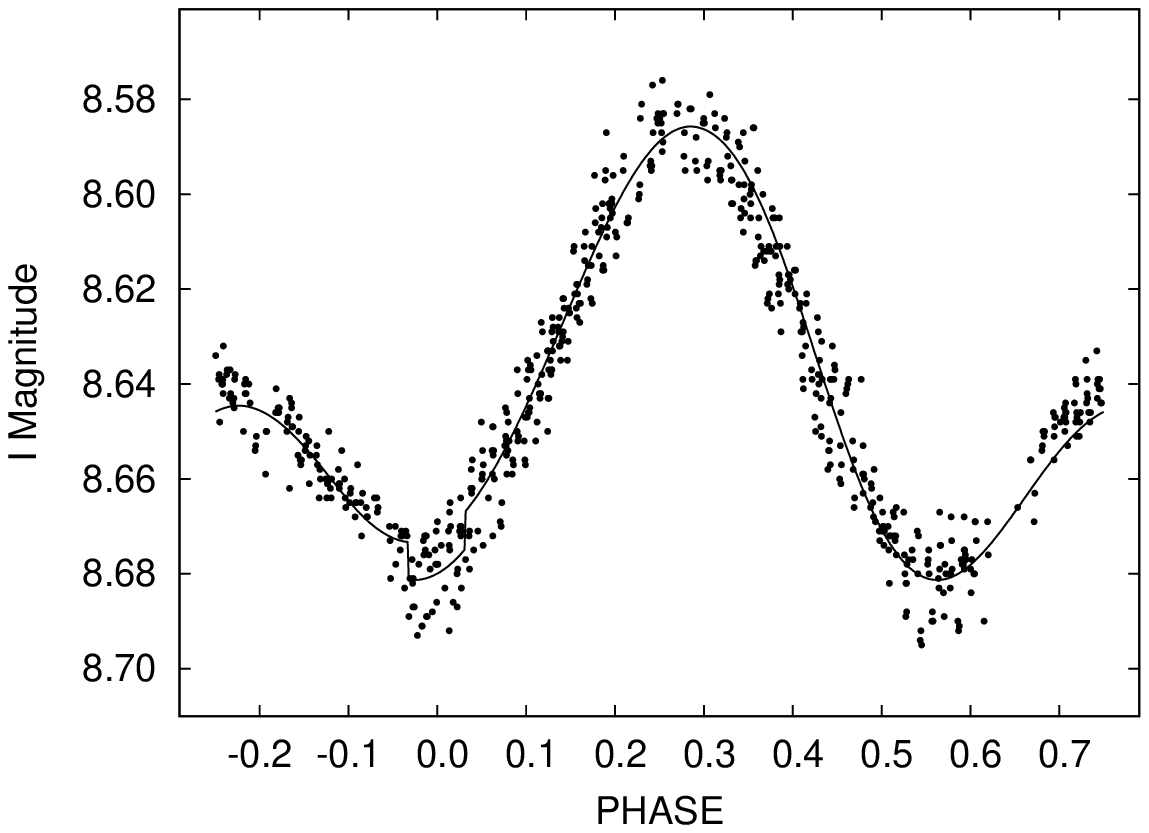}
\caption{Same as Figure~\ref{v471_u} but for the $I_{C}$ band observations of this paper.} \label{v471_I}
\end{figure}

\begin{figure}
\plottwo{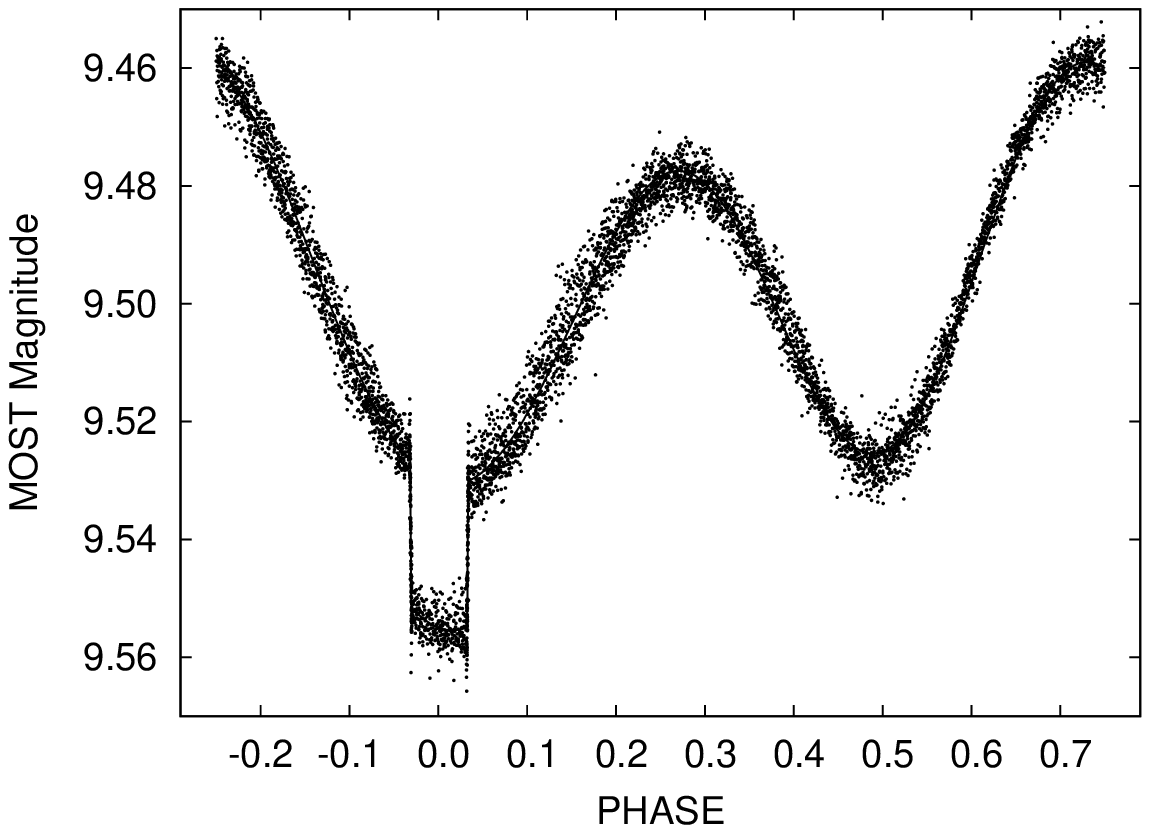}{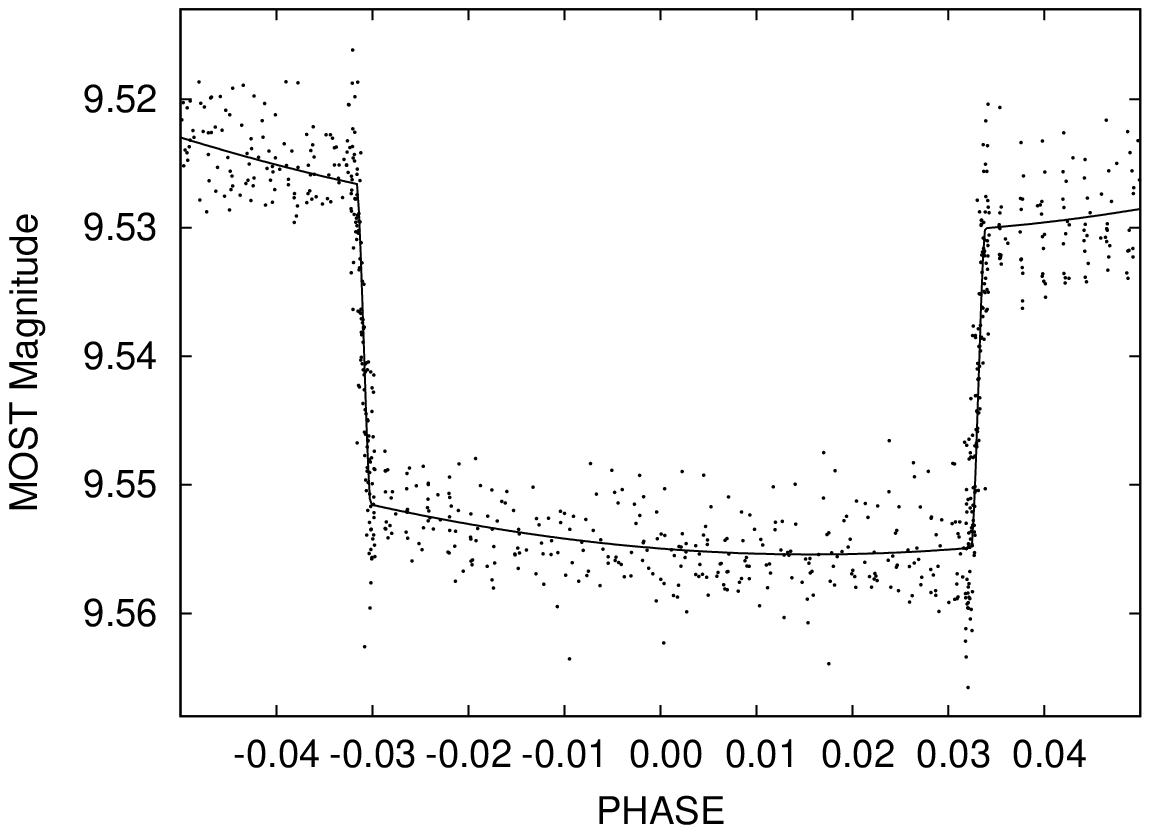}
\caption{The MOST observations of December, 2005 by \citet{kaminski} phased with the ephemeris of the light curve-only solution
(column 4 of Table~\ref{tbl-sol}). Variation due to spots is much smaller than in 1976 or 1998. 
The points are the light curve data processed in our solutions. They are averages of 10 original points, except in the very
brief partial eclipse phases where they are original points.
The right panel shows the eclipse in more detail.} \label{v471_MOST}
\end{figure}

\begin{figure}
\plottwo{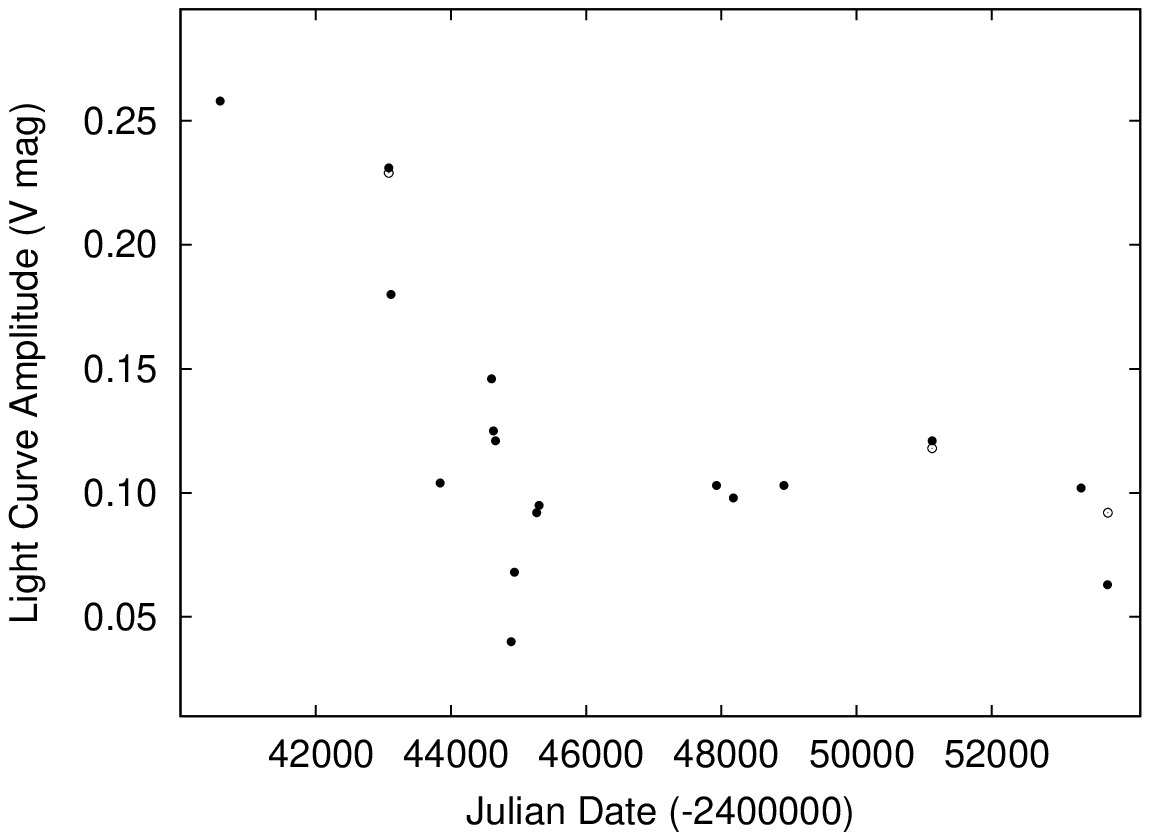}{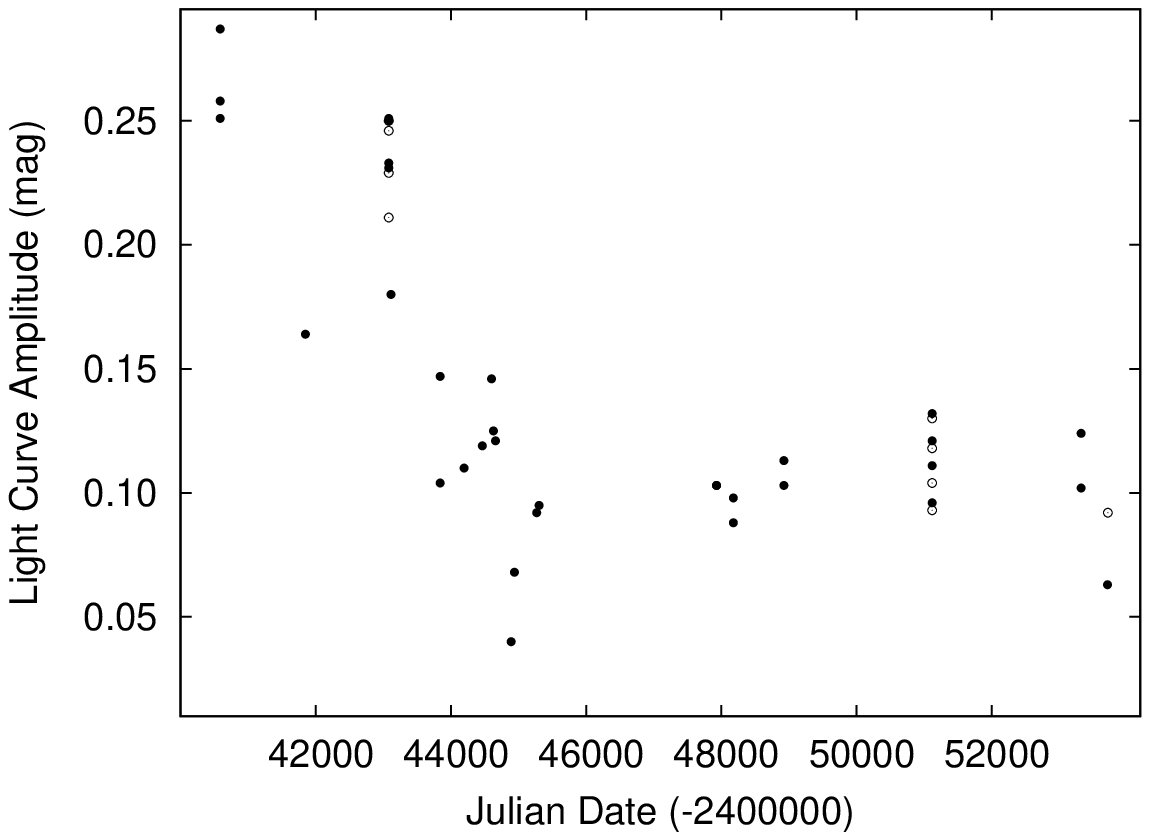}
\caption{Light curve amplitude vs. time over recent decades, showing decline 
that was not simply linear but a fast decline followed by an era of near-constancy. 
Left panel: Amplitude vs. time for $V$, $y$, and MOST mission light curves, which have similar effective wavelengths. The main variation
phenomena are modulation by starspots and tides, both being on the red dwarf. Dots are observed amplitudes read from light curve figures (mostly
from the literature with a few from our 1998 KPNO observations). Open circles are analytic amplitudes that correspond to
our solution parameters.
Right panel: Same as in left panel but for all bands ($u$, $v$, $b$, $y$, $B$, $V$, $R_C$, $I_C$, MOST).} \label{ampl}
\end{figure}

\begin{figure}
\plottwo{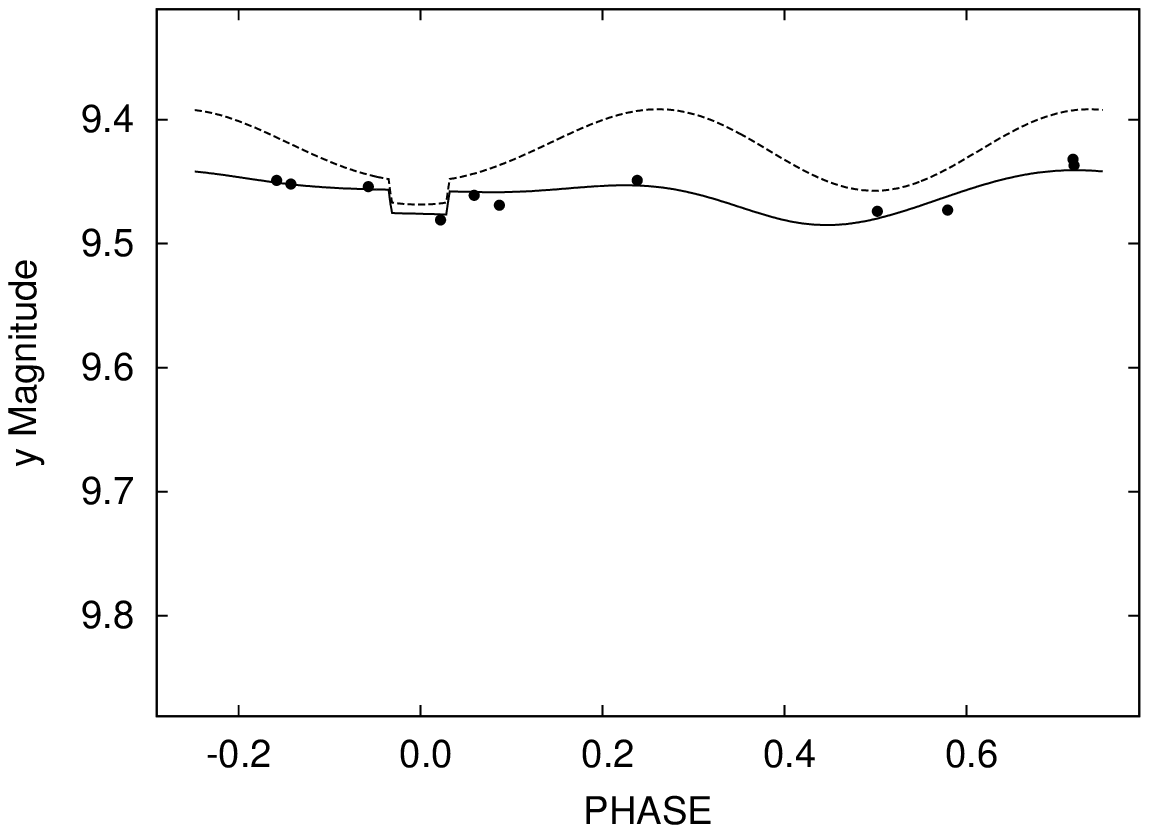}{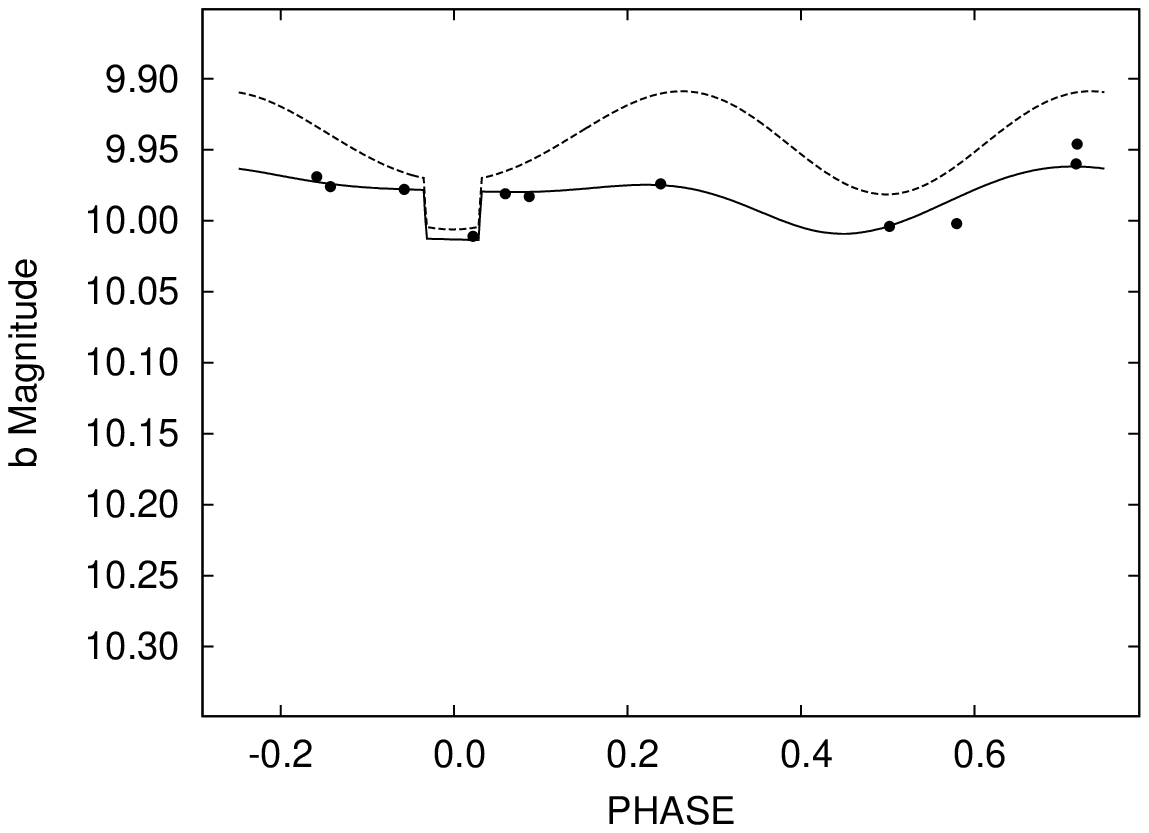}
nts of \caption{Results showing that a curious flat light curve of 1981 can be explained 
by partial cancelation between tidal and spot variation. 
Left panel: A $y$ light curve by \citet{rucinski83} (dots) with remarkably little variation, as mentioned by Rucinski and discussed in \S\ref{curious}.
The lower curve is from a Least Squares solution that followed exploration of spot configurations that nearly cancel variation due to tides.
The upper curve is for complete absence of starspots, so with only tidal distortion and a small reflection effect outside eclipse.
The large amount of white space allows the $u$, $v$, $b$, and $y$ panels (see also next figure) to have the same differential magnitude scale, so as to allow
direct comparison of amplitudes.
Right panel: Same as left panel but for the \citet{rucinski83} $b$ curve} \label{rucin_b_y}
\end{figure}

\begin{figure}
\plottwo{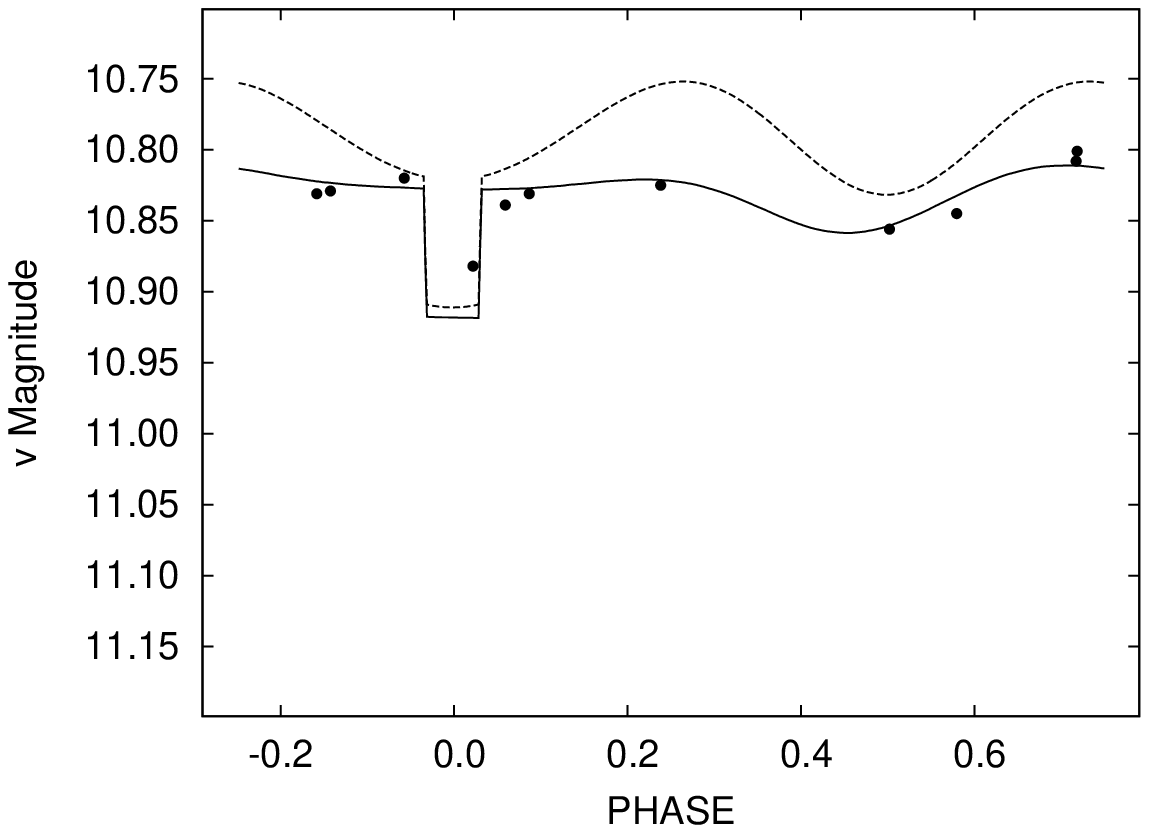}{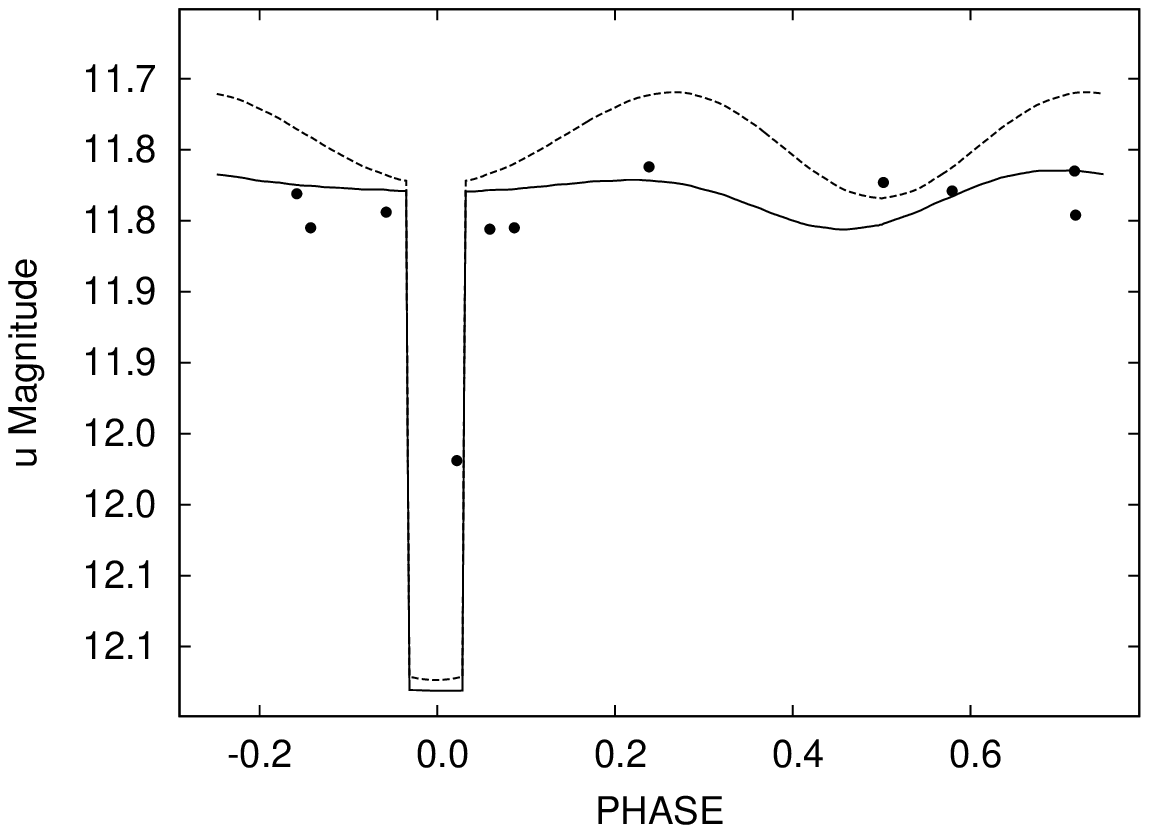}
\caption{Same as Figure~\ref{rucin_b_y} but for the \citet{rucinski83} $v$ and $u$ curves. Note the strong depth increase toward short wavelengths for eclipse of the very hot white dwarf.} \label{rucin_u_v} 
\end{figure}

\begin{figure}
\plottwo{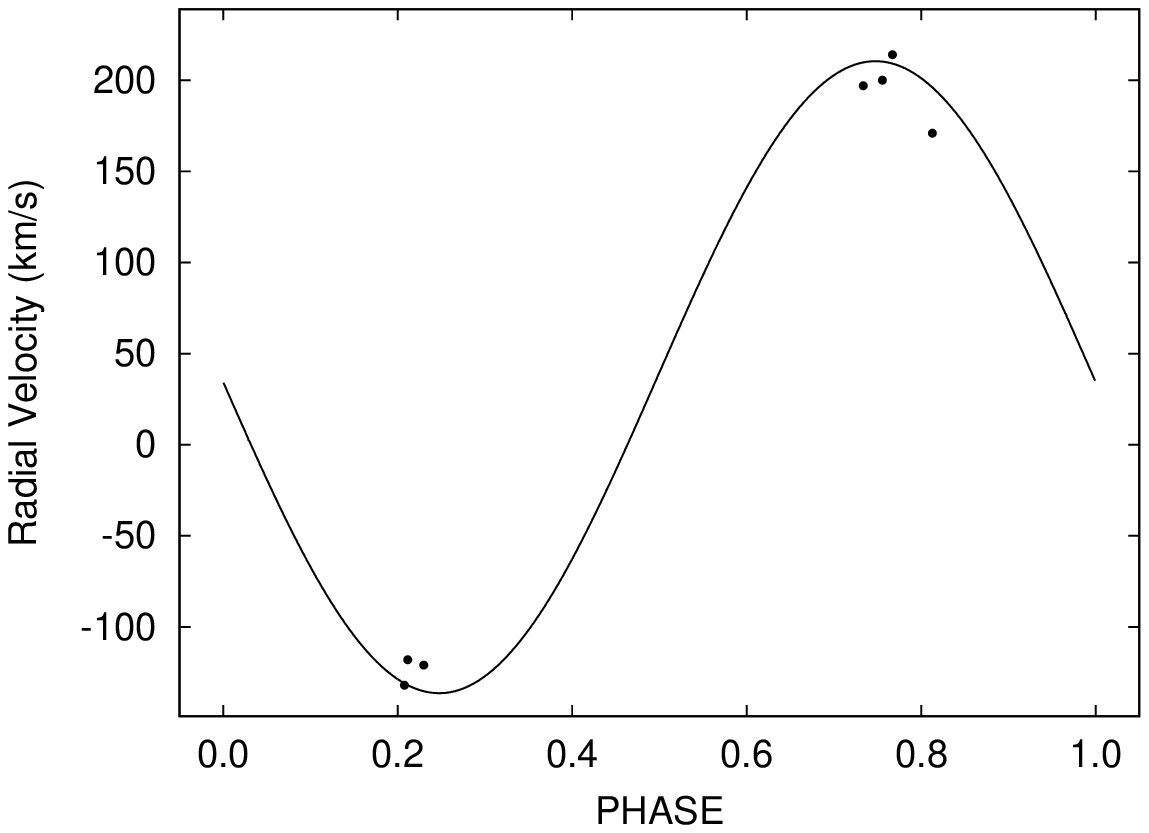}{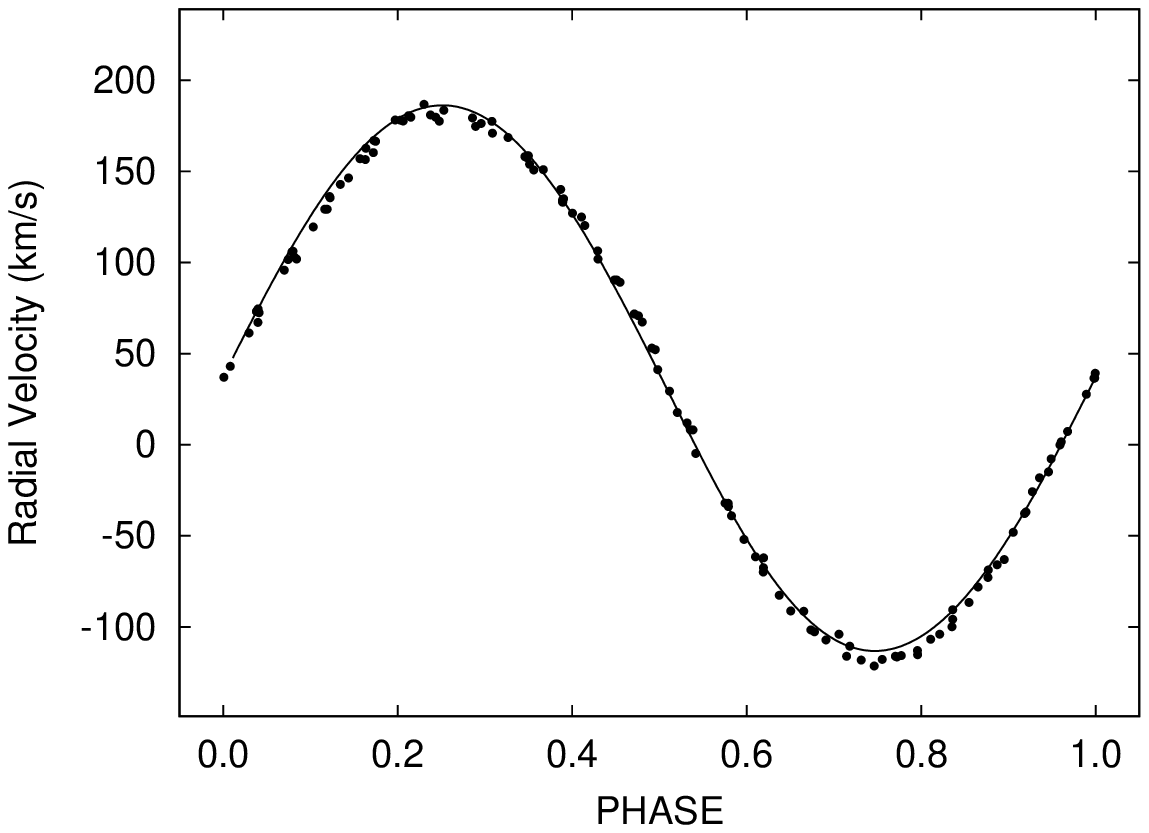}
\caption{Left panel: White dwarf radial velocities from 1994-95 by \citet{obrien} and fixed-radii solution (last column of Table~\ref{tbl-sol}) curve. Only the seven
velocities in the interval HJED 2449643-2449651 are shown, with the phased curve computed for the cycle that starts near HJED 2449648.427.
Right panel: KPNO red dwarf radial velocities (this paper) and computed curve for the cycle that starts near HJED 2451126.0.} \label{rv1}
\end{figure}

\begin{figure}
\plottwo{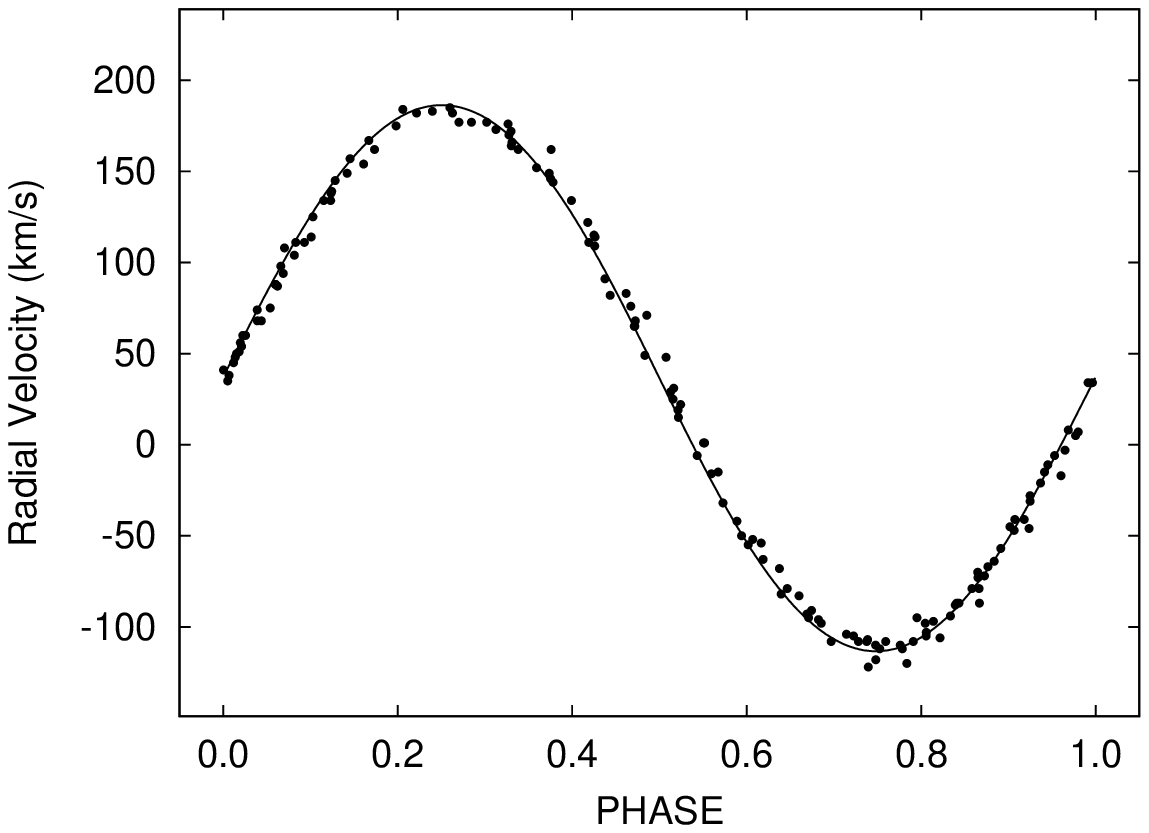}{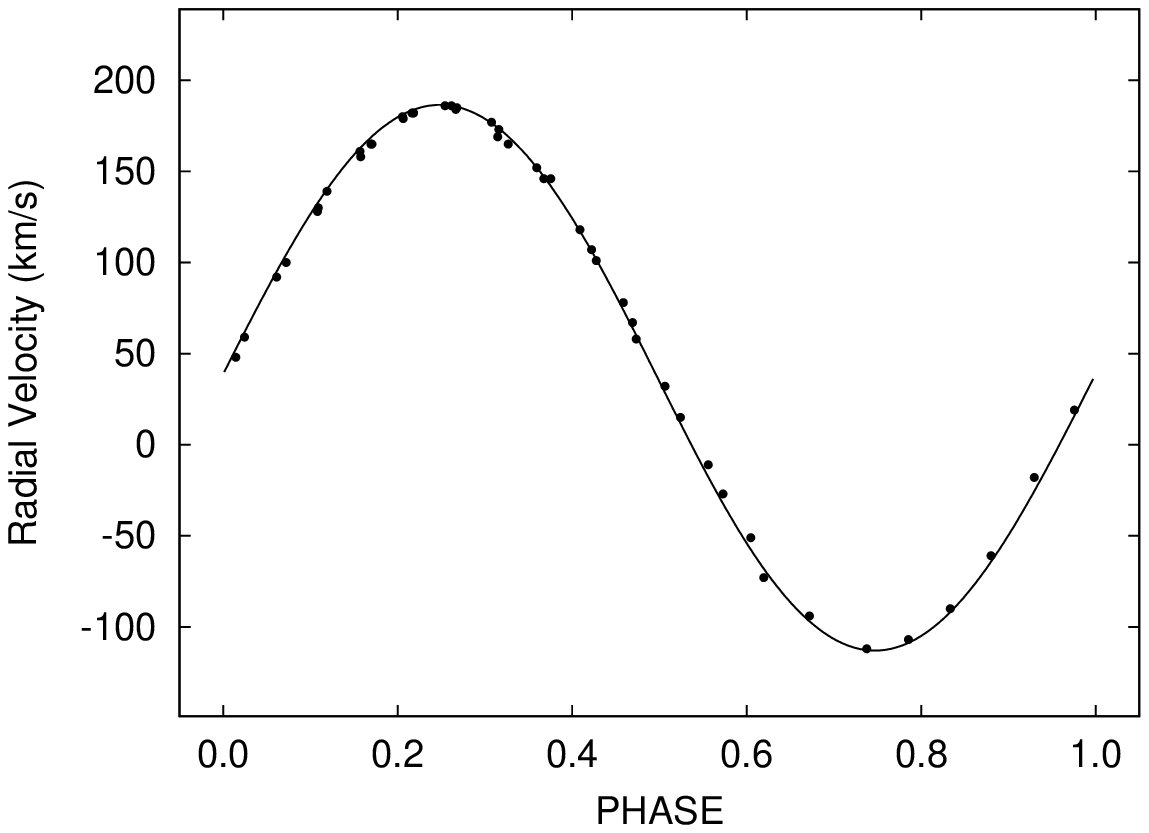}
\caption{Left panel: Red dwarf RVs by \citet{bois88} for velocities between HJED 2443185 and HJED 2444281
and curve for the fixed-radii solution 
for the cycle that starts near HJED 2443828.4.
Right panel: same as the left panel for velocities between HJED 2445644 and HJED 2445648, with the computed curve for 
the cycle that starts near HJED 2445646.3.} \label{rv2}
\end{figure}

\begin{figure}
\plottwo{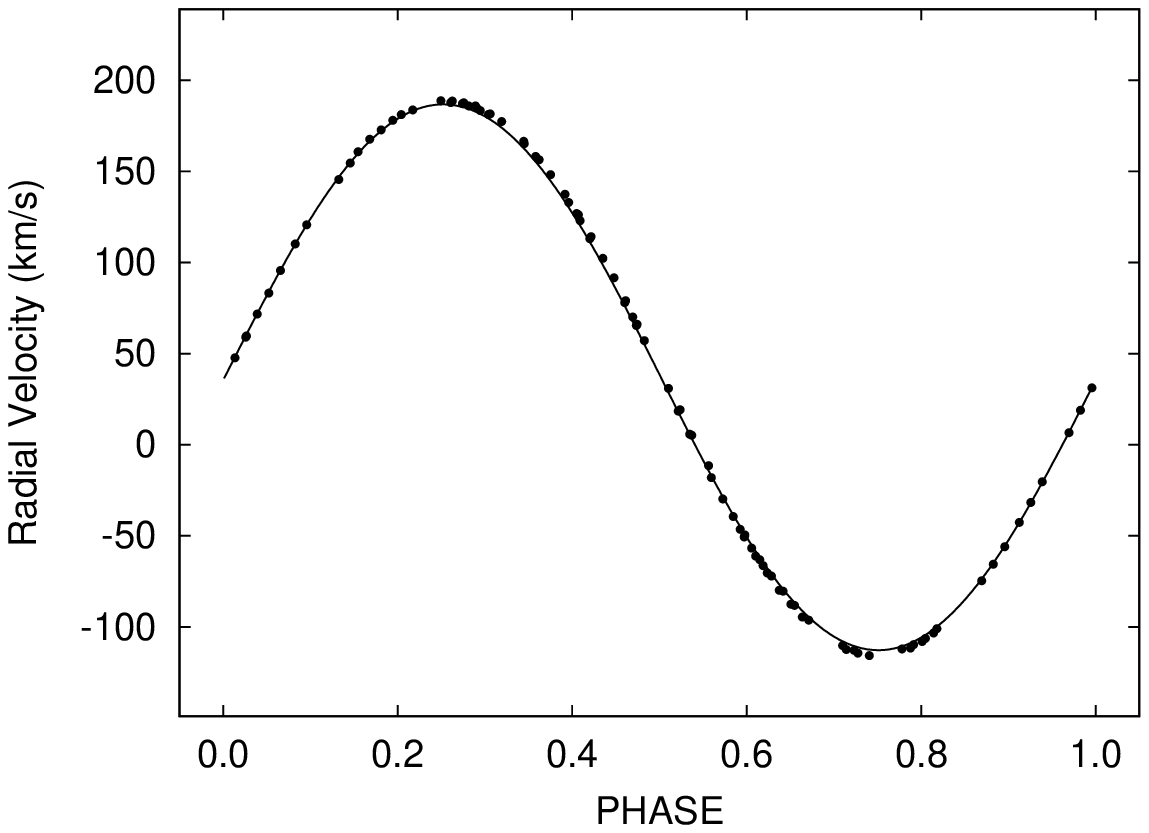}{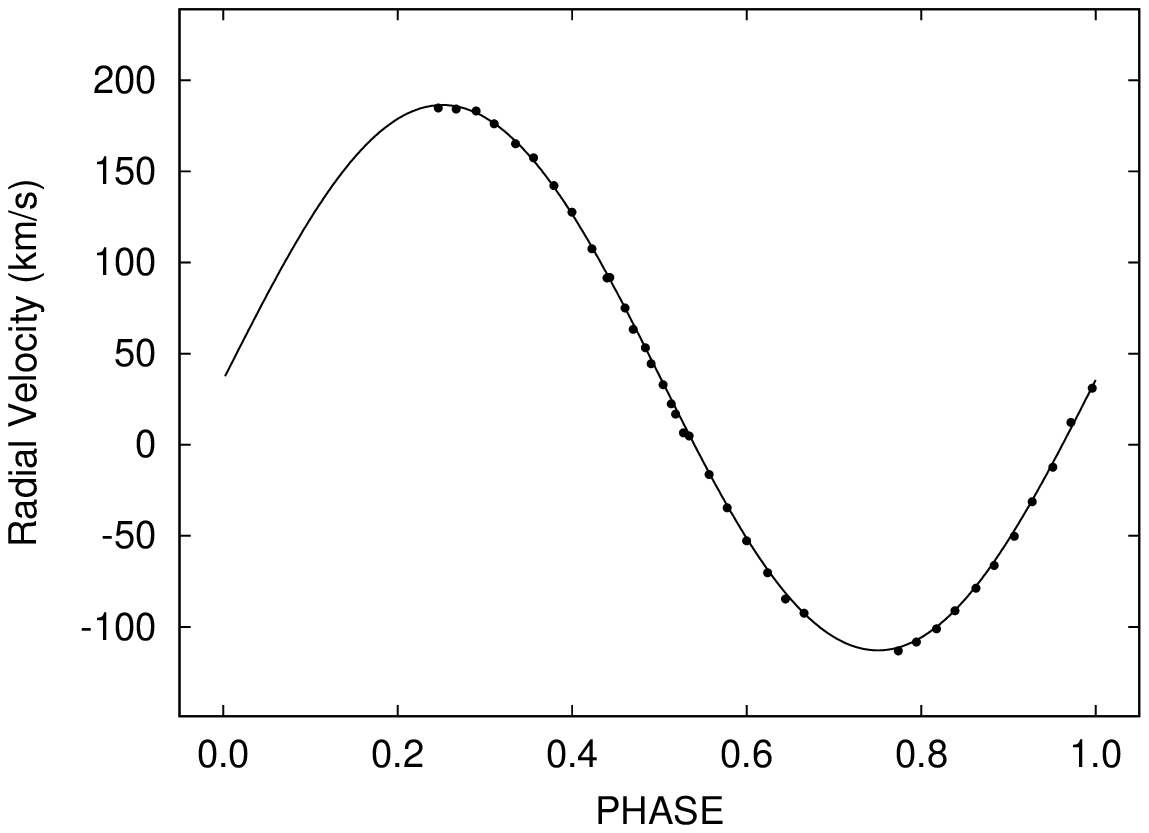}
\caption{Left panel: Red dwarf RVs \citep{hussain}  
and curve for the fixed-radii solution 
at the cycle that starts near HJED 2452603.0.
Right panel: Red RVs velocities by \citet{kaminski}, with a computed curve for 
the cycle that starts near HJED 2453717.8.} \label{rv3}
\end{figure}

\begin{figure}
\plottwo{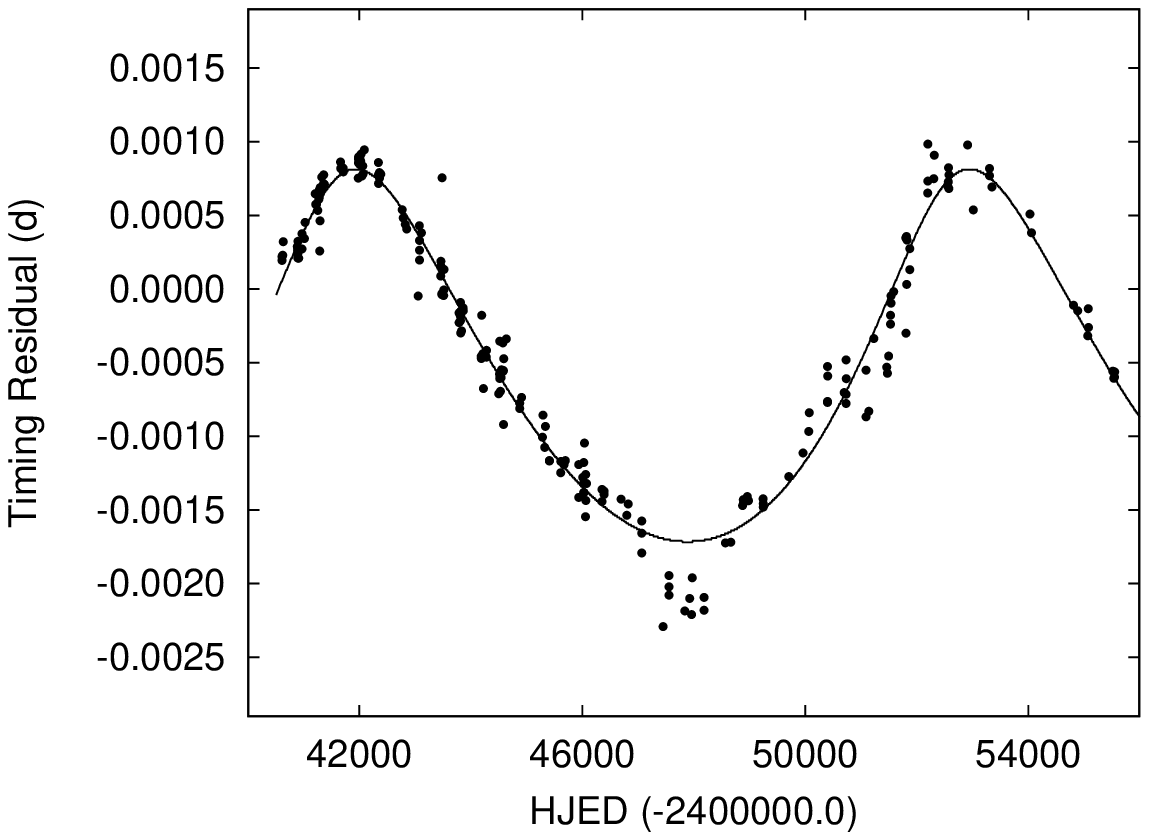}{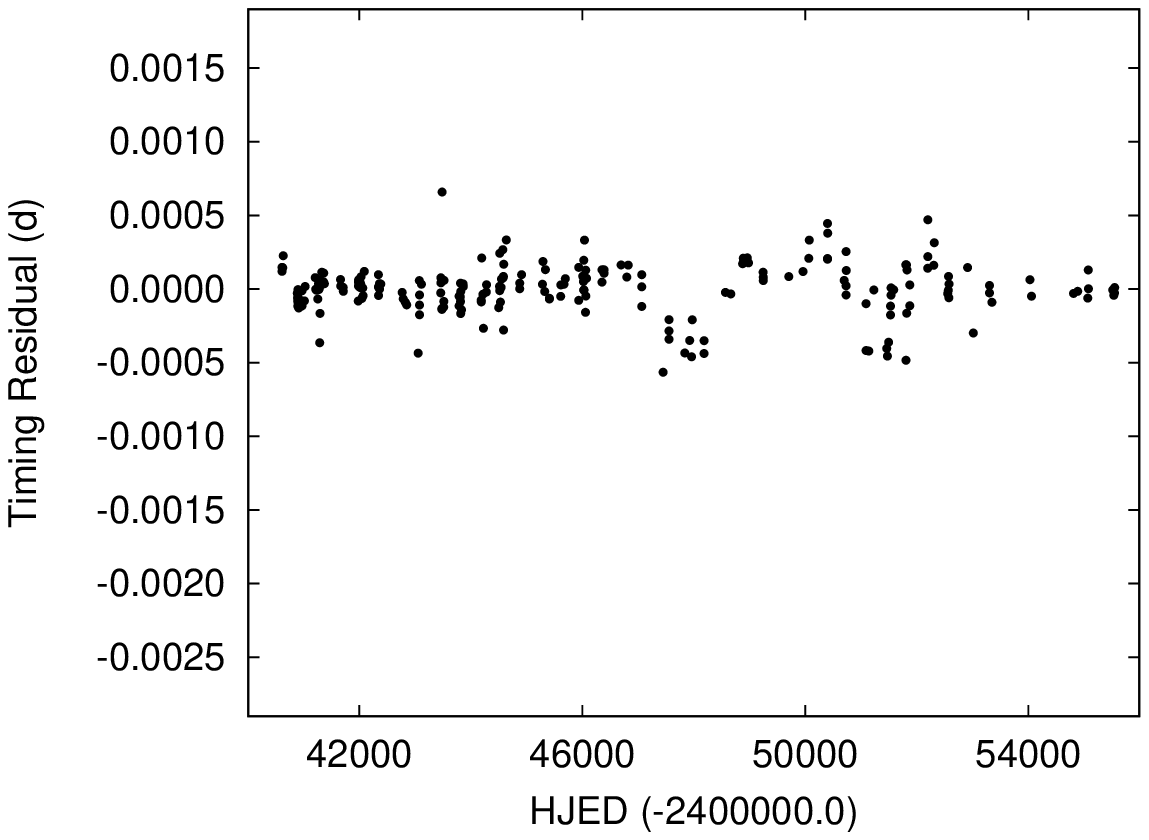}
\caption{Residuals for the timing-only solution. The left panel shows 
residuals without a modeled 3b light-time effect. The continuous curve shows the corresponding idealized
behavior. The right panel is for the full ephemeris/light-time solution. Unexplained features of
order 2000 days wide that have not attracted comment in the literature are 
centered on roughly JD 244800 and JD 2453000.} \label{eclresno3b}
\end{figure}

\begin{figure}
\plottwo{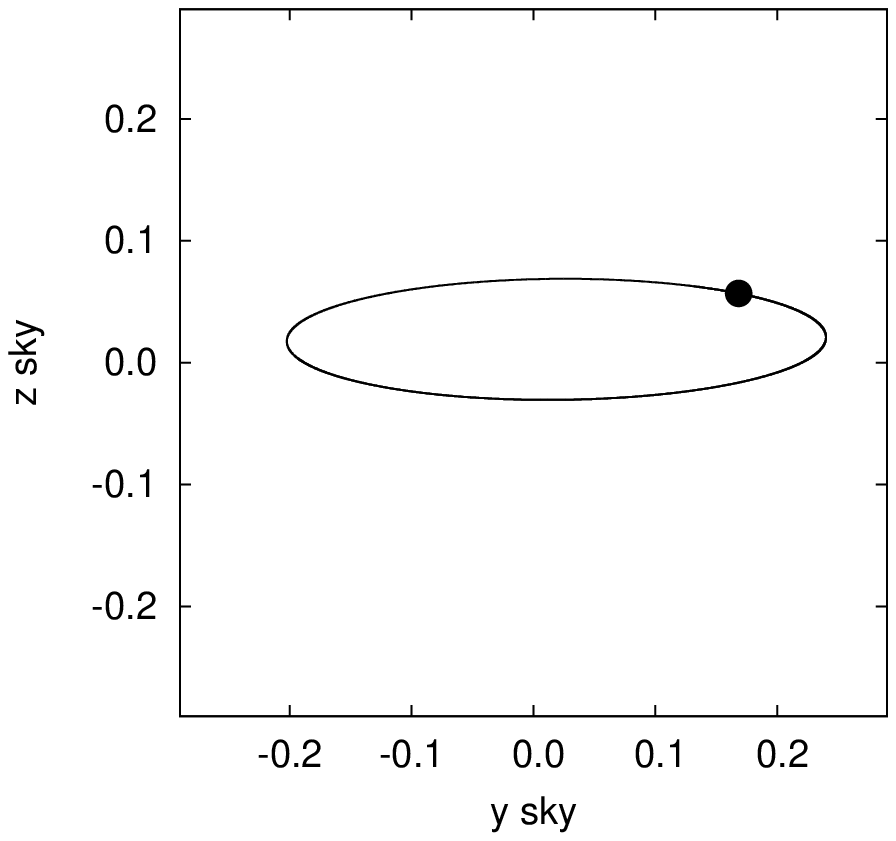}{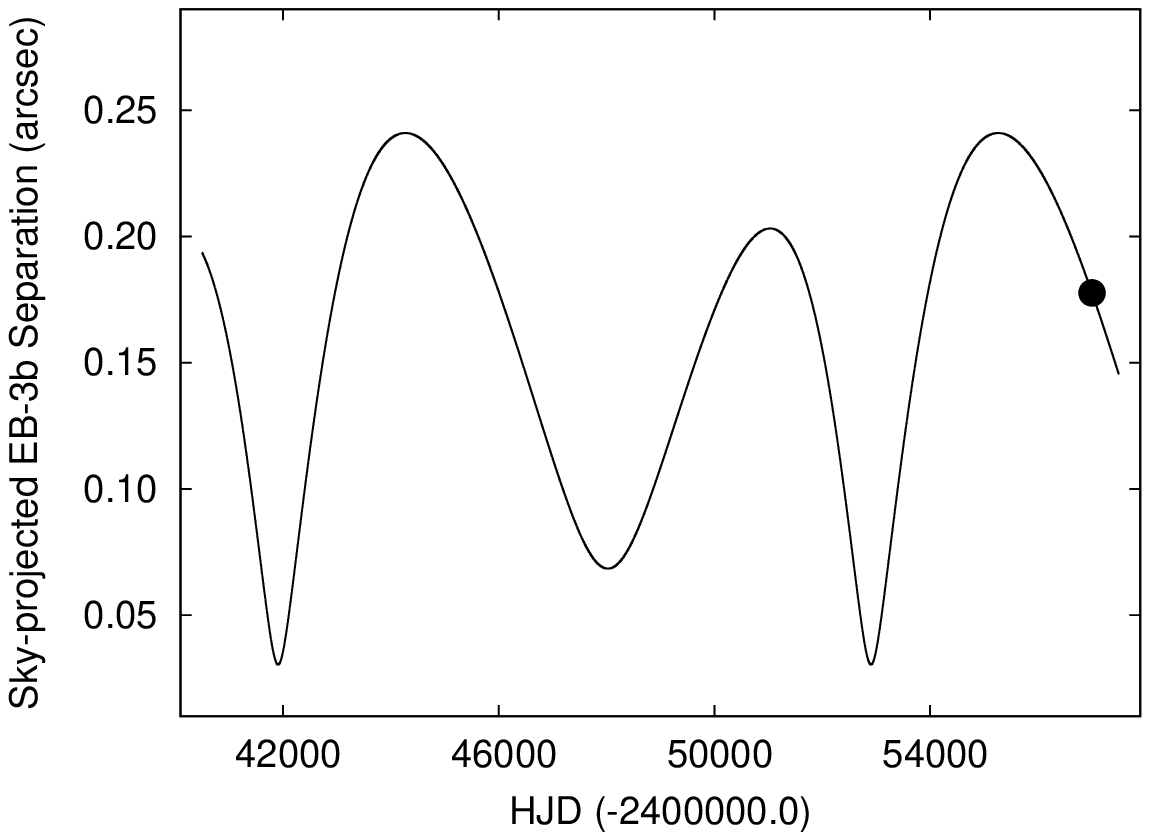}
\caption{The computed EB-3b sky coordinates (left panel) and separation vs. time (right panel) for parameters of the all-data solution
(column 3 of Table~\ref{tbl-sol}) and a 3b inclination of $78\degr$. The dot indicates the position of the 3b on December 11, 2014 \citep[the date of
the AO observation by][]{hardy}.} 
\label{sep_vs_time}
\end{figure}

\begin{figure}
\plotone{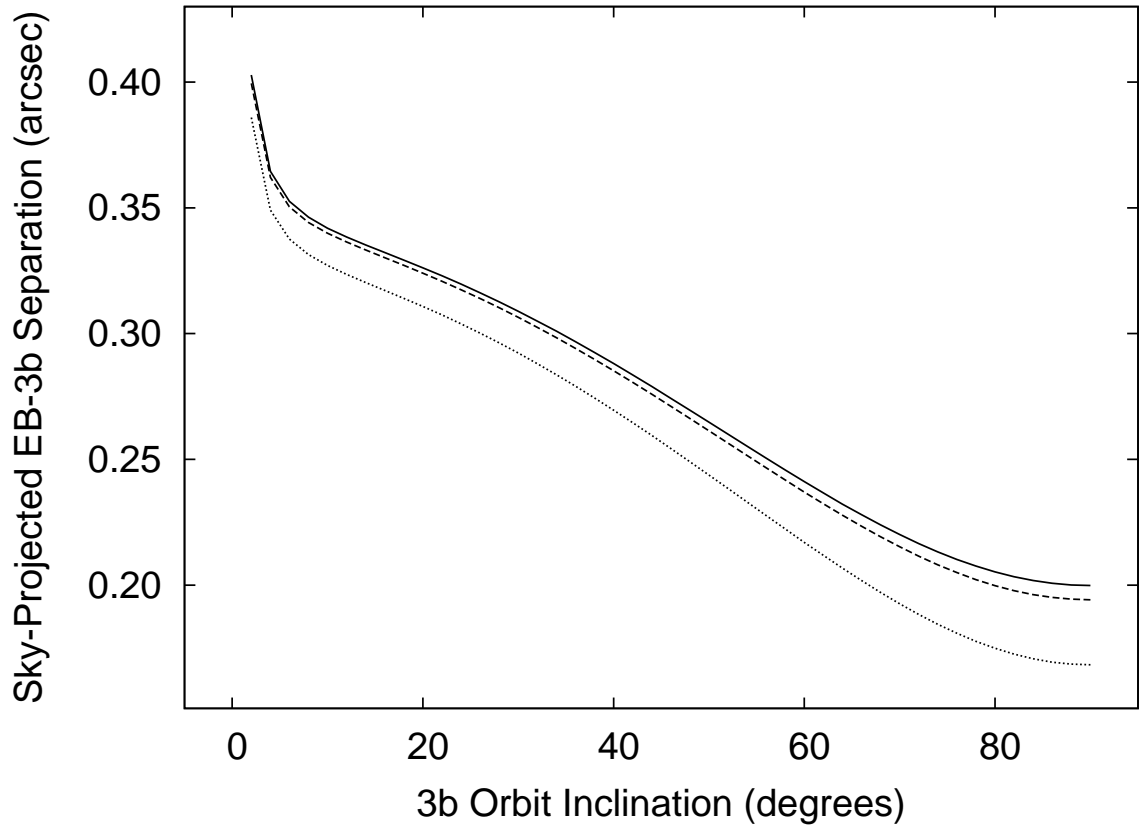}
\caption{The computed EB-3b sky separation vs. 3b orbit inclination on December 11, 2014 \citep[the reference date for
Figure 3 of][]{hardy}. The continuous, dashed and dotted lines are for
parameters of the timing-only (column 3 of Table~\ref{tbl-sol}), no-MOST (column 2 of Table~\ref{tbl-sol}), and all-data (column 6 of
Table~\ref{tbl-sol}) solutions, respectively.} \label{sep_vs_i3b}
\end{figure}

\clearpage

\begin{figure}
\plotone{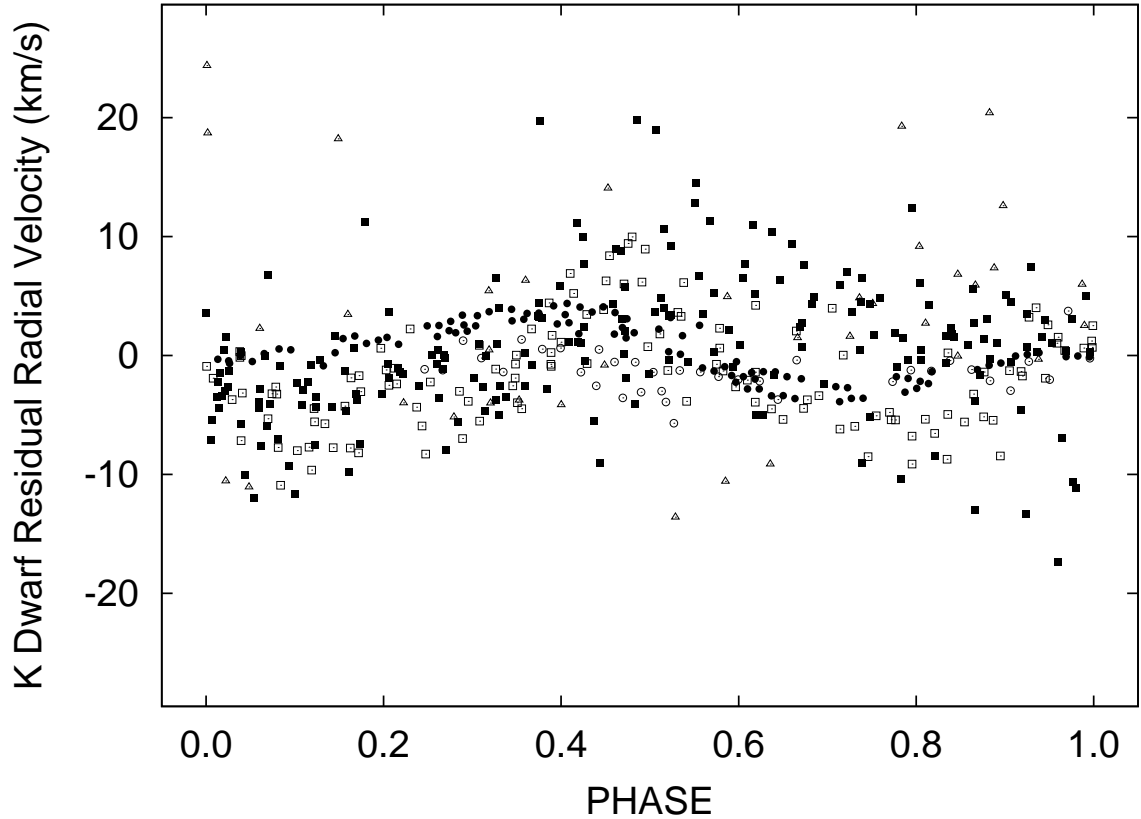}
\caption{Red dwarf radial velocity residuals for the all-data solution (column 6 of Table~\ref{tbl-sol}). 
Essential overlap of residuals for the five curves, even with relatively large scatter
of the two older sets, shows essential agreement in amplitude among the curves, although there are 
small amplitude differences and subtle shape differences.  
Open triangles: \citet{young76}; filled squares: \citet{bois88}; open squares: this paper (Table~\ref{tbl-v471rv});
filled circles: \citet{hussain}; open circles: \citet{kaminski}. The \citet{young76} and \citet{bois88} data have low
weights and influence the overall results only slightly. The \citet{hussain} and \citet{kaminski} velocities have highest weights.} \label{rv2res-all}
\end{figure}

\begin{figure}
\plottwo{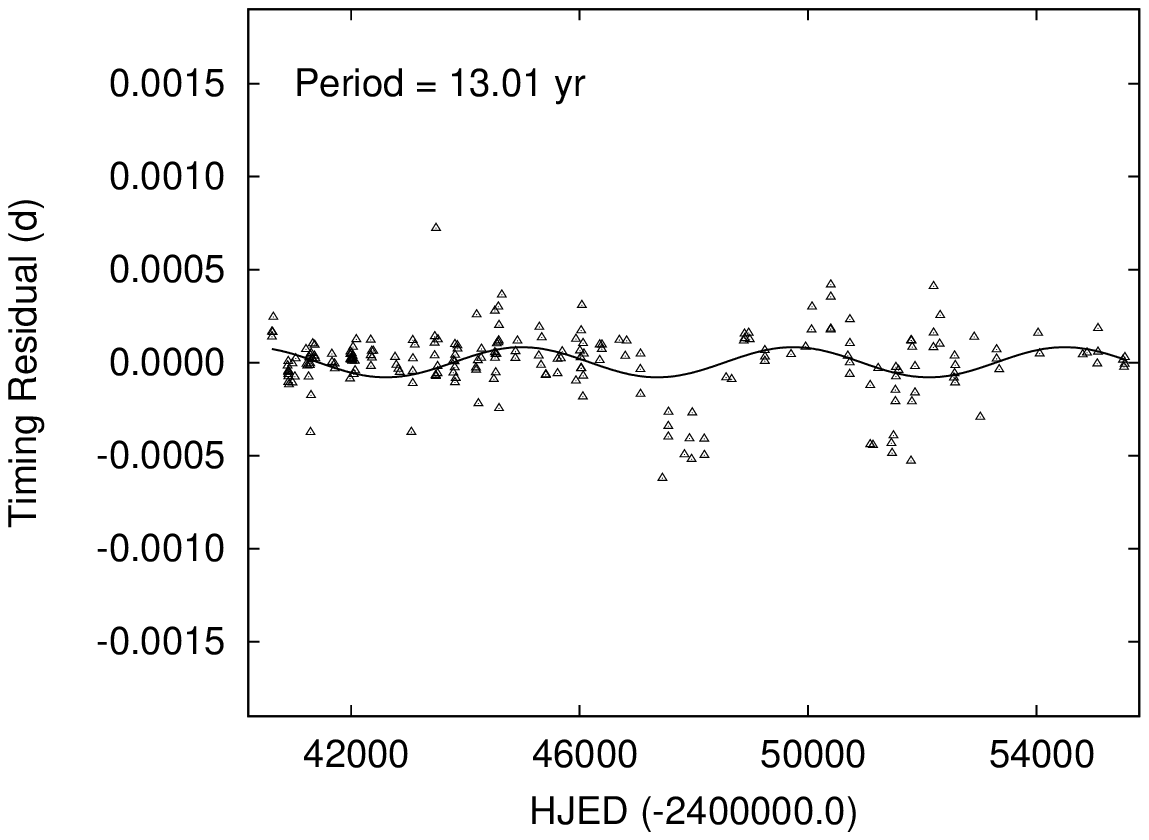}{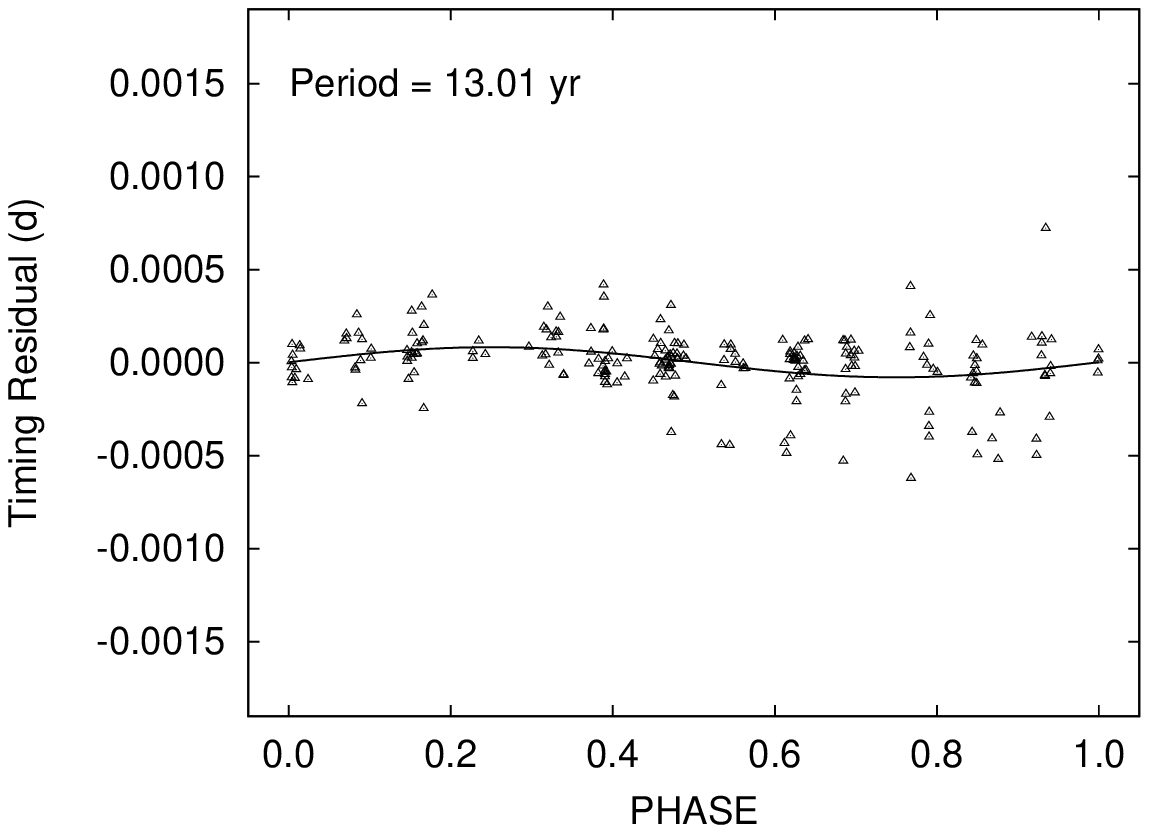}
\caption{Residuals for the timing-only solution with a fitted sinusoid of period 13.01 yr (left panel) and 
the corresponding phased curve (right panel).} \label{eclresp13}
\end{figure}

\begin{figure}
\plottwo{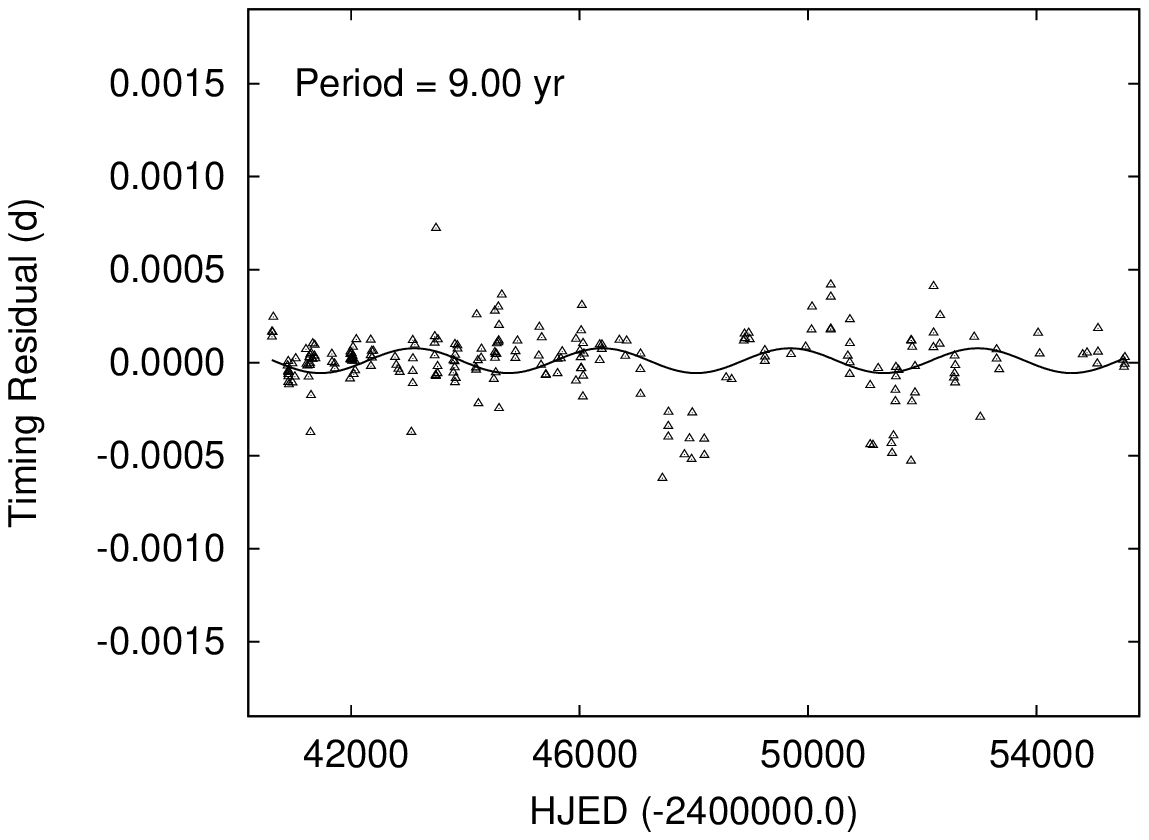}{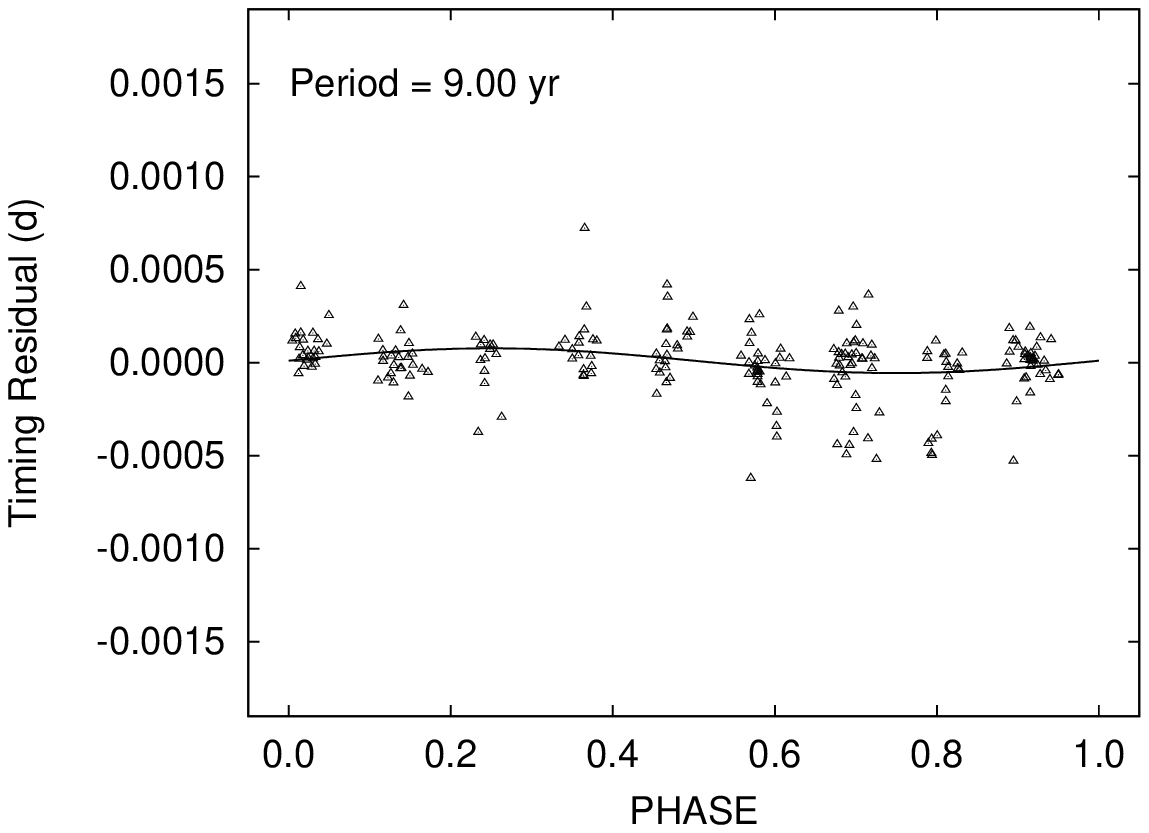}
\caption{Same as Fig.~\ref{eclresp13} for a period of 9.00 yr.}  \label{eclresp9}
\end{figure}

\begin{figure}
\plottwo{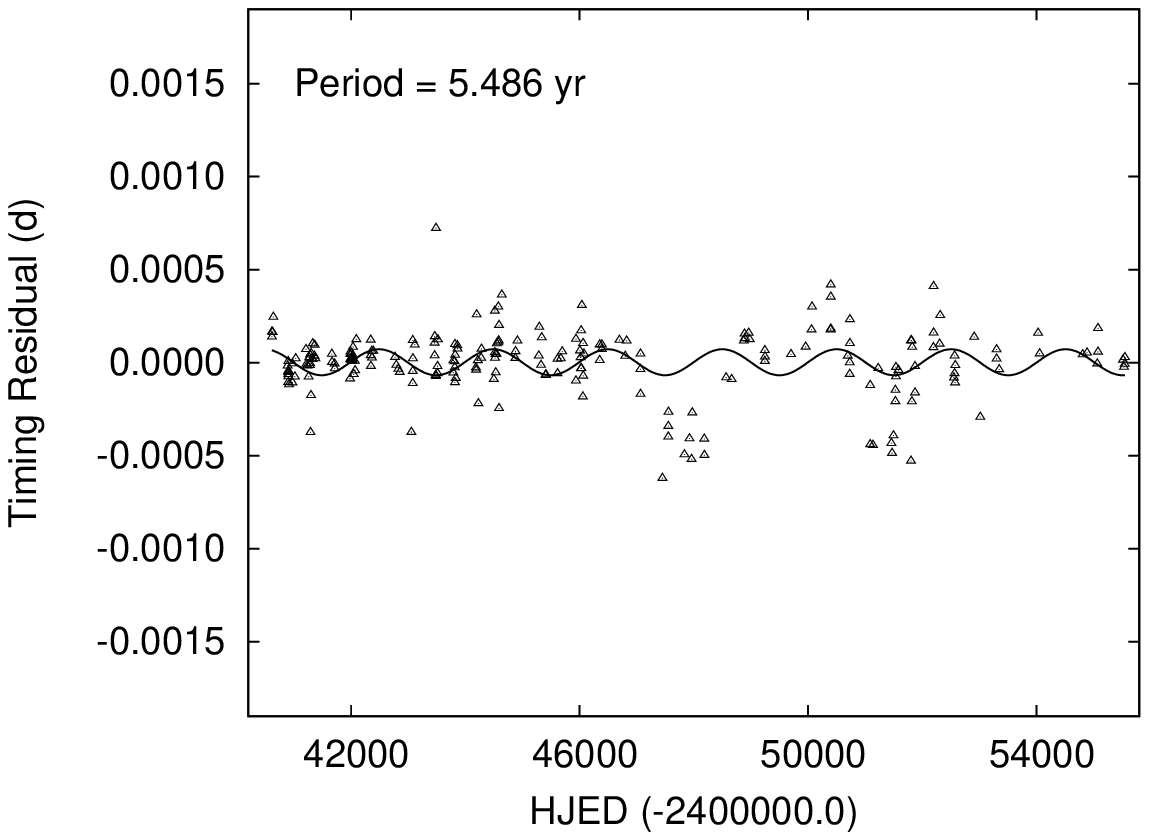}{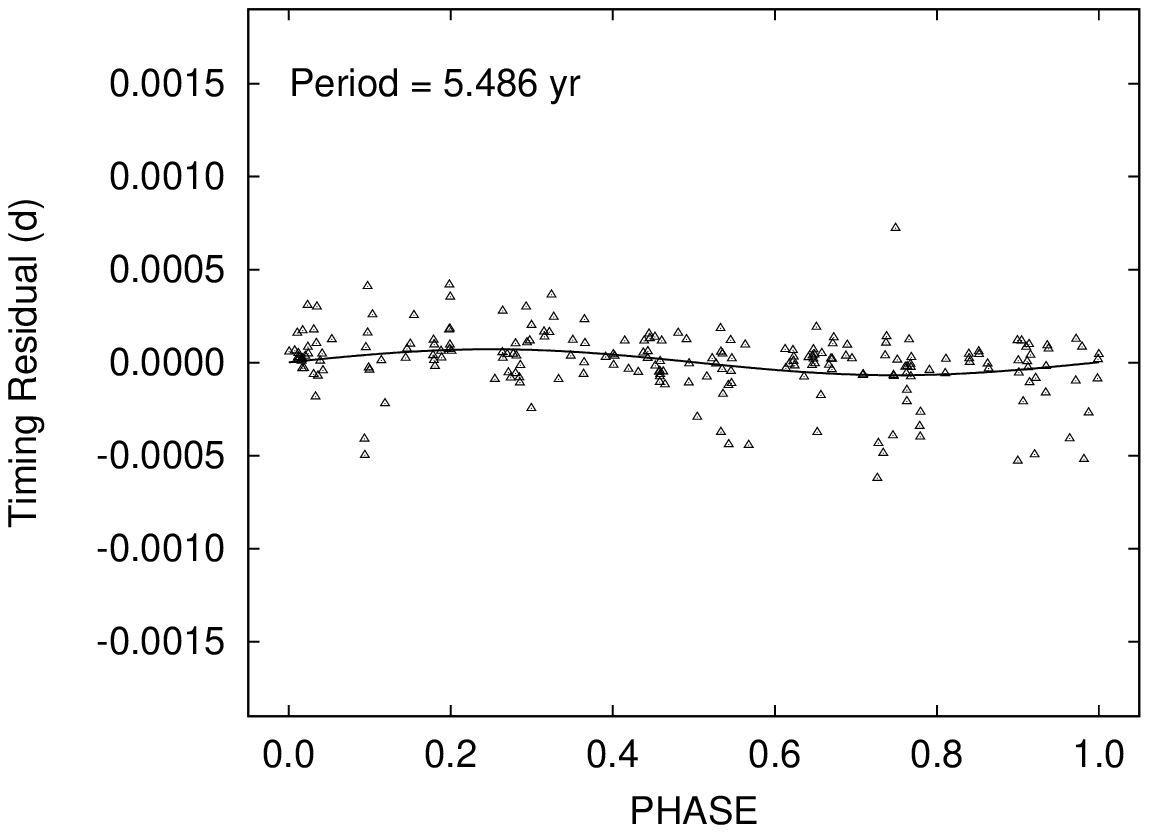}
\caption{Same as Fig.~\ref{eclresp13} for a period of 5.486 yr.}  \label{eclresp5}
\end{figure}

\begin{figure}
\plottwo{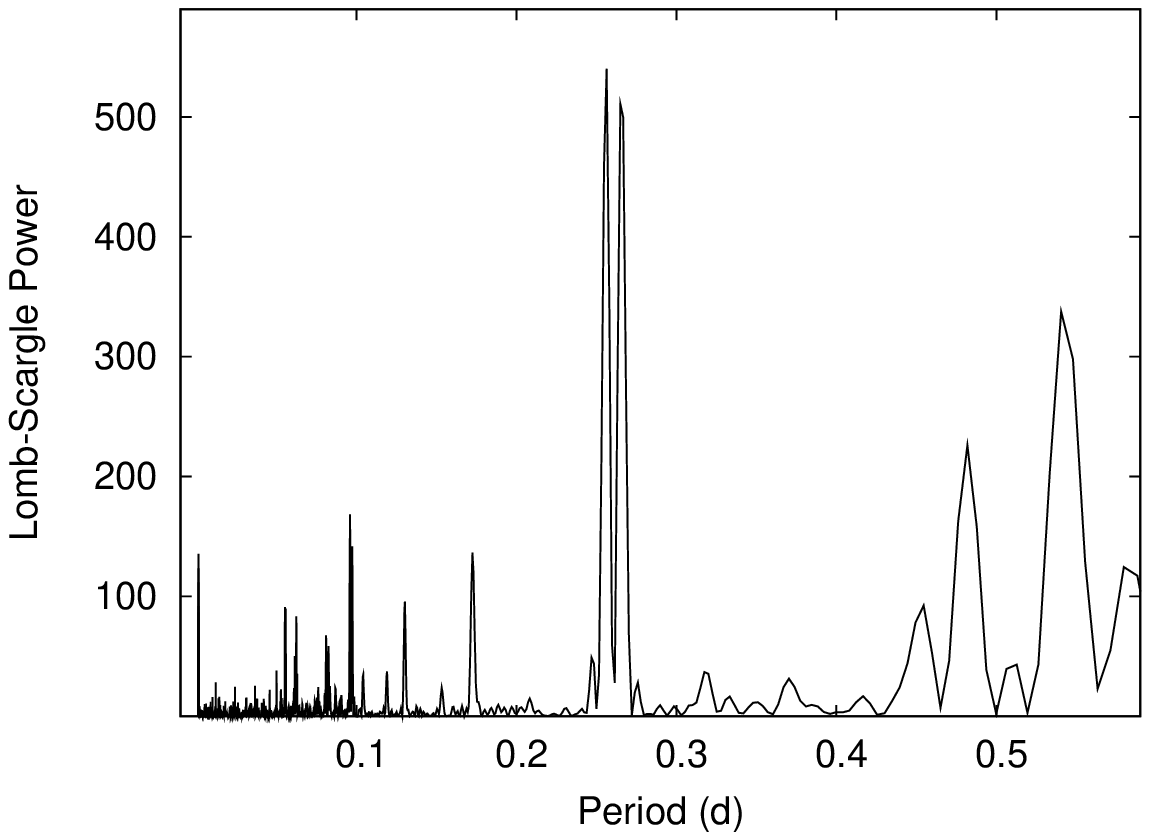}{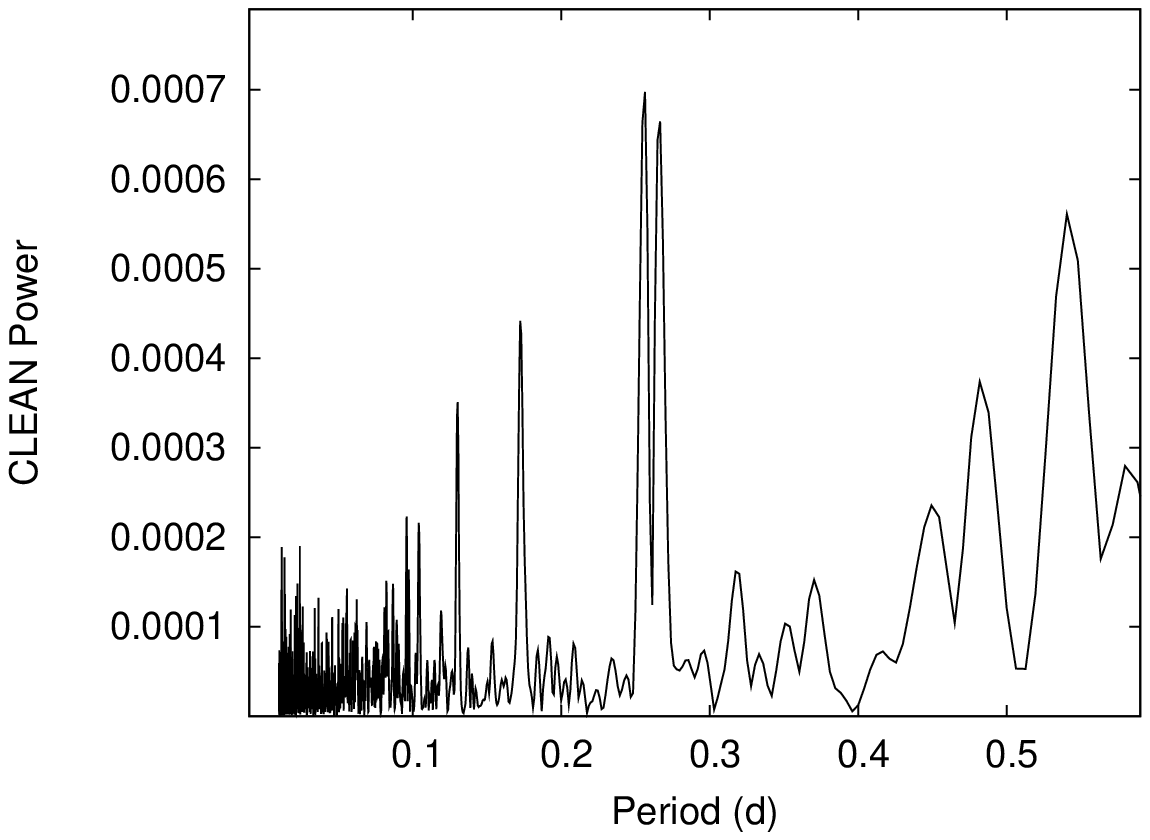}
\caption{MOST residual magnitude Lomb-Scargle (left) and CLEAN (right) periodograms.} \label{LS-CLEAN}
\end{figure}

\begin{figure}
\plottwo{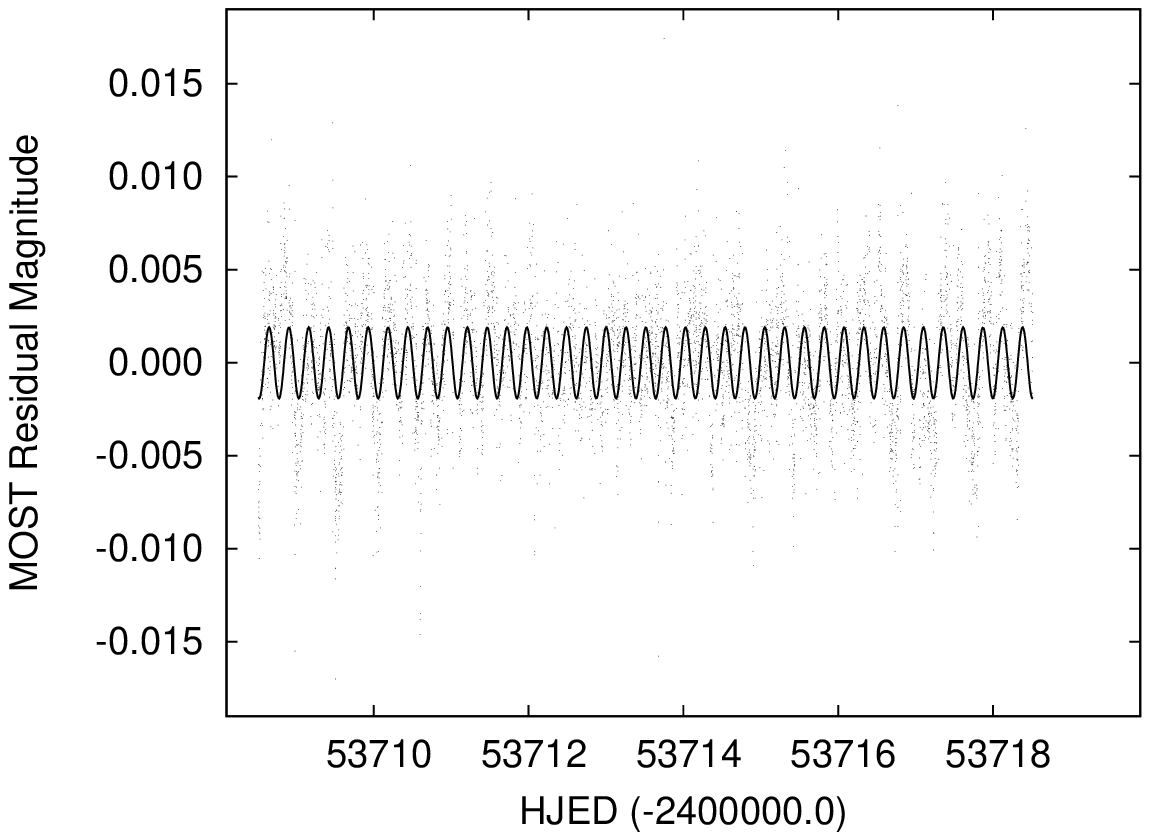}{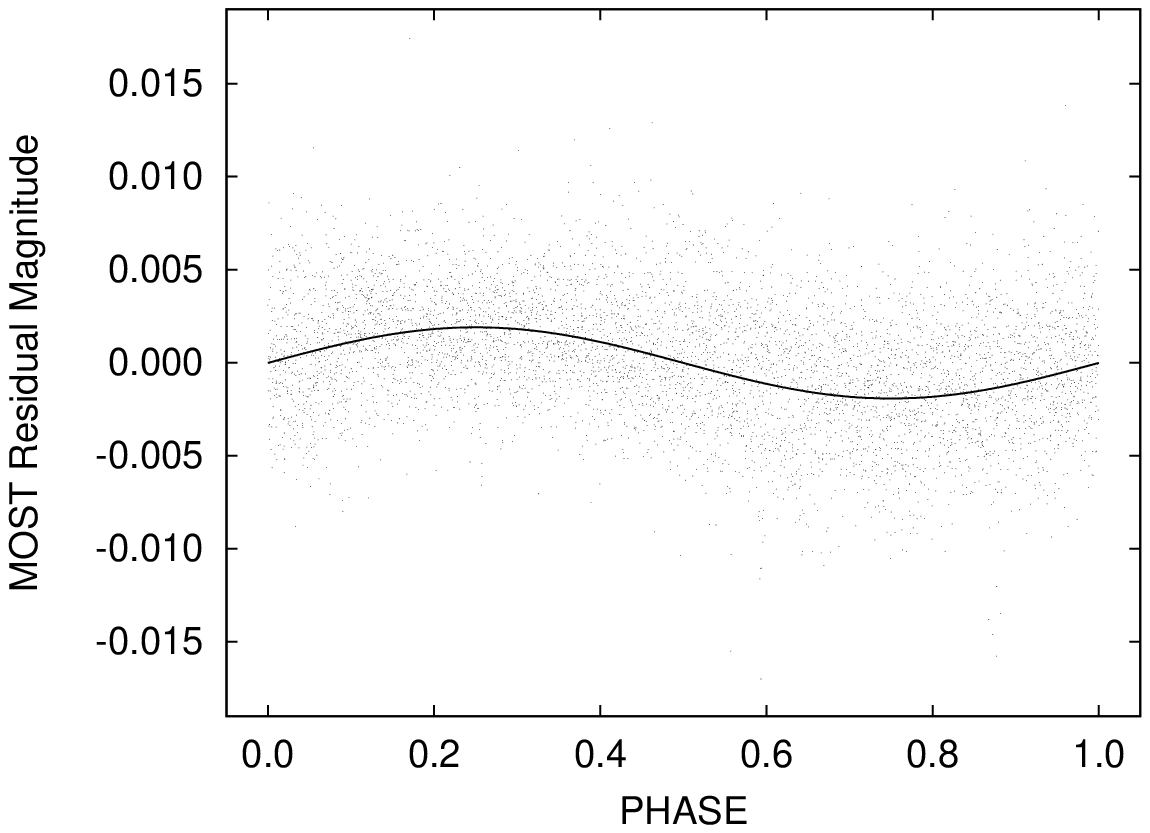}
\caption{Residual magnitude and fitted sinusoid (left panel) for the MOST light 
curve for a detected $0.^d26568$ periodicity, near half the orbit period.
The phased curve is in the right panel. Given the large quantity of points,
the variation cannot be not a statistical artifact.} \label{mostres1}
\end{figure}

\begin{figure}
\plottwo{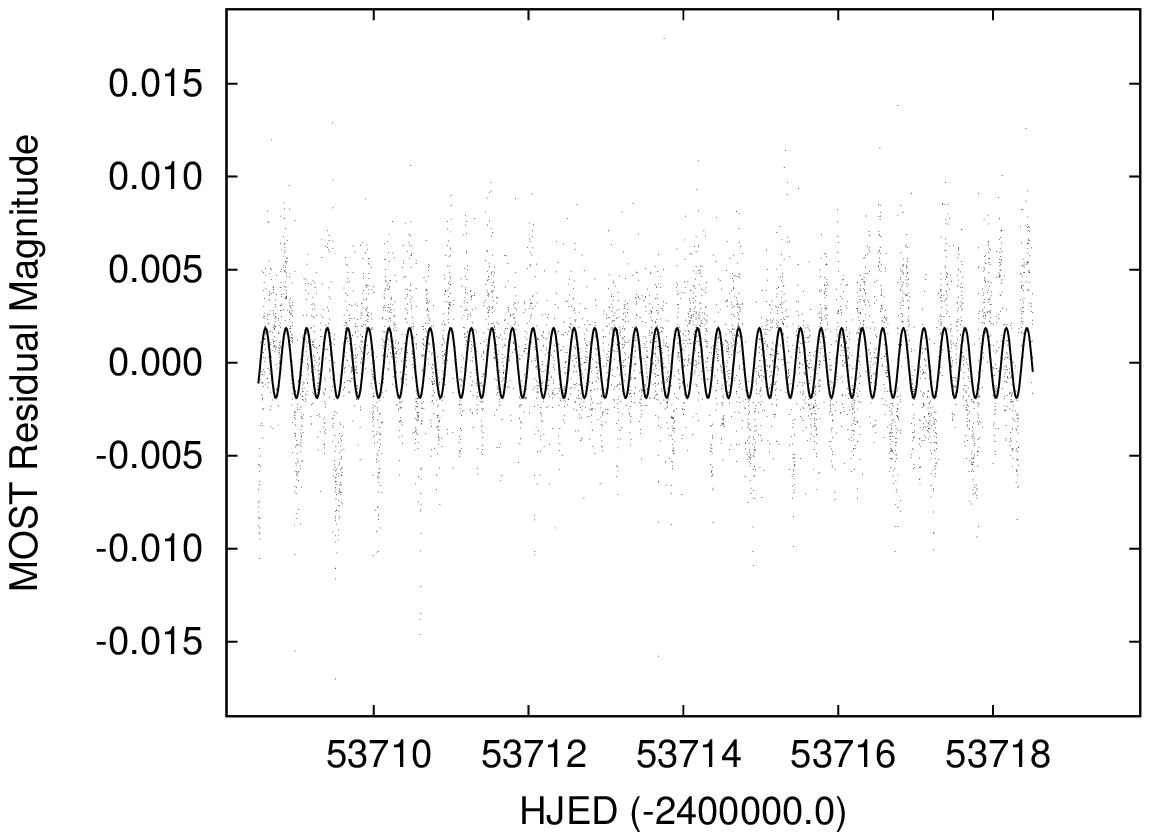}{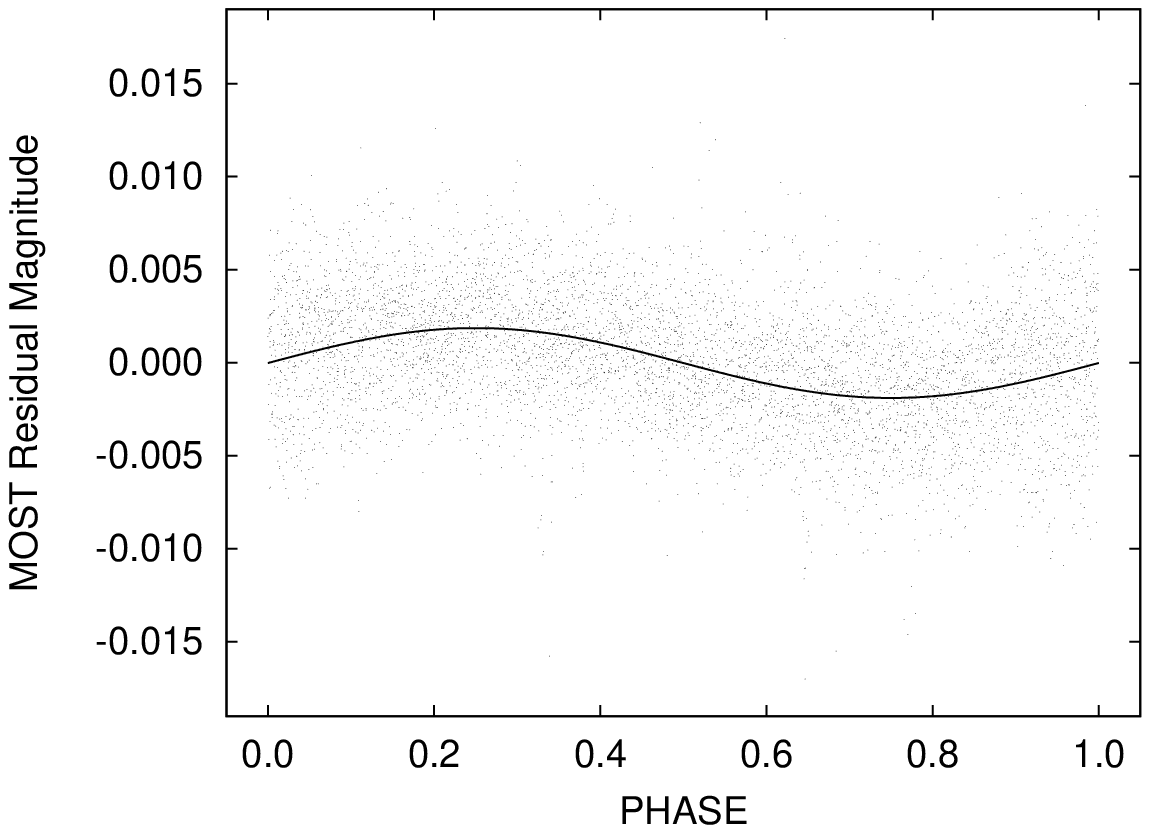}
\caption{Same as Figure \ref{mostres1} but for period $0.^d256040$.} \label{mostres2}
\end{figure}

\begin{figure}
\plottwo{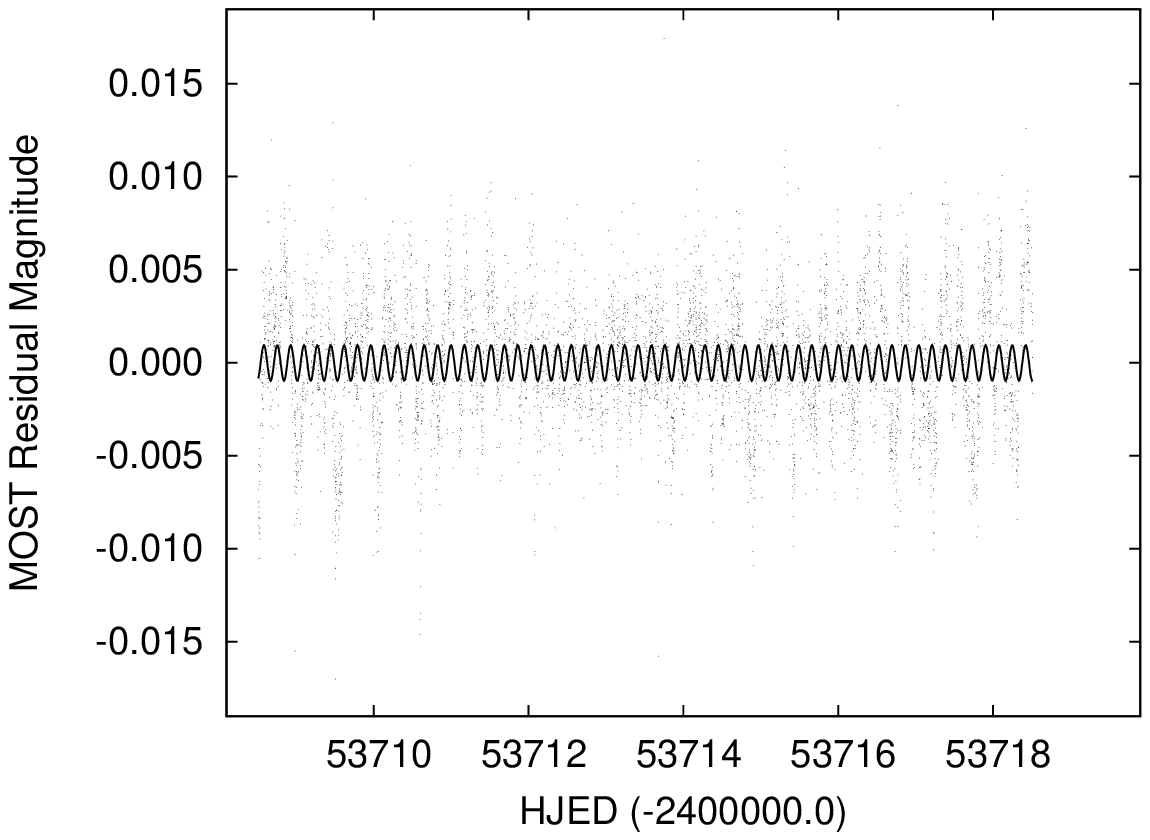}{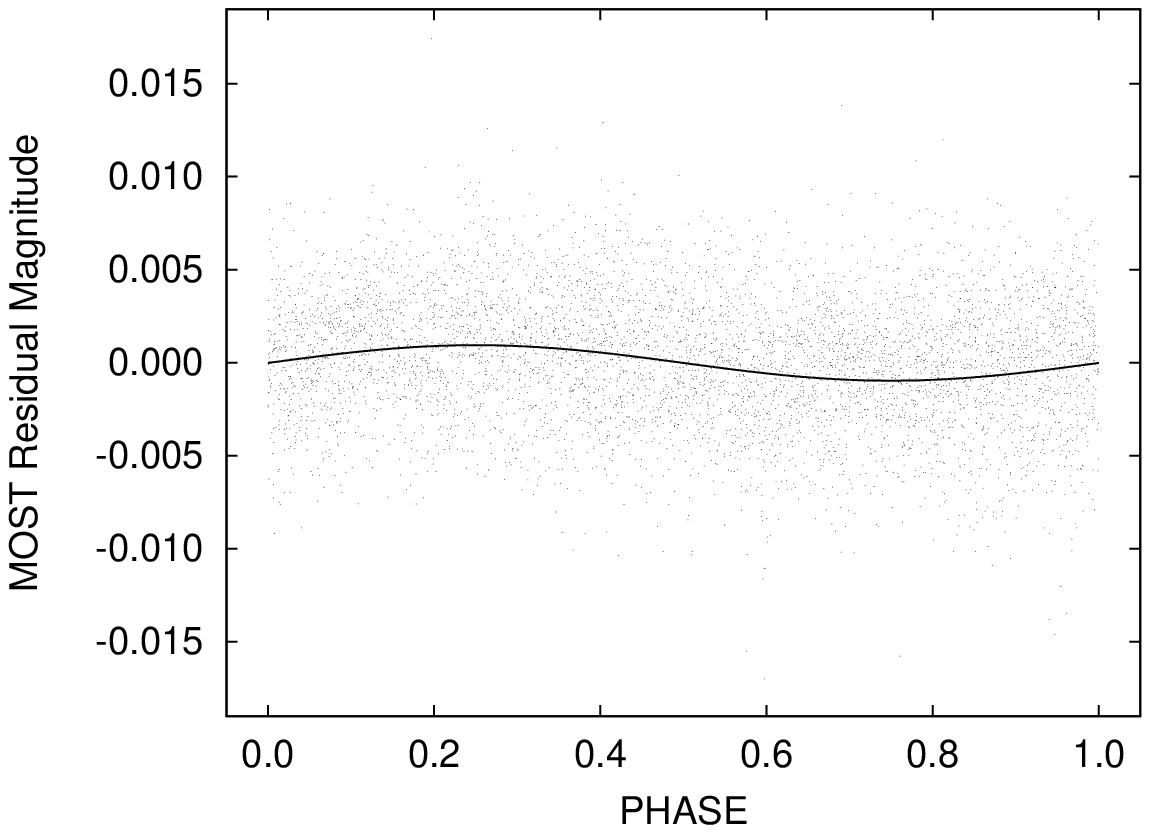}
\caption{Same as Figure \ref{mostres1} but for detected periodicity $0.^d172546$, near 1/3 the orbit period.} \label{mostres3}
\end{figure}

\begin{figure}
\plottwo{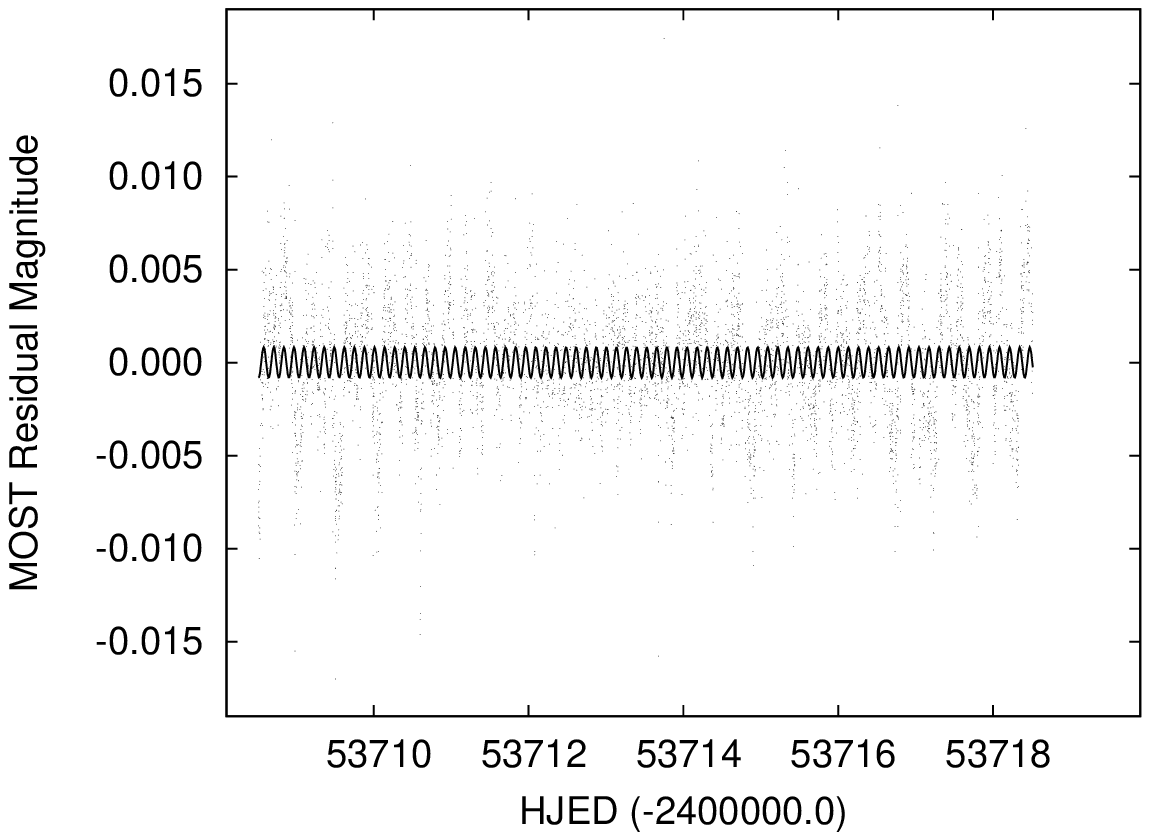}{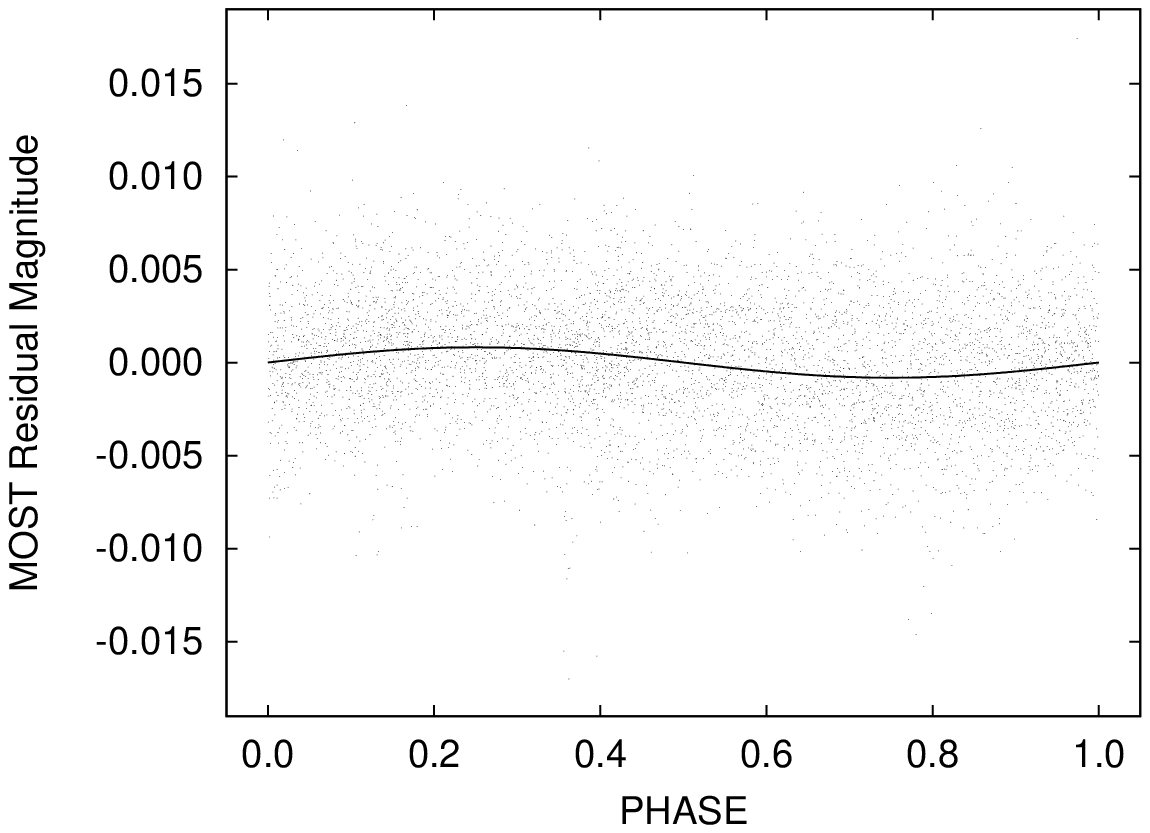}
\caption{Same as Figure \ref{mostres1} but for detected periodicity $0.^d130151$, near 1/4 the orbit period.} \label{mostres4}
\end{figure}

\begin{figure}
\plottwo{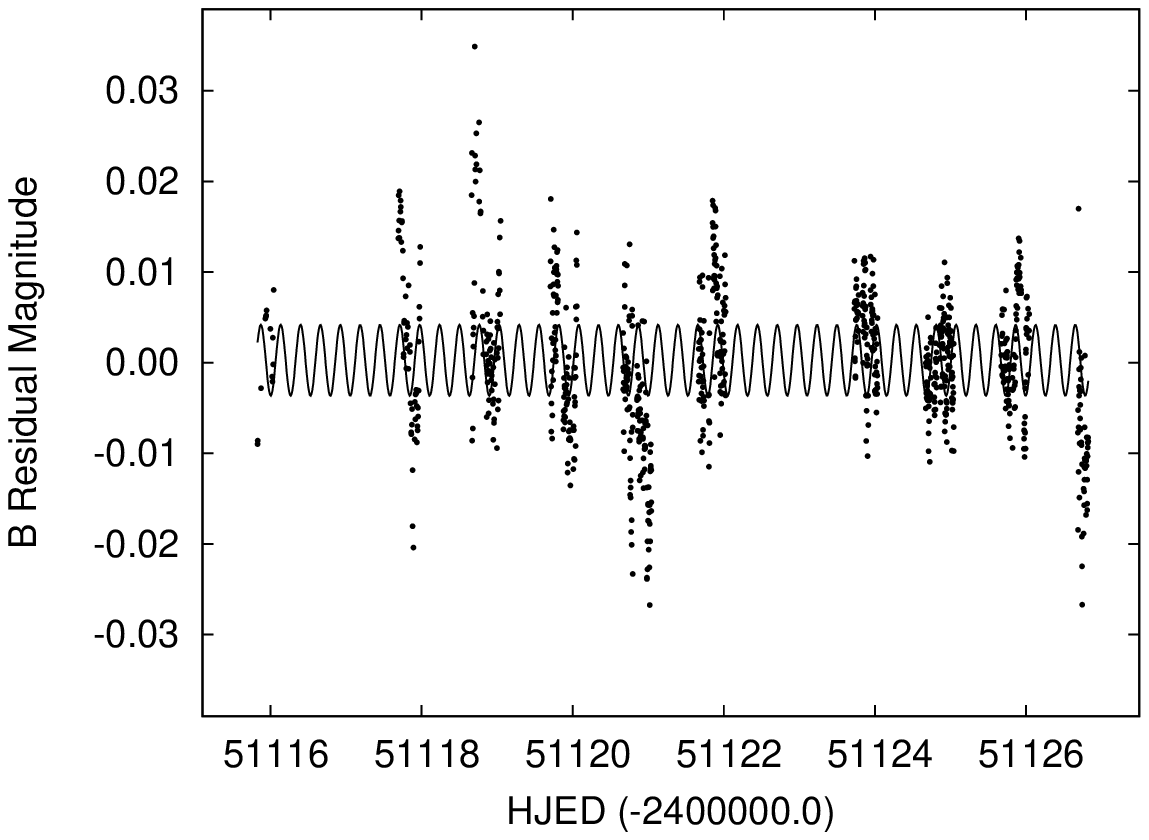}{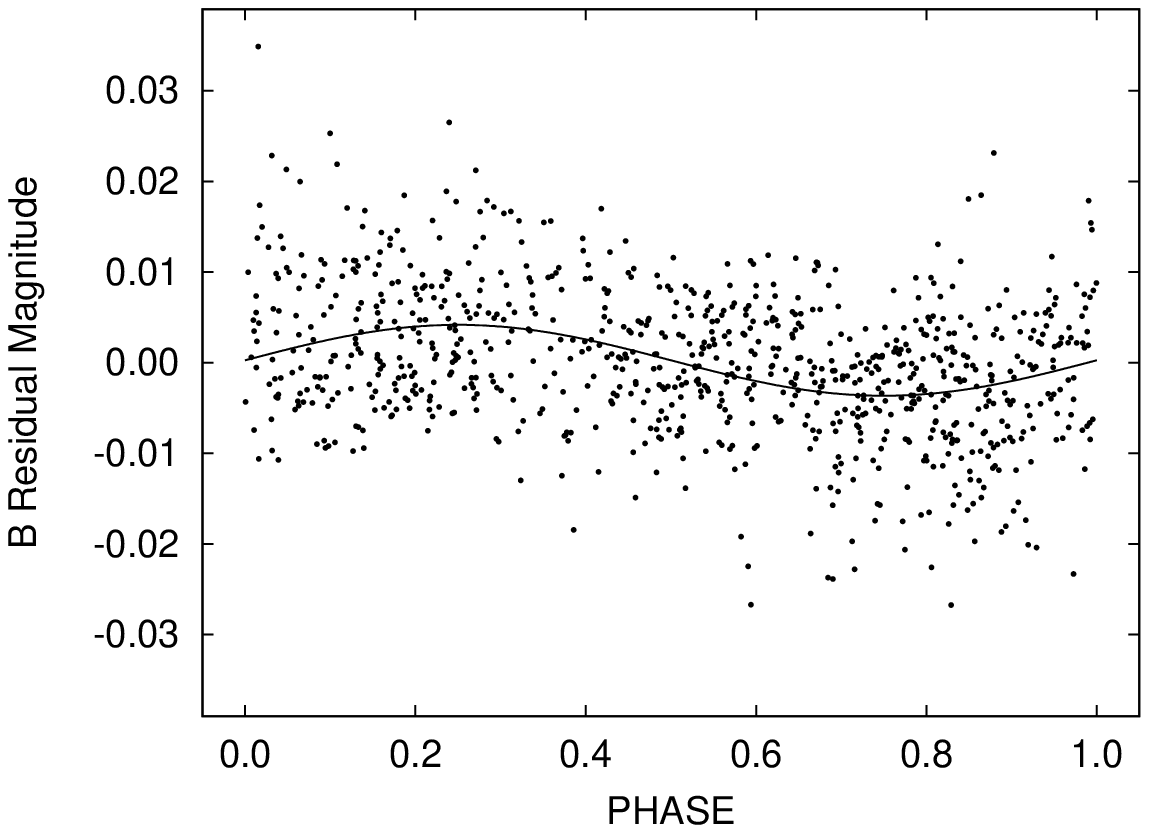}
\caption{$B$ residual magnitude and fitted 0.26294-d period sinusoid, close to half the orbit period.} \label{Bres}
\end{figure}

\begin{figure}
\plottwo{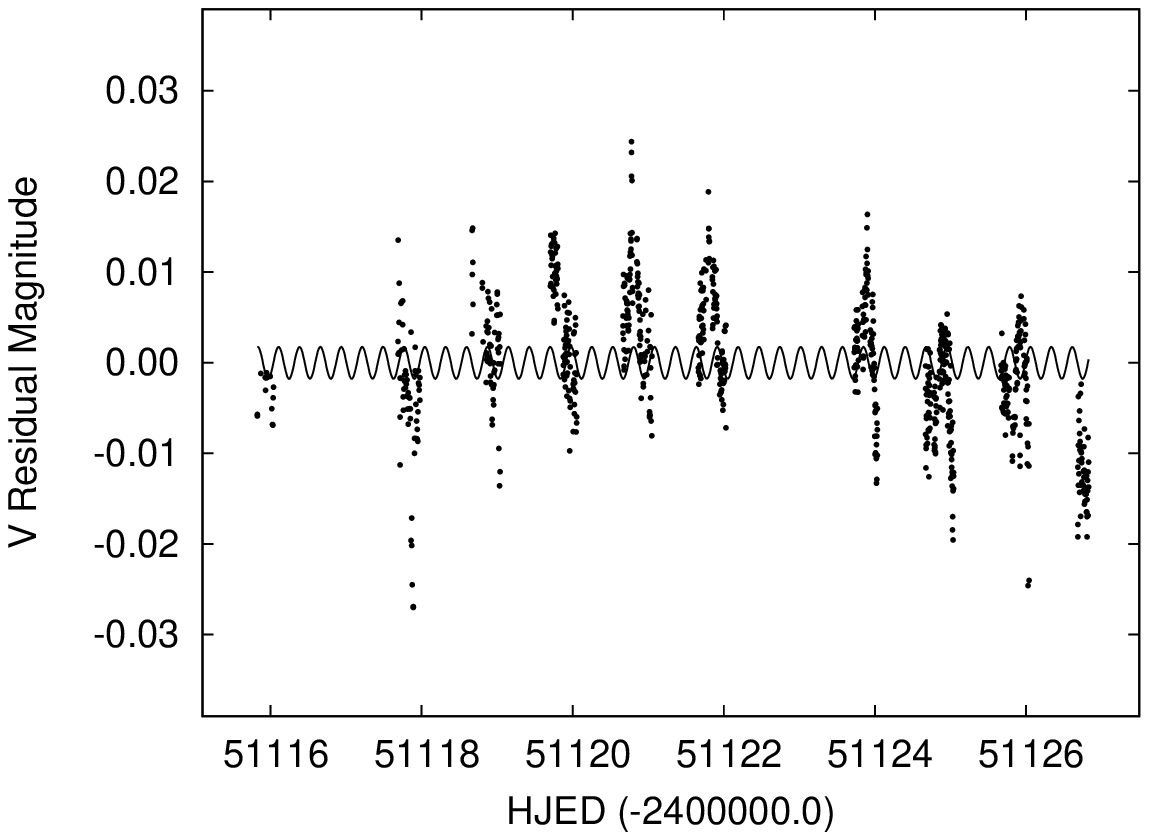}{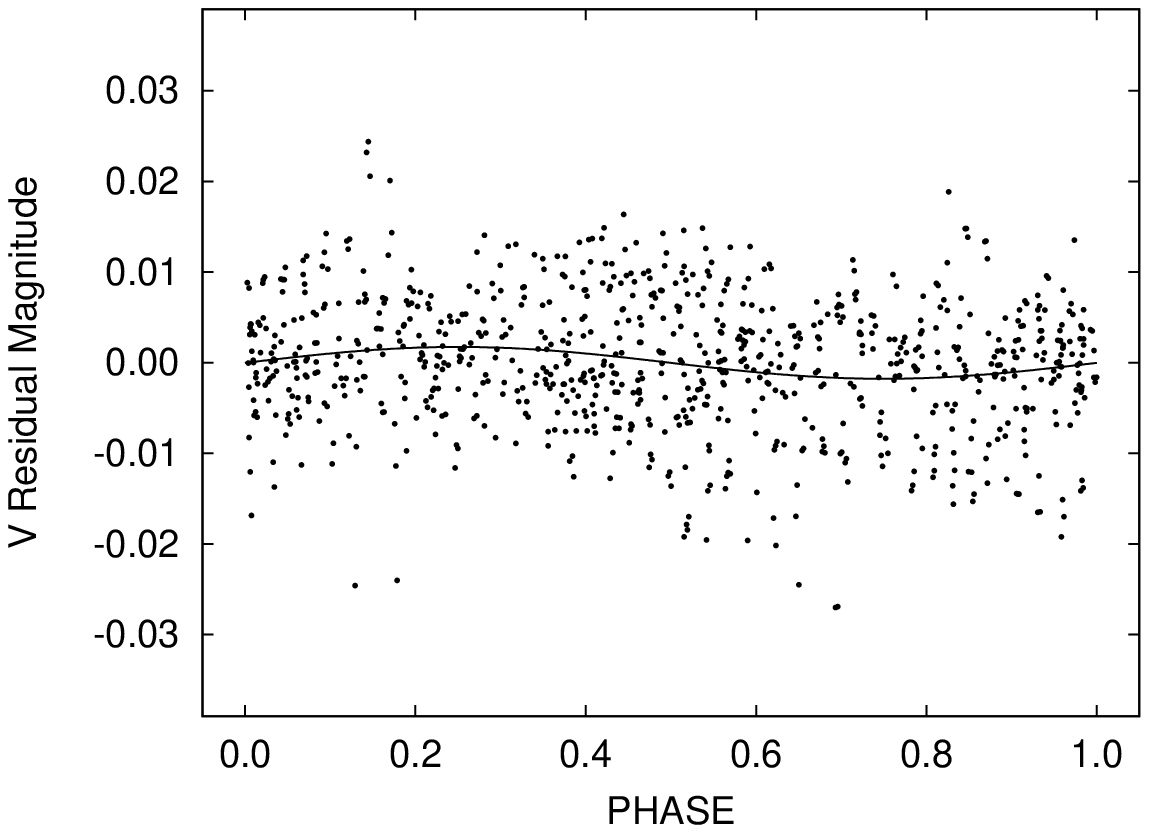}
\caption{Same as Figure \ref{Bres} but for V band and period $0.^d27644$.} \label{Vres}
\end{figure}

\begin{figure}
\plottwo{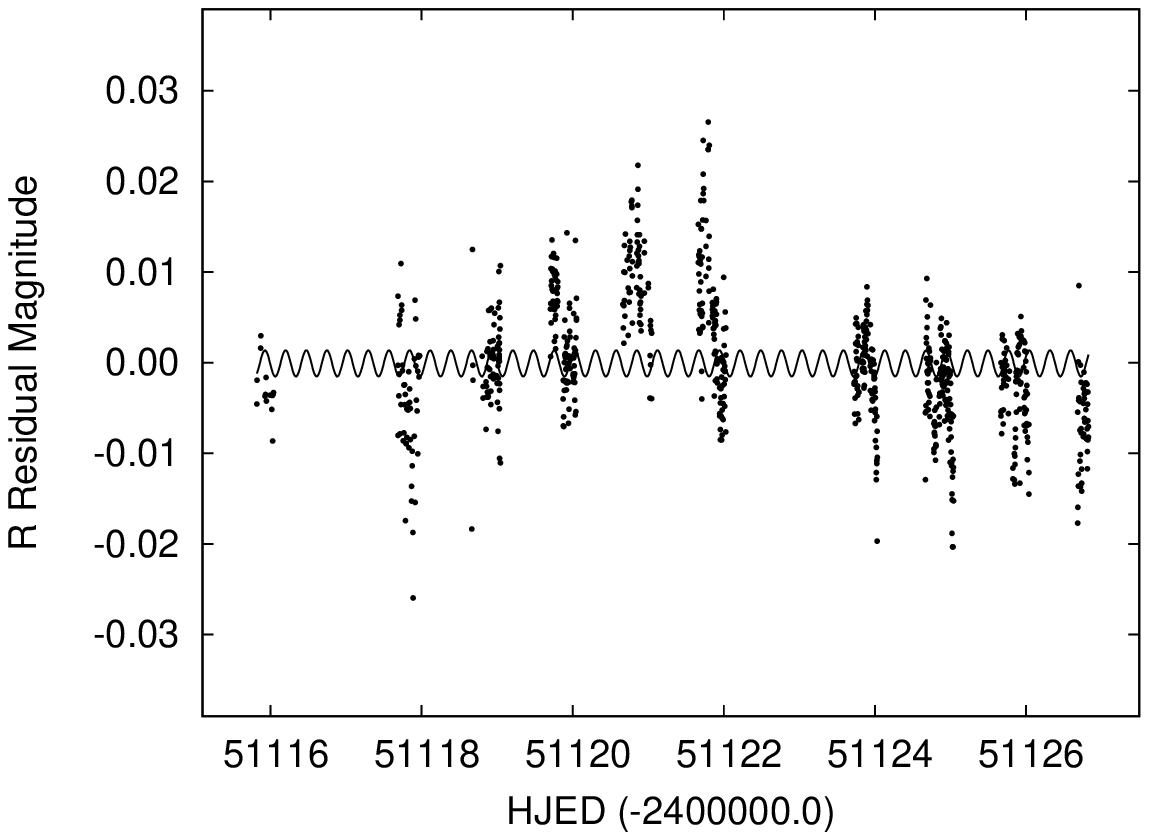}{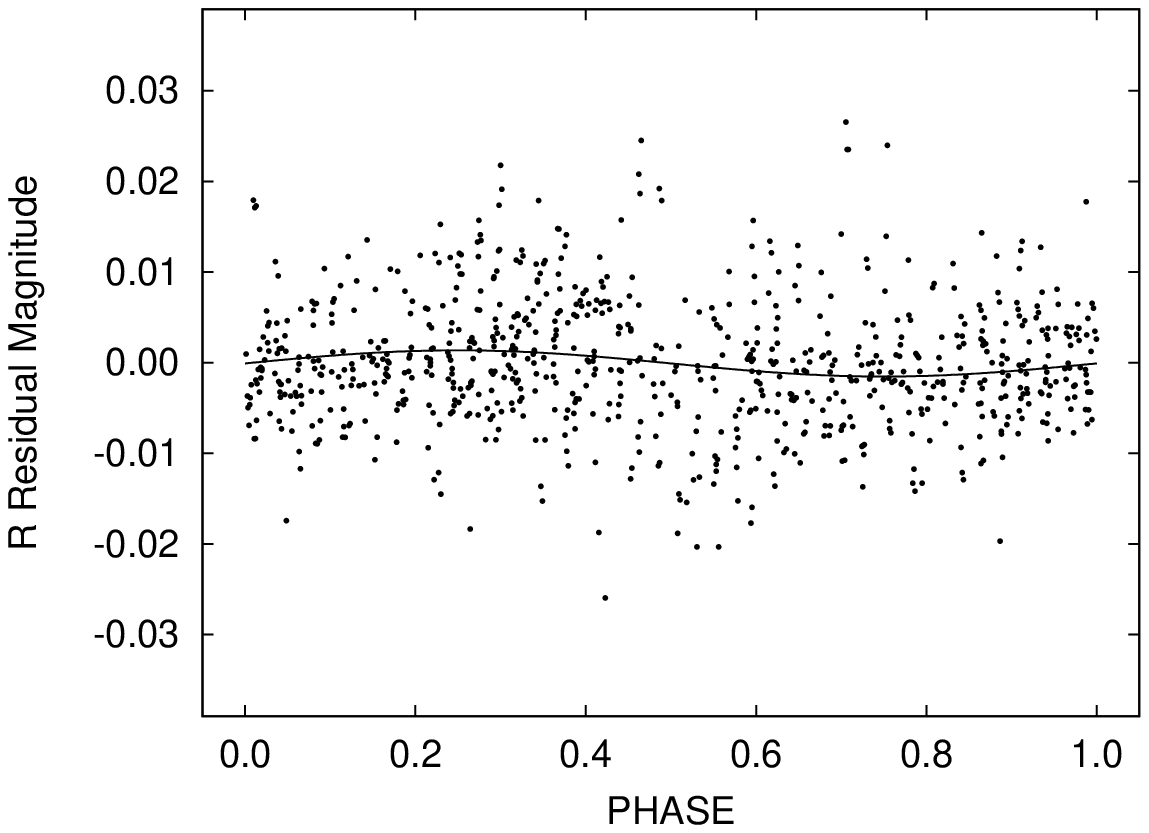}
\caption{Same as Figure \ref{Bres} but for $R_C$ band and period $0.^d2734$.} \label{Rres}
\end{figure}

\begin{figure}
\plottwo{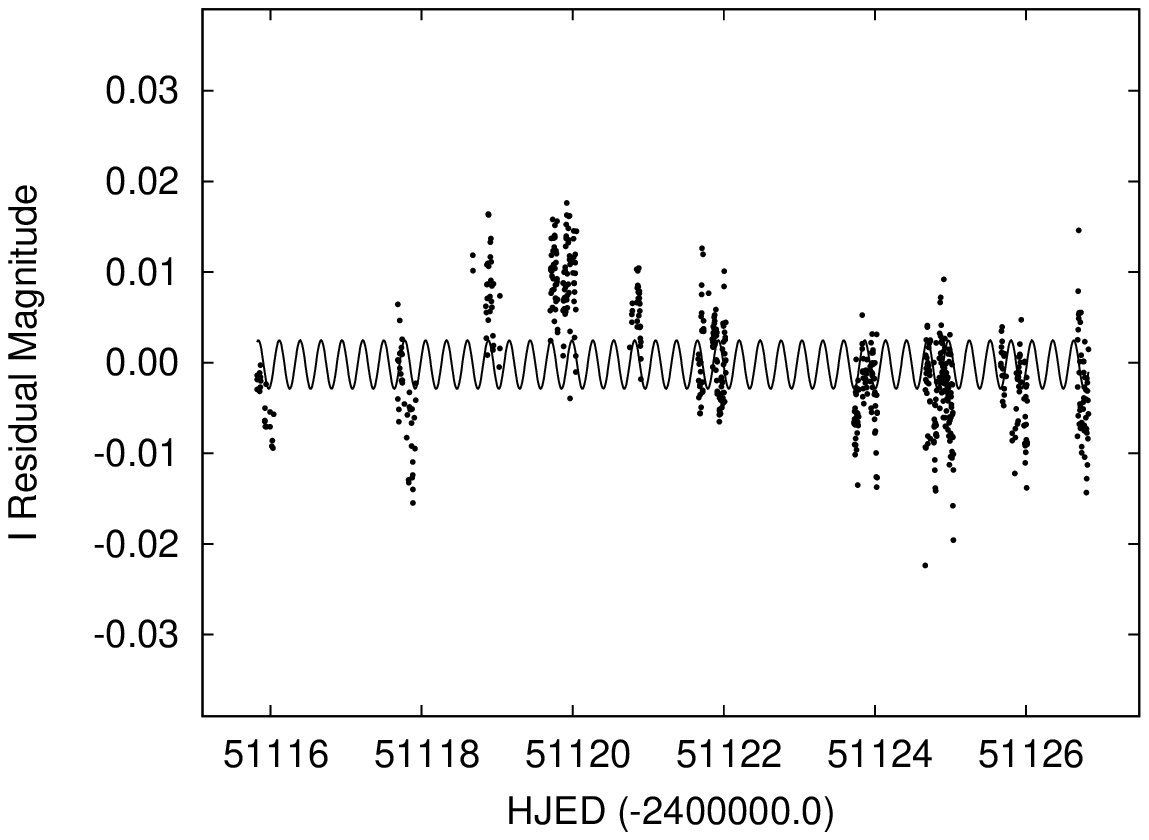}{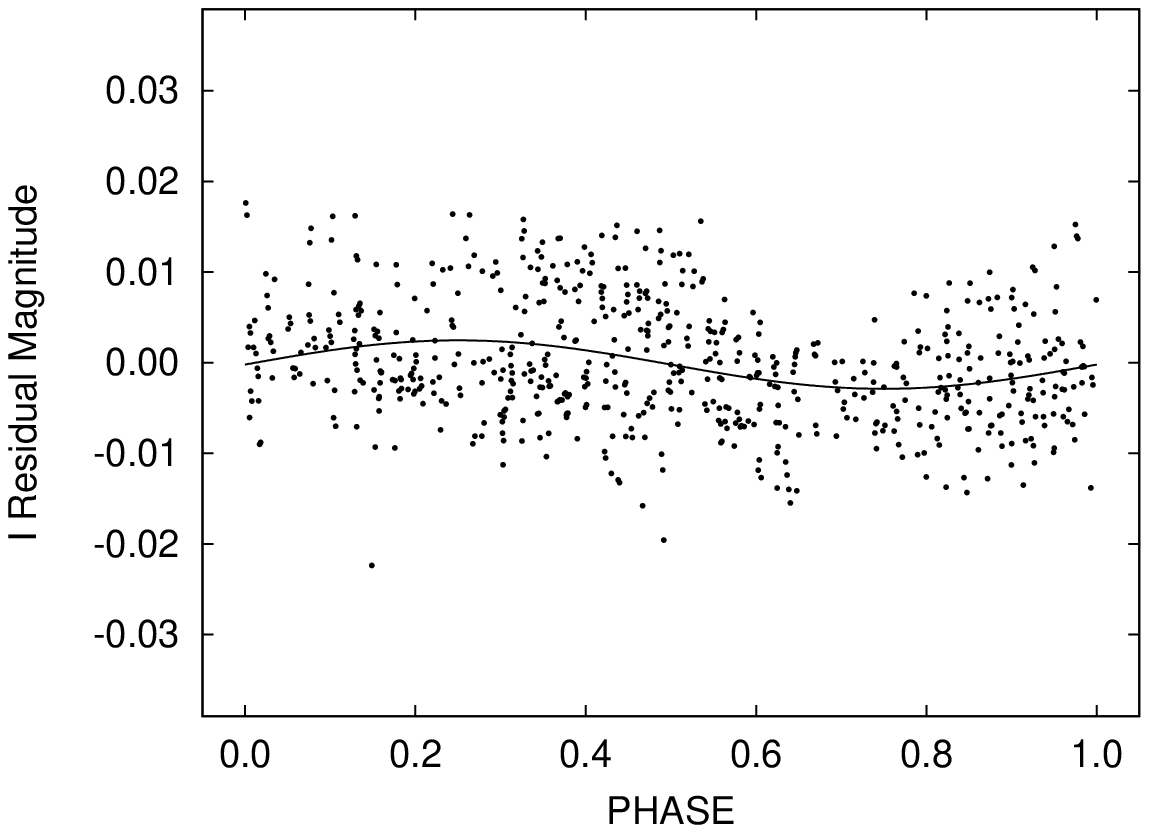}
\caption{Same as Figure \ref{Bres} but for $I_C$ band and period $0.^d27675$.} \label{Ires}
\end{figure}

\begin{figure}
\plottwo{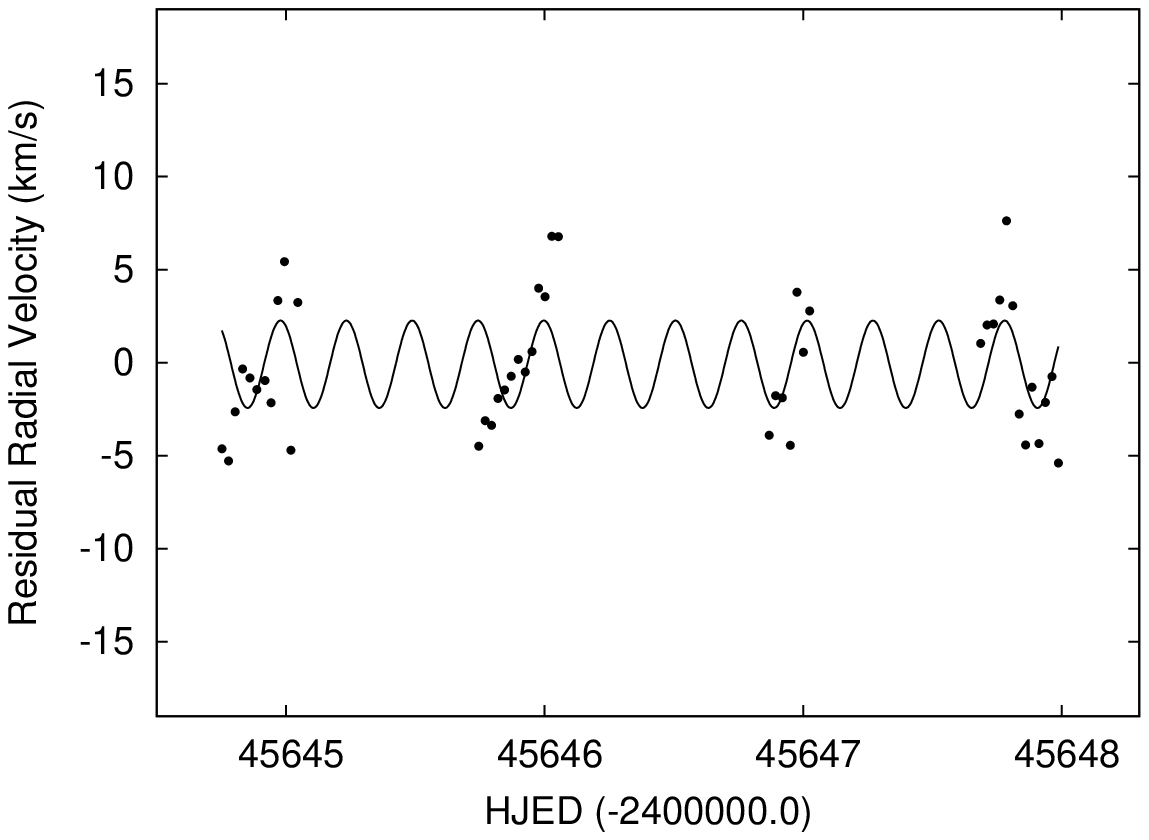}{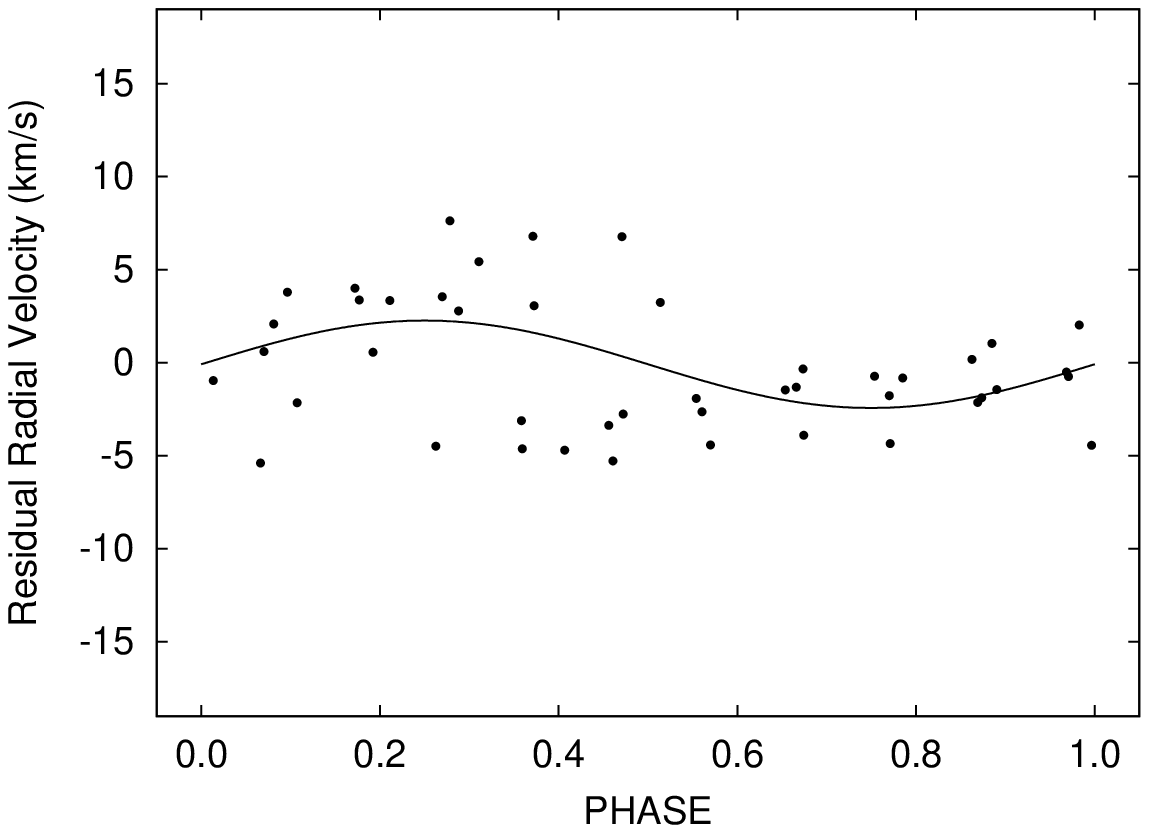} 
\caption{Radial velocity residuals \citep{bois88} and fitted 0.2546-d period sinusoid, close to half the orbit period.} \label{boisres}
\end{figure}

\begin{figure}
\plottwo{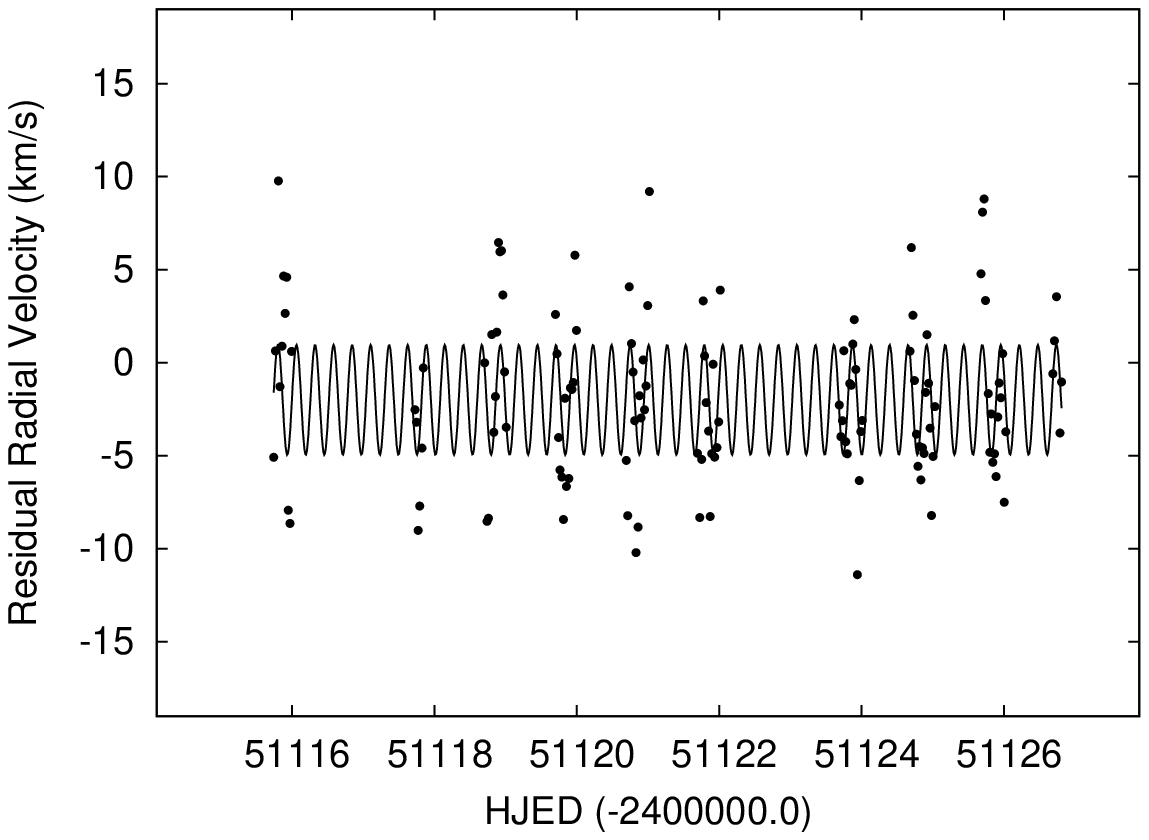}{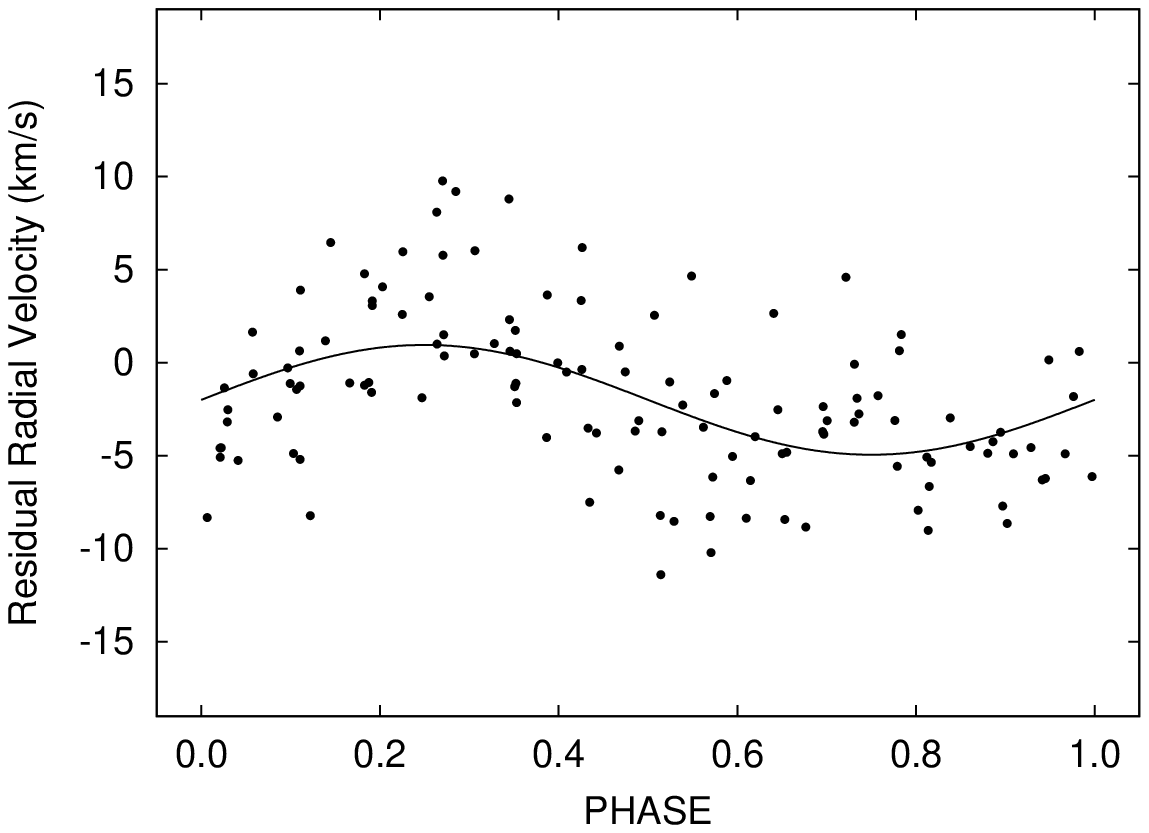}
\caption{Radial velocity residuals (KPNO) and fitted 0.26029-d period sinusoid.} \label{kpnorvres}
\end{figure}

\begin{figure}
\plottwo{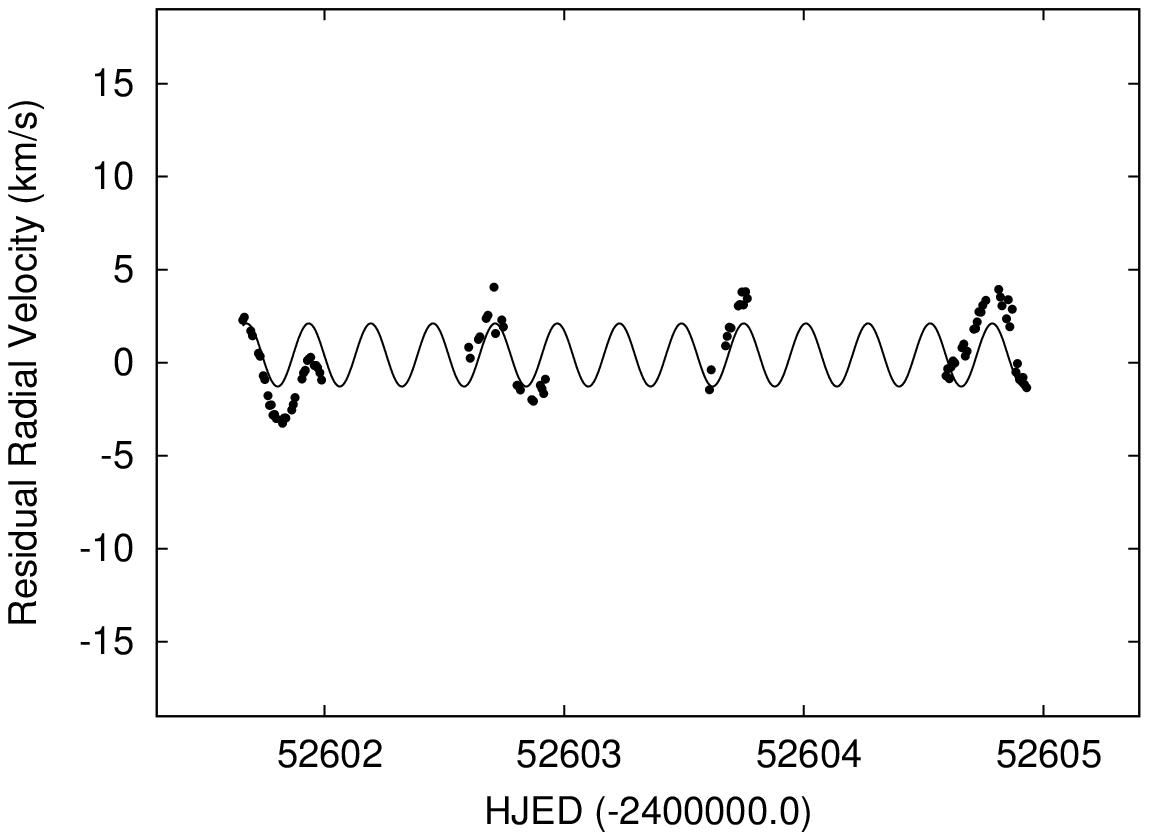}{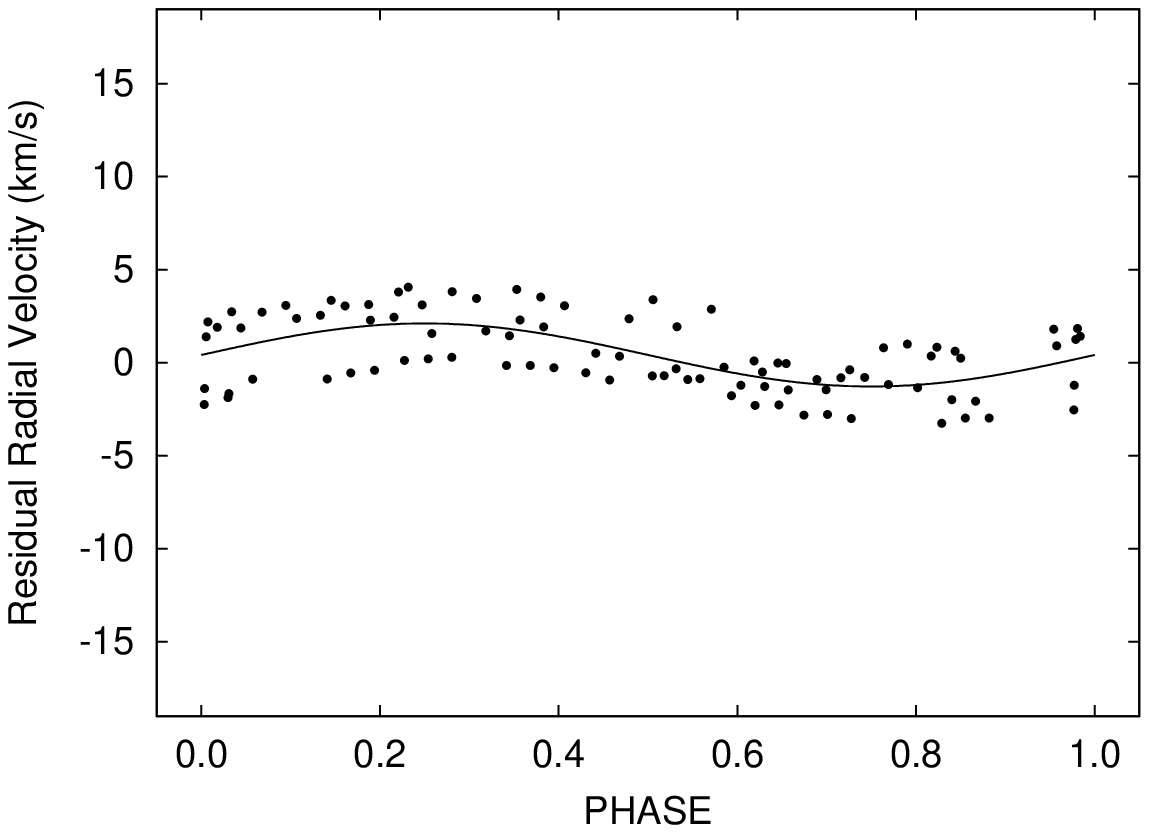}
\caption{Radial velocity residuals \citep{hussain} and fitted 0.2593-d period sinusoid.} \label{hussainres}
\end{figure}

\end{document}